\documentclass[11pt, oneside, reqno]{amsart}   	% use "amsart" instead of "article" for AMSLaTeX format
\usepackage{amsaddr}
\usepackage{geometry}                		% See geometry.pdf to learn the layout options. There are lots.
\usepackage[parfill]{parskip}    		% Activate to begin paragraphs with an empty line rather than an indent
\usepackage{graphicx}				% Use pdf, png, jpg, or eps§ with pdflatex; use eps in DVI mode
\usepackage{amssymb}
\usepackage{amsmath}
\usepackage{amsfonts}
\usepackage{mathbbol}
\numberwithin{equation}{section}
\newcommand\abs[1]{\left|#1\right|}
\title{Mass gap in U(1) Higgs-Yukawa model on a unit lattice}
\author{Abhishek Goswami}
\address{Department of Mathematics, SUNY at Buffalo, Buffalo, NY 14260, USA}
\curraddr{}
\email{goswami3@buffalo.edu}

\date{}							% Activate to display a given date or no date

\begin{document}

\begin{abstract}
A non perturbative proof of the mass generation of fermions via the Higgs mechanism is given. This is done
by showing exponential decay of the two point fermionic correlation function in a weakly coupled
U(1) Higgs-Yukawa theory on a unit lattice in $d=4$. This decay implies that the Higgs boson, the photon and
the fermion all have a non zero physical mass and the theory is said to have a mass gap.
\\
\smallskip
\\
\textbf{Keywords.} Higgs mechanism; correlation functions; exponential decay
\\
\textbf{MSC Class.} 81T08; 81T13; 81T25
\end{abstract}
\maketitle

\section{Introduction}

The recent discovery of the Higgs particle at the CERN Large Hadron Collider (LHC) \cite{ATLAS}, \cite{CMS}
is one of the biggest achievements of the modern physics. It completes the Standard Model of particle physics and 
thus, our understanding of the mass generation of $W^{\pm}$ and \textit{Z} bosons through the Higgs mechanism. 
While the Higgs was first introduced in the Standard Model to explain the mass generation of 
gauge bosons, it is widely accepted that even fermions also get their mass due to interaction with the Higgs via Yukawa coupling.
One of the major success at the LHC since the discovery of the Higgs particle is a direct evidence of Yukawa coupling of the Higgs to 
the top quarks in agreement with the Standard Model (for example, observation of the production of the Higgs 
through $t\bar{t}H$ mechanism \cite{CMS2}). 
However, a rigorous proof of the mass generation of fermions is absent in
the literature. Here, we build upon our work on the mass generation in a weakly coupled U(1) Higgs theory \cite{G1}
and study the fermion field weakly coupled to the Higgs via Yukawa coupling constant. This is known as U(1) Higgs -Yukawa model.

In our earlier work, we revisited the proof of the Higgs mechanism on a unit lattice due to Balaban,
Imbrie, Jaffe and Brydes \cite{BBIJ} by applying a new power series cluster expansion developed by Balaban, Feldman, 
Kn$\ddot{\text{o}}$rrer and Trubowitz \cite{BFKT} and showed the existence of a mass gap for the observable 
electromagnetic field strength $F_{\mu\nu}$. We concluded that our method was conceptually simple and
also provided a clean alternative to decoupling expansions for Gaussian measures (see for example, \cite{AR1}). 
Then we constructed a similar power series expansion for the fermionic systems \cite{G2} 
and applied it to prove exponential decay of the two point fermion correlation function in a massive Gross Neveu
model on a unit lattice. Here, we use the two expansions together and
continue our work on the Higgs mechanism to show the mass generation of fermions. 
We do that by establishing a mass gap in U(1) Higgs -Yukawa model on a unit lattice in $d = 4$.

The strong gauge coupling regime of lattice U(1) Higgs-fermion theories has been extensively studied.
For example, Lee and Shrock \cite{LS1}, \cite{LS2} analytically and numerically investigated chiral phase transition 
in lattice U(1) Higgs-fermion models in $d = 4$ both with and without Yukawa coupling. 
However, it is the weak gauge coupling that is closer to the continuum theory and hence, our results 
in this paper are closer to a continuum model.

\subsection{The model} 
We work on a 4 dimensional Euclidean unit lattice which is a torus $\mathbb{T} = (\mathbb{Z}/L \mathbb{Z})^{4}$, where
\textit{L} is a large positive integer. At every site $x \in \mathbb{T}$, we have a Higgs doublet; $\phi(x) = \phi_{1}(x) + i \phi_{2}(x)$.
A bond \textit{b} is a link between adjacent sites $(x, x + e_{\mu})$. $A_{\mu}(x) = A(x, x+e_{\mu})$ is an oriented U(1) gauge field 
defined on the link $(x, x + e_{\mu})$.  A plaquette $p \in \mathbb{T}$ is a unit square formed by
the bonds $(x, x+e_{\nu}), (x+e_{\nu}, x+e_{\mu}+e_{\nu}), (x+e_{\mu}+e_{\nu}, x+e_{\mu})
\hspace{0.05 cm} \text{and} \hspace{0.05 cm} (x+e_{\mu}, x)$ and $(dA)(p) = \sum_{b \in \partial p} A_{b}$  is the field strength. 
$\mathbb{T}^{\ast}$ and $\mathbb{T}^{\ast\ast}$ denote the set of all bonds and plaquettes respectively.

At every $x \in \mathbb{T}$, 
define independent spinors $\psi_{\alpha}(x)$ and $\bar{\psi}_{\beta}(x)$ where $1 \leqslant \alpha \leqslant 4$
and $1 \leqslant \beta \leqslant 4$ are spinor indices. Here, $\psi_{\alpha}(x)$ and $\bar{\psi}_{\beta}(x)$ form the basis  
of a vector space \textit{V} over field $\mathbb{C}$ with anti-commuting property,
\begin{equation}
\begin{aligned}
\{\psi_{\alpha}(x), \bar{\psi}_{\beta}(x)\} = 0, \hspace{0.5 cm}
\psi_{\alpha}(x) \bar{\psi}_{\beta}(x) = - \bar{\psi}_{\beta}(x) \psi_{\alpha}(x).
\end{aligned}
\end{equation}
$\psi_{\alpha}(x)$ and $\bar{\psi}_{\beta}(x)$ are known as \textit{Grassmann variables}. 
Let $\gamma_{1}, \gamma_{2}, \gamma_{3}, \gamma_{4}$ be Dirac matrices and
let $\gamma_{5} = - \gamma_{1}\gamma_{2}\gamma_{3}\gamma_{4}$. Define projection operators 
\begin{equation}
P_{R} = \frac{1}{2}(\mathbb{1} + \gamma_{5})  \hspace{1 cm} P_{L} = \frac{1}{2}(\mathbb{1} - \gamma_{5})
\end{equation}
such that $P_{L}P_{R} = 0, P_{L} + P_{R} = \mathbb{1}, P_{L}^{2} = P_{L}$ and $P_{R}^{2} = P_{R}$. Let 
(see for example, \cite{Smit})
\begin{equation}
\begin{aligned}
\psi_{L}(x) &= P_{L} \hspace{0.05 cm} \psi(x), \hspace{1 cm} \psi_{R}(x) = P_{R} \hspace{0.05 cm} \psi(x), \\
\bar{\psi}_{L}(x) &= \bar{\psi}(x) \hspace{0.05 cm} P_{R} \hspace{1 cm} \bar{\psi}_{R}(x) = \bar{\psi}(x) \hspace{0.05 cm} P_{L}.
\end{aligned}
\end{equation}
$\psi_{L}(x), \psi_{R}(x), \bar{\psi}_{L}(x)$ and $\bar{\psi}_{R}(x)$ are known as \textit{Weyl} fermions. 
 
While $\psi_{R}(x), \bar{\psi}_{R}(x)$ do not interact with the gauge field $A_{\mu}(x), \psi_{L}(x), \phi(x)$ 
acquire a phase factor $e^{i e_{0} A_{\mu}(x)}$ upon translating along the link $(x, x + e_{\mu})$ 
and $\bar{\psi}_{L}(x)$ acquires a phase factor $e^{- i e_{0} A_{\mu}(x)}$ under the same operation.
$e_{0}$ is the gauge coupling constant and $e_{0} \ll 1$.

\textbf{Definition.} Define covariant Dirac operator
\begin{equation}
\mathfrak{D}_{A} = \gamma \cdot \nabla_{A} - \frac{1}{2} \Delta_{A}, \hspace{0.5 cm} 
\nabla_{A, \mu} = \frac{1}{2} (\partial_{A, \mu} - \partial^{T}_{-A, \mu}), \hspace{0.5 cm} \Delta_{A} = - \partial^{T}_{-A, \mu} \partial_{A, \mu}
\end{equation}
where for some $\psi$,
\begin{equation}
\begin{aligned}
(\partial_{A, \mu} \psi)(x) &= e^{i e_{0} A_{\mu}(x)} \psi(x+e_{\mu}) - \psi(x) \\
(\partial^{T}_{-A, \mu} \psi)(x) &= e^{- i e_{0} A_{\mu}(x)} \psi(x-e_{\mu}) - \psi(x)
\end{aligned}
\end{equation}
and
\begin{equation}
(\mathfrak{D}_{A}\psi) (x) = - \sum_{\mu=1}^{4} \Big[ \Big(\frac{1-\gamma_{\mu}}{2}\Big) e^{i e_{0} A_{\mu}(x)} \psi(x+e_{\mu})
+  \Big(\frac{1+ \gamma_{\mu}}{2}\Big) e^{-i e_{0} A_{\mu}(x)} \psi(x-e_{\mu}) - \psi(x) \Big].  
\end{equation}
Define Dirac operator,
\begin{equation}
\mathfrak{D} = \gamma \cdot \nabla - \frac{1}{2} \Delta, \hspace{0.5 cm} 
\nabla_{\mu} = \frac{1}{2} (\partial_{\mu} - \partial^{T}_{\mu}), \hspace{0.5 cm} \Delta = - \partial^{T}_{\mu} \partial_{\mu}
\end{equation}
where for some $\psi$,
\begin{equation}
\begin{aligned}
(\partial_{\mu} \psi)(x) = \psi(x+e_{\mu}) - \psi(x), \hspace{0.5 cm} (\partial^{T}_{\mu} \psi)(x) =  \psi(x-e_{\mu}) - \psi(x)
\end{aligned}
\end{equation}
and
\begin{equation}
(\mathfrak{D}\psi) (x) = - \sum_{\mu=1}^{4} \Big[ \Big(\frac{1-\gamma_{\mu}}{2}\Big) \psi(x+e_{\mu})
+  \Big(\frac{1+ \gamma_{\mu}}{2}\Big) \psi(x-e_{\mu}) - \psi(x) \Big].  
\end{equation}
The terms $\Delta_{A}$ and $\Delta$ in (1.4) and (1.7) are due to Wilson's method to fix the fermion species doubling phenomenon.

\textbf{Remark 1}. Wilson introduced the Laplacian in the Dirac operator to solve fermion doubling problem. 
However, it is only a problem while taking a continuum limit. Here, we are working on a fixed unit lattice
and hence, fermion doubling is not an issue. The Laplacian makes the Dirac operator more regular. 

\textbf{Action.} The action for U(1) Higgs-Yukawa model is
\begin{equation}
\mathcal{A}(\bar{\psi}, \psi, \phi, A) = S_{NC} (A) + S_{H}(\phi, A) + S_{Y} (\bar{\psi}, \psi, \phi, A). \nonumber
\end{equation}
$S_{NC} (A)$ is the non compact lattice action given by
 \begin{equation}
S_{NC}(A) =\frac{1}{2} \sum_{p \subset \mathbb{T}^{\ast\ast}}\hspace{0.1cm} (dA)^{2}(p)  
\end{equation}   
$S_{H}(\phi, A)$ is the lattice Higgs action given by   
\begin{equation}
S_{H}(\phi,A) = \frac{1}{2} \sum_{x \in \mathbb{T}} \sum_{\mu=1}^{4} |e^{i e_{0}A_{\mu}(x)} \phi(x+e_{\mu}) - \phi(x)|^{2}
  + \sum_{x \in \mathbb{T}} \hspace{0.1 cm}\big(\lambda|\phi(x)|^{4} - \frac{1}{4} \mu^{2}|\phi(x)|^{2} \big)
\end{equation}
where $\lambda$ is the Higgs self coupling constant with $\lambda \ll 1$ and $\mu$ is a fixed positive constant.

$S_{Y} (\bar{\psi}, \psi, \phi, A)$ is the action of lattice fermions interacting with the Higgs particle via Yukawa
coupling constant $g$ and $g \ll 1$. It is given by   
 \begin{equation}
\begin{aligned}
S_{Y} (\bar{\psi}, \psi, \phi, A) &=  \sum_{x \in \mathbb{T}}
\bar{\psi}_{L}(x) \mathfrak{D}_{A} \psi_{L}(x) + \sum_{x \in \mathbb{T}}
\bar{\psi}_{R}(x) \mathfrak{D} \psi_{R}(x)  \\
&+  \sum_{x \in \mathbb{T}} \Big[ g \hspace{0.05 cm} \big(\bar{\psi}_{L}(x) \phi(x)\big) \psi_{R}(x) + 
   g \hspace{0.05 cm} \bar{\psi}_{R}(x) \big(\bar{\phi}(x) \psi_{L}(x)\big) \Big].
\end{aligned}
\end{equation}
\textbf{Remark 2}. We make no attempt to have a chiral invariant theory and we are not trying to
reproduce the situation that occurs in the Standard Model. 
We work with non-chiral fermions. Now due to the interaction vertex $i \bar{\psi} e_{0} A \psi$, 
the radiative corrections generate a fermion mass of order $\mathcal{O}(e_{0}^{2})$.
In our model, $e_{0} \ll 1$ which makes this fermion mass tiny. Whereas the fermion mass
generated due to Yukawa coupling is of $\mathcal{O}(1)$ (Eq. \ref{24}) which is a robust effect.
Our goal here is to show rigorously that this Higgs effect exists.

Thus the action for U(1) Higgs-Yukawa model is
\begin{equation}
\begin{aligned}
\mathcal{A}(\bar{\psi}, \psi, \phi, A) &= \frac{1}{2} \sum_{p \in \mathbb{T}^{\ast\ast}}\hspace{0.1cm} (dA)^{2}(p) 
 + \frac{1}{2} \sum_{x \in \mathbb{T}} \sum_{\mu=1}^{4}
 |e^{i e_{0}A_{\mu}(x)}\phi(x+e_{\mu}) - \phi(x)|^{2} \\ &+ \sum_{x \in \mathbb{T}}
\bar{\psi}_{L}(x) \mathfrak{D}_{A} \psi_{L}(x) + \sum_{x \in \mathbb{T}}
\bar{\psi}_{R}(x) \mathfrak{D} \psi_{R}(x) + \sum_{x \in \mathbb{T}} \hspace{0.1 cm} 
\big(\lambda|\phi(x)|^{4} - \frac{1}{4} \mu^{2}|\phi(x)|^{2} \big) \\ &+ 
 \sum_{x \in \mathbb{T}} \Big[ g \hspace{0.05 cm} \big(\bar{\psi}_{L}(x) \phi(x)\big) \psi_{R}(x) + 
  g \hspace{0.05 cm} \bar{\psi}_{R}(x) \big(\bar{\phi}(x) \psi_{L}(x)\big) \Big].
\end{aligned}
\end{equation}
This is constructed to be gauge invariant.

\textbf{Partition function.} For $x \in \mathbb{T}$ and for $b \in \mathbb{T}^{\ast}$, 
let $d\psi_{\alpha}(x), d\bar{\psi}_{\beta}(x), d\phi_{1}(x), d\phi_{2}(x), dA_{b}$ denote Lebesgue measure on $\mathbb{R}$. 
Denote $\prod_{x}  d\phi_{1}(x)d\phi_{2}(x) = \mathcal{D}\phi$, 
$\prod_{x,\alpha} d\psi_{\alpha}(x) = \mathcal{D}\psi$, $\prod_{x,\beta}  d\bar{\psi}_{\beta}(x)  = \mathcal{D}\bar{\psi}$
and $\prod_{b} dA_{b} = \mathcal{D}A$. Then the non compact (NC) partition function for U(1) Higgs-Yukawa model is given by
\begin{equation}
\mathrm{Z}^{\text{NC}} = \int \mathcal{D}\phi \hspace{0.05 cm} \mathcal{D}A \hspace{0.05 cm} \mathcal{D}\psi
\mathcal{D}\bar{\psi} \hspace{0.1 cm} e^{- \mathcal{A}(\bar{\psi}, \psi, \phi, A) - G(A)}
\end{equation}
where $G(A) = \sum_{x \in \mathbb{T}} \frac{1}{2} (\delta A)^{2} (x) + \text{const}$ is the gauge fixing term
and $\delta$ is the adjoint of differential operator \textit{d} with respect to $l^{2}$ inner product.
Gauge fixing is needed to make the integral (1.14) converge.
See the work on the Higgs mechanism \cite{G1}, \cite{BBIJ} for the treatment of $G(A)$.
We make use of the equivalence of periodic compact (i.e. gauge field integral over the interval 
$\big[-\frac{\pi}{e_{0}}, \frac{\pi}{e_{0}}\big]$) formalism and non compact formalism \cite{BBIJ}. 
The equivalence is  
\begin{equation}
\mathrm{Z}^{\text{NC}} = \Big(\frac{2\pi}{e_{0}}\Big)^{-|\mathbb{T}| + 1} \mathrm{Z}^{\text{C}} \nonumber
\end{equation}
and rewrite the partition function (see \cite{G1}, \cite{BBIJ} for details)   
\begin{equation}
\begin{aligned}
\mathrm{Z}^{\text{NC}} = \Big(\frac{2\pi}{e_{0}}\Big)^{-|\mathbb{T}| + 1}  \sum_{v : dv=0} \int^{\frac{\pi}{e_{0}}}_{-\frac{\pi}{e_{0}}}
 \mathcal{D}A \hspace{0.05 cm} \int \mathcal{D}\phi \mathcal{D}\psi \mathcal{D}\bar{\psi}
 \hspace{0.1 cm}  e^{ -\frac{1}{2} \sum_{p}|(dA + v)(p)|^{2}  - S_{H}(\phi, A) -  S_{Y}(\bar{\psi}, \psi, \phi, A)}.
\end{aligned}
\end{equation}
The curvature $dA$ acquires a dependence on vortex field $v \in \frac{2\pi}{e_{0}} \mathbb{Z}$.   
 
\textbf{Truncated correlation.}
To study the fermion-fermion truncated correlation function, first we need to define a generating functional 
using some external sources. At every $x \in \mathbb{T}$, 
let $\{\text{J}_{\alpha}, \bar{\text{J}}_{\beta}\}$ be basis for a vector space $V^{\prime}$ over field $\mathbb{C}$
such that
\begin{equation}
\begin{aligned}
&\{\psi_{\alpha}(x), \bar{\psi}_{\alpha}(x)\} = 0, \hspace{0.6 cm} 
\{\text{J}_{\alpha}(x), \bar{\text{J}}_{\alpha}(x)\} = 0, \hspace{0.6 cm} \{\psi_{\alpha}(x), \text{J}_{\alpha}(x)\} = 0, \\
& \{\bar{\psi}_{\alpha}(x), \bar{\text{J}}_{\alpha}(x)\} = 0, \hspace{0.6 cm}
\{\bar{\psi}_{\alpha}(x), \text{J}_{\alpha}(x)\} = 0, \hspace{0.6 cm} \{\psi_{\alpha}(x), \bar{\text{J}}_{\alpha}(x)\} = 0. 
\end{aligned}
\end{equation}
Here, $\{\text{J}_{\alpha}, \bar{\text{J}}_{\beta}\}$ are external sources. Denote
\begin{equation}
\langle \bar{\text{J}}, \psi \rangle = \sum_{\alpha, x} \bar{\text{J}}_{\alpha}(x) \psi_{\alpha}(x) \hspace{1 cm}
 \langle \bar{\psi}, \text{J} \rangle = \sum_{\alpha, x} \bar{\psi}_{\alpha}(x) \text{J}_{\alpha}(x)    \nonumber
\end{equation}
The generating functional is given by  
\begin{equation}
\begin{aligned}
\mathrm{Z} [\text{J}, \bar{\text{J}}] = & \Big(\frac{2\pi}{e_{0}}\Big)^{-|\mathbb{T}| + 1} \sum_{v : dv=0}
 \int^{\frac{\pi}{e_{0}}}_{-\frac{\pi}{e_{0}}} \mathcal{D}A \hspace{0.05 cm} \int \mathcal{D}\phi 
\hspace{0.05 cm} \mathcal{D}\psi \mathcal{D}\bar{\psi} \\
 & e^{ -\frac{1}{2} \sum_{p}|(dA + v)(p)|^{2}  - S_{H}(\phi, A) -  S_{Y}(\bar{\psi}, \psi, \phi, A)} \hspace{0.05 cm}
e^{e_{0} \langle \bar{\psi}, \text{J} \rangle +  e_{0} \langle \bar{\text{J}}, \psi \rangle}
\end{aligned}
\end{equation}
The truncated two point fermion correlation function is defined as  
\begin{equation}
\begin{aligned}
\langle \bar{\psi}_{\beta}(x_{2}) \psi_{\alpha}(x_{1}) \rangle - 
\langle  \bar{\psi}_{\beta}(x_{2})\rangle \langle \psi_{\alpha}(x_{1}) \rangle
&= \langle \bar{\psi}_{\beta}(x_{2}) \psi_{\alpha}(x_{1})\rangle^{T}      \\
&= \frac{1}{e_{0}^{2}} \left. \frac{\delta}{\delta \bar{\text{J}}_{\alpha}(x_{1})} \text{log}\hspace{0.05 cm} 
\mathrm{Z} [\text{J},\bar{\text{J}}] \frac{\delta}{\delta \text{J}_{\beta}(x_{2})} \right\rvert_{\text{J} = \bar{\text{J}} = 0} 
\end{aligned}
\end{equation}
 
\subsection{Classical Higgs mechanism}
Note that in the action (1.13), there are no explicit mass terms for the fields $\phi, A, \psi, \bar{\psi}$ 
(a mass term for fermion is like $\bar{\psi} \psi = \bar{\psi}_{L} \psi_{R} + \bar{\psi}_{R} \psi_{L}$).
First we discuss the classical appearance of the mass terms in the action. Classical mass refer to the mass term
in the zeroth order of perturbation theory, i.e., without quantum corrections. 
Since, $\mathcal{A}(\psi, \bar{\psi}, \phi, A)$ is gauge invariant (U(1) symmetry), change variables as
\begin{equation}
\begin{aligned}
& \phi \rightarrow e^{i\theta} \rho, \hspace{1 cm} \bar{\phi} \rightarrow e^{-i\theta} \rho, \hspace{1 cm} 
 A \rightarrow A - \frac{d\theta}{e_{0}}, \\
& \psi_{L} \rightarrow e^{i\theta} \psi_{L},  \hspace{1 cm}  \bar{\psi}_{L} \rightarrow e^{-i\theta} \bar{\psi}_{L}, 
 \hspace{1 cm}  \psi_{R} \rightarrow \psi_{R}, \hspace{1 cm} \bar{\psi}_{R} \rightarrow \bar{\psi}_{R}.
\end{aligned}
\end{equation}
Let $\psi_{L}^{\theta}, \bar{\psi}_{L}^{-\theta}, \phi^{\theta}, \bar{\phi}^{-\theta}, A^{\theta}$ denote the gauge transformed 
variables. The action (1.13) is independent of $\theta$. Since, 
$\mathfrak{D}_{A^{\theta}} \psi_{L}^{\theta} = (\mathfrak{D}_{A} \psi_{L})^{\theta}$ and
$\bar{\psi}_{L}^{-\theta} (\mathfrak{D}_{A} \psi_{L})^{\theta} = \bar{\psi}_{L} \mathfrak{D}_{A} \psi_{L}$,
$d(A^{\theta}) = dA$ since $d^{2} \theta = 0$ as $\theta$ is a 0-form and due to the absolute value of the Higgs. 

The expectation of the Higgs is 
\begin{equation}
\langle \phi(x) \rangle = \int \mathcal{D} \psi \mathcal{D}\bar{\psi}
\int \mathcal{D}A  \int d\rho \rho(x) e^{- \mathcal{A}}  \int_{- \pi}^{\pi} d\theta \hspace{0.05 cm}\rho(x) e^{i\theta(x)} = 0.
\end{equation}
\textbf{Remark 3}. In physics literature, the expectation of the Higgs is taken to be non zero to explain the Higgs 
mechanism. This is because the mechanism is explained using spontaneous symmetry breaking (SSB) of the 
U(1) symmetry in the action $\mathcal{A}$. After SSB, a specific value (unitary gauge) is chosen for $\theta(x)$  to 
eliminate the unphysical angular degree of freedom (Goldstone boson). The remaining massive degree of freedom
$\rho(x)$ is then referred to as the Higgs. In our model we do not assume SSB instead we average over all the 
possible values of $\theta(x)$ which is $(2\pi)^{|\mathbb{T}|}$ and drop this constant. 

Higgs potential in the new variables is
\begin{equation}
V(\rho) = \lambda \rho(x)^{4} - \frac{1}{4} \mu^{2} \rho(x)^{2}
\end{equation}
with minimum at $\rho_{0} =  \frac{\mu}{\sqrt{8\lambda}}$.
The  weight factor $e^{- V(\rho)}$ pushes the measure $d\rho$ towards $\rho_{0}$. Since, $\mu$ is fixed and $\lambda$ is
extremely small, this forces $\rho_{0}$ away from 0. Thus we expand $\mathcal{A}$ about $\rho_{0}$ and get a mass term 
for gauge field $m_{A} = e_{0} \hspace{0.05 cm} \rho_{0}$ and a mass term for fermion $m_{f} = g \hspace{0.05 cm} \rho_{0}$.
Substituting $\rho \rightarrow \rho_{0} + \rho$ the total action is
\begin{equation}
\begin{aligned}
 \mathcal{A}(\psi, \bar{\psi}, \rho, A) =  \mathcal{S}(\rho, A) &+ \sum_{x \in \mathbb{T}}
\bar{\psi}_{L}(x) \mathfrak{D}_{A} \psi_{L}(x) + \sum_{x \in \mathbb{T}} \bar{\psi}_{R}(x) \mathfrak{D} \psi_{R}(x) \\
&+  \sum_{x \in \mathbb{T}} \hspace{0.05 cm} g \hspace{0.05 cm} (\bar{\psi}_{L}(x) \rho(x) \psi_{R}(x) + \bar{\psi}_{R}(x) \rho(x) \psi_{L}(x)) \\
&+  \sum_{x \in \mathbb{T}} \hspace{0.05 cm} m_{f} \hspace{0.05 cm} (\bar{\psi}_{L}(x) \psi_{R}(x) + \bar{\psi}_{R}(x) \psi_{L}(x))
\end{aligned}
\end{equation}
where   
\begin{equation}
\begin{aligned}
\mathcal{S}(\rho, A) &= \frac{1}{2} \sum_{p \in \mathbb{T}^{\ast\ast}} (dA + v)^{2}(p) + \frac{1}{2}m_{A}^{2}
\sum_{b \in \mathbb{T}^{\ast}}A_{b}^{2}  
+ \frac{1}{2}  \sum_{x \in \mathbb{T}} \sum_{\mu=1}^{4} (\rho(x+e_{\mu}) - \rho(x))^{2} + \frac{1}{2}\mu^{2}\sum_{x \in \mathbb{T}} \rho^{2}(x) \\
&+ \rho_{0}^{2}  \sum_{x \in \mathbb{T}} \sum_{\mu=1}^{4} (1 -\text{cos}\hspace{0.05 cm} e_{0}A_{\mu}(x)
 - \frac{1}{2}e_{0}^{2}A_{\mu}(x)^{2})
+ \rho_{0} \sum_{x \in \mathbb{T}} \sum_{\mu=1}^{4} (\rho(x+e_{\mu}) + \rho(x))(1 - \text{cos}\hspace{0.05 cm} e_{0}A_{\mu}(x)) \\ 
&+ \sum_{x \in \mathbb{T}} \sum_{\mu=1}^{4} \rho(x+e_{\mu})\rho(x)(1 - \text{cos}\hspace{0.05 cm} e_{0}A_{\mu}(x))  
+ \sum_{x \in \mathbb{T}}\Big(\lambda \rho^{4}(x) + \sqrt{2\lambda} \mu\rho^{3}(x) - \text{log}\left[ 1+\frac{\rho(x)}{\rho_{0}}\right]\Big). 
\end{aligned}
\end{equation}   
The gauge field and fermion fields are now massive with (classical) masses
\begin{equation}
\label{24}
m_{A} = \frac{\mu \hspace{0.05 cm} e_{0}}{\sqrt{8\lambda}},  \hspace{1 cm}
m_{f} = \frac{\mu \hspace{0.05 cm} g}{\sqrt{8\lambda}}.
\end{equation}
There is also a mass term for the Higgs 
\begin{equation}
m_{H} = \mu.  \nonumber
\end{equation}
This is the classical Higgs mechanism.

\subsection{Main result} After the above transformations (1.19), (1.22), rewrite the 
the generating functional (1.17) as 
\begin{equation}
\begin{aligned}
\mathrm{Z} [\text{J}, \bar{\text{J}}] =  \Big(\frac{2\pi}{e_{0}}\Big)^{-|\mathbb{T}| + 1} \sum_{v : dv=0}
 \int^{\frac{\pi}{e_{0}}}_{-\frac{\pi}{e_{0}}} \mathcal{D}A \hspace{0.05 cm} \int_{-\rho}^{\infty} \mathcal{D}\rho 
\int \mathcal{D}\psi \mathcal{D}\bar{\psi} \hspace{0.05 cm}
  e^{- \mathcal{A}(\psi, \bar{\psi}, \rho, A) + e_{0}\langle \bar{\psi}, \text{J} \rangle + e_{0}\langle \bar{\text{J}}, \psi \rangle}
\end{aligned}
\end{equation}
\textbf{Theorem 1} Given a U(1) Higgs-Yukawa model on a unit lattice (1.25) with masses $\mu, m_{A}, m_{f}$ fixed  
and $\text{min} (\mu, m_{A}) > 4\sqrt{2}$, $m_{f} > 12$. Let the coupling constants $e_{0}, g$ and $\lambda$ 
be sufficiently weak depending on the masses with $g = \mathcal{O}(e_{0})$ and $ \frac{e_{0}^{2}}{\lambda} \sim \mathcal{O}(1)$. 
Then there exist positive constants \textit{c} and \textit{m} such that the correlation of fermions on two sites 
$x$ and $y$ has an exponential decay as
\begin{equation}
\begin{aligned}
\langle \bar{\psi}_{\beta}(x) \psi_{\alpha}(y)\rangle^{T} \leqslant c \hspace{0.05 cm} e^{- m |x-y|}.     \nonumber
\end{aligned}
\end{equation}

\textbf{Remark 4}. This decay implies that every particle in the theory has a physical mass at least as big as $m$,
i.e. mass generation has actually taken place. This is known as the mass gap. Note that in a perturbative approach to the 
mass generation, a statement about the physical mass can only be made up to a specific order of quantum correction
in the perturbation theory. Whereas proving a mass gap existence settles the question of mass generation for
this lattice model.   

\textbf{Remark 5}. For mass generation of fermions establishing a mass gap in a Higgs-Yukawa theory 
is not enough. Higgs field is a doublet with two degrees of freedom. To see Higgs phenomenon Higgs must be
coupled to a gauge field. The presence of the gauge field (hence the U(1) symmetry) 
is needed to suppress the angular degree of freedom (Goldstone mode).

\textbf{Outline of proof}. The structure of the proof is same as our proof of the Higgs mechanism \cite{G1}.
We begin by expanding the lattice into regions where all the fields are bounded and regions where they are not.
We call these regions as small field  region and large field region respectively. Note that the criterion for boundedness only applies 
to bosonic fields. Then in the small field region we proceed to construct a localized integral having Gaussian measure
with unit covariance and potential terms. As emphasized in \cite{G1} this procedure does not include any decoupling expansion 
for the Gaussian measure. Instead we make some change of variables and bring the covariance to the potential
term. This is done in section 2.

The small field integral contains both bosonic and fermionic type of fields. Therefore, in section 3 we first construct a 
power series representation for such type of general integrals. Then we apply this series expansion
to our specific model and get bounds on the small field integral. In section 4 we study the large field region and get
a representation of the generating functional as a sum of local pieces. Then in section 5, we combine various 
overlapping regions of the lattice into connected components, rewrite the generating functional as sum over these 
components and show convergence. Finally, in section 6 we show exponential decay of the correlation functions.

\section{The expansion}
First rewrite the fermoinic quadratic terms in the action (1.22) as
\begin{equation}
\begin{aligned}
\bar{\psi}_{L} & \mathfrak{D}_{A} \psi_{L} + \bar{\psi}_{R} \mathfrak{D}\psi_{R}
+ m_{f}(\bar{\psi}_{L}\psi_{R} + \bar{\psi}_{R}\psi_{L}) \\ 
&= \bar{\psi}_{L} \mathfrak{D} \psi_{L} + 
\bar{\psi}_{L} (\mathfrak{D}_{A} - \mathfrak{D}) \psi_{L} + \bar{\psi}_{R} \mathfrak{D} \psi_{R}
+ m_{f}(\bar{\psi}_{L}\psi_{R} + \bar{\psi}_{R}\psi_{L}) \\
&= \bar{\psi}_{\beta} \mathfrak{D} \psi_{\alpha} + m_{f} \bar{\psi}_{\beta}\psi_{\alpha} 
+ \bar{\psi}_{L} (\mathfrak{D}_{A} - \mathfrak{D}) \psi_{L}  \\
&= \langle \bar{\psi}_{\beta}, (\mathfrak{D} + m_{f}) \psi_{\alpha}\rangle 
+V^{\prime}_{\epsilon} (\psi, \bar{\psi})  \\
&= \langle \bar{\psi}_{\beta}, S^{-1} \psi_{\alpha} \rangle + V^{\prime}_{\epsilon} (\psi, \bar{\psi})
\end{aligned}
\end{equation}
where $V^{\prime}_{\epsilon} (\psi, \bar{\psi}) = 
\bar{\psi}_{L} (\mathfrak{D}_{A} - \mathfrak{D}) \psi_{L}$ will be treated like other terms in the potential
and $S = (\mathfrak{D} + m_{f})^{-1} = (\gamma \cdot \nabla - \frac{1}{2} \Delta + m_{f})^{-1}$ is the fermionic covariance operator.
We represent all the bosonic fields on the lattice as $\Phi = (\rho, A)$. Define the bosonic covariance operator
\begin{equation}
\langle\Phi, C\Phi\rangle = \langle\rho, (- \Delta + \mu^{2})^{-1} \rho\rangle + \langle A, (\delta d + m_{A}^{2})^{-1} A\rangle
\end{equation} 
where  
\begin{equation}
\langle \rho, -\Delta\rho \rangle = \sum_{ x \in \mathbb{T}} \sum_{\mu=1}^{4} (\rho(x+e_{\mu}) - \rho(x))^{2}  
\hspace{0.4 cm} \text{and}  \hspace{0.4 cm}
\langle A, -\delta d \hspace{0.1 cm} A \rangle =  \sum_{p \subset \mathbb{T}^{\ast\ast}} (dA)^{2}.
\end{equation}
Denote $\mathrm{D} = S^{-1}$ and $\text{T} = C^{-1}$. Rewrite the action $ \mathcal{A}(\psi, \bar{\psi}, \rho, A)$ 
as the sum of a Gaussian part and higher order interaction part
\begin{equation}
\begin{aligned}
 \mathcal{A}(\psi, \bar{\psi}, \Phi) &= \frac{1}{2}\langle\Phi, C^{-1}\Phi\rangle + 
 \langle\bar{\psi}_{\beta}, S^{-1}\psi_{\alpha} \rangle  + V(\psi, \bar{\psi}, \Phi) \\
&= \frac{1}{2}\langle\Phi, \text{T}\Phi\rangle + 
 \langle\bar{\psi}_{\beta}, \mathrm{D} \psi_{\alpha} \rangle  + V(\psi, \bar{\psi}, \Phi) 
\end{aligned}
\end{equation}
where the term $V^{\prime}_{\epsilon} (\psi, \bar{\psi})$ in included in the potential $V$.
The generating functional is given by
\begin{equation}
\begin{aligned}
\mathrm{Z} [\text{J}, \bar{\text{J}}] = \Big(\frac{2\pi}{e_{0}}\Big)^{-|\mathbb{T}| + 1} \sum_{v : dv=0}
 \int \mathcal{D}\Phi \hspace{0.05 cm}  \int  \mathcal{D}\bar{\psi} \mathcal{D}\psi \hspace{0.05 cm}
  e^{ - \frac{1}{2}\langle\Phi, \text{T}\Phi\rangle - 
 \langle\bar{\psi}_{\beta}, \mathrm{D}\psi_{\alpha} \rangle  - V(\psi, \bar{\psi}, \Phi) } \hspace{0.05 cm}
e^{- e_{0} \langle \bar{\psi}, \text{J}\rangle - e_{0} \langle \bar{\text{J}}, \psi\rangle }.
\end{aligned}
\end{equation}
Fermion fields $\bar{\psi}_{\beta}, \psi_{\alpha}$ are Grassmann elements whereas $\Phi$ is a scalar function that
is unbounded. Therefore, to construct a localized small field integral we expand the generating functional into regions where
all fields are finite and the regions where $\Phi$ is unbounded. We discuss this construction in four steps.
\begin{enumerate}
  \item Let $p_{\lambda} = |\text{log} \hspace{0.05 cm} \lambda|^{9}$ and $r_{\lambda} = |\text{log} \hspace{0.05 cm} \lambda|^{2}$.
  Divide the lattice $\mathbb{T}$ into blocks $\Box$ of length $[r_{\lambda}]$, where $[\cdot]$ denotes the integer part.
 Note that $\mathbb{T} = (\mathbb{Z}/L \mathbb{Z})^{4}$, so we assume $L \gg [r_{\lambda}]$.
  Define
   \begin{equation}
\Phi(u) = \begin{cases} 
\rho(x)  & \hspace{0.1 cm}  \text{if} \hspace{0.1 cm} u = x  \\
A_{\mu}(x)  & \hspace{0.1 cm} \text{if} \hspace{0.1 cm} u = (x, x+e_{\mu}).
\end{cases}
\end{equation}
  Define the characteristic function
  \begin{equation}
\chi_{\Box}(\Phi) = \begin{cases} 
1  & \hspace{0.1 cm} \sup_{u \in \Box} |\Phi(u)| < p_{\lambda},  \\
0  & \hspace{0.1 cm} \text{otherwise}
\end{cases}
\hspace{0.5 cm} \zeta_{\Box}(\Phi) = 1- \chi_{\Box}(\Phi). 
\end{equation}
Then the decomposition of unity is
 \begin{equation}
1 = \prod_{\Box} (\chi_{\Box} + \zeta_{\Box}) = \sum_{\mathrm{Q}\subset \mathbb{T}} 
\prod_{\Box \in \mathbb{T} - \mathrm{Q}} \chi_{\Box} \prod_{\Box \in \mathrm{Q}} \zeta_{\Box}
= \sum_{\mathrm{Q}\subset \mathbb{T}} \chi_{\mathbb{T} - \mathrm{Q}} \zeta_{\mathrm{Q}}.
\end{equation}
Thus, a $\Box \in \mathrm{Q}$  if there is at least one $u \in \Box$ with $|\Phi(u)| > p_{\lambda}$.
Set $v(p) = 0$ if $p \in \mathrm{Q}^{c}$.
Define 
\begin{equation}
\Lambda_{0} = \mathbb{T} - \mathrm{Q}. \nonumber
\end{equation}
Insert (decomposition of unity) 1 in the generating functional and rewrite 
\begin{equation}
\begin{aligned}
\mathrm{Z} [\text{J}, \bar{\text{J}}] = \Big(\frac{2\pi}{e_{0}}\Big)^{-|\mathbb{T}| + 1} &
\sum_{\mathrm{Q} \subset \mathbb{T}} \sum_{v : dv=0}
 \int \mathcal{D}\Phi \hspace{0.05 cm}  \int \mathcal{D}\bar{\psi} \mathcal{D}\psi \hspace{0.05 cm} \\
 & e^{ - \frac{1}{2}\langle\Phi, \text{T}\Phi\rangle - 
 \langle\bar{\psi}_{\beta}, \mathrm{D} \psi_{\alpha} \rangle  - V(\psi, \bar{\psi}, \Phi) } \hspace{0.05 cm}
e^{- e_{0} \langle\bar{\psi}, \text{J}\rangle - e_{0} \langle\bar{\text{J}}, \psi\rangle} 
\chi_{\Lambda_{0}}(\Phi) \zeta_{\mathrm{Q}}(\Phi).
\end{aligned}
\end{equation}

\item Conditioning

Contract $\Lambda_{0}$ by a layer of $[r_{\lambda}]$ cubes to get $\Lambda_{1}$, such that, 
$d(\mathrm{Q}, \Lambda_{1}) = [r_{\lambda}]$. Denote 
\begin{equation}
\tilde{\mathrm{Q}}  = \mathbb{T} - \Lambda_{1}. \nonumber
\end{equation}
Let $1_{\Lambda_{1}}$ and $1_{\tilde{\mathrm{Q}}}$ be  characteristic functions restricting 
operator to $\Lambda_{1}$ and $\tilde{\mathrm{Q}}$ respectively. 
Define  $\mathrm{D}_{\Lambda_{1}} = 1_{\Lambda_{1}}\mathrm{D}1_{\Lambda_{1}}$, 
 $\mathrm{D}_{\Lambda_{1}\tilde{\mathrm{Q}} } = 1_{\Lambda_{1}}\mathrm{D}1_{\tilde{\mathrm{Q}} }$,  
 $\mathrm{D}_{\tilde{\mathrm{Q}} } = 1_{\tilde{\mathrm{Q}} } \mathrm{D}1_{\tilde{\mathrm{Q}} }$,
$\text{T}_{\Lambda_{1}} = 1_{\Lambda_{1}}\text{T}1_{\Lambda_{1}}$, 
 $\text{T}_{\Lambda_{1}\tilde{\mathrm{Q}} } = 1_{\Lambda_{1}}\text{T}1_{\tilde{\mathrm{Q}} }$
 and $\text{T}_{\tilde{\mathrm{Q}} } = 1_{\tilde{\mathrm{Q}} }\text{T}1_{\tilde{\mathrm{Q}} }$. 
Then rewrite
\begin{equation}
\begin{aligned}
 \langle\bar{\psi}_{\beta}, \mathrm{D}\psi_{\alpha} \rangle &= \langle\bar{\psi}_{\beta}, \mathrm{D}_{\Lambda_{1}}\psi_{\alpha} \rangle 
+   \langle\bar{\psi}_{\beta}, \mathrm{D}_{\Lambda_{1}\tilde{\mathrm{Q}} }\psi_{\alpha} \rangle + 
 \langle\bar{\psi}_{\beta}, \mathrm{D}_{\tilde{\mathrm{Q}} \Lambda_{1}}\psi_{\alpha} \rangle +  
\langle\bar{\psi}_{\beta}, \mathrm{D}_{\tilde{\mathrm{Q}} }\psi_{\alpha} \rangle \\
\frac{1}{2}\langle\Phi, \text{T}\Phi\rangle &= \frac{1}{2}\langle\Phi, \text{T}_{\Lambda_{1}}\Phi\rangle + 
\langle\Phi, \text{T}_{\Lambda_{1}\tilde{\mathrm{Q}} }\Phi\rangle + 
\frac{1}{2}\langle\Phi, \text{T}_{\tilde{\mathrm{Q}} }\Phi\rangle.
\end{aligned}
\end{equation}
Denote $V_{s} = e_{0} \langle\bar{\psi}, \text{J}\rangle + e_{0} \langle\bar{\text{J}}, \psi\rangle$ and using (2.10) rewrite (2.9) as
\begin{equation}
\begin{aligned}
\mathrm{Z} [\text{J}, \bar{\text{J}}]  &= \Big(\frac{2\pi}{e_{0}}\Big)^{-|\mathbb{T}| + 1} \sum_{\mathrm{Q} \subset \mathbb{T}} 
\sum_{v: dv=0} \int\mathcal{D}\Phi_{\tilde{\mathrm{Q}} } \hspace{0.1 cm}
  \int \mathcal{D}\bar{\psi}_{\tilde{\mathrm{Q}}} \mathcal{D}\psi_{\tilde{\mathrm{Q}}} \\ 
  & \hspace{0.1 cm}
e^{-\frac{1}{2}\langle\Phi, \text{T}_{\tilde{\mathrm{Q}} }\Phi\rangle - \langle\bar{\psi}_{\beta}, \mathrm{D}_{\tilde{\mathrm{Q}} }\psi_{\alpha} \rangle 
 - V(\tilde{\mathrm{Q}} , \psi, \bar{\psi}, \Phi)} e^{- V_{s}(\tilde{\mathrm{Q}})}  
  \zeta_{\mathrm{Q}}(\Phi) \chi_{\Lambda_{0} - \Lambda_{1}}(\Phi) \\ &
  \int\mathcal{D}\Phi_{\Lambda_{1}}  \int \mathcal{D}\bar{\psi}_{\Lambda_{1}}  \mathcal{D}\psi_{\Lambda_{1}}  
   \hspace{0.1 cm} 
e^{-\frac{1}{2}\langle\Phi_{\Lambda_{1}}, \text{T}\Phi_{\Lambda_{1}}\rangle
 - \langle\Phi_{\Lambda_{1}}, \text{T}\Phi_{\tilde{\mathrm{Q}} }\rangle -V(\Lambda_{1}, \psi, \bar{\psi}, \Phi)} e^{- V_{s}(\Lambda_{1})} \\
 & e^{- \langle\bar{\psi}_{\beta, \Lambda_{1}}, \mathrm{D}\psi_{\alpha, \Lambda_{1}} \rangle 
-  \langle\bar{\psi}_{\beta, \Lambda_{1}}, \mathrm{D}\psi_{\alpha, \tilde{\mathrm{Q}} } \rangle
-  \langle\bar{\psi}_{\beta, \tilde{\mathrm{Q}} }, \mathrm{D}\psi_{\alpha, \Lambda_{1}} \rangle } \chi_{\Lambda_{1}}(\Phi).
\end{aligned}
\end{equation} 
Define \textit{shift} as 
\begin{equation}
\begin{aligned}
&\Psi_{\alpha} = 
S_{\Lambda_{1}}\mathrm{D}_{\Lambda_{1}\tilde{\mathrm{Q}} }\bar{\psi}_{\alpha, \tilde{\mathrm{Q}} } \hspace{1 cm}
\bar{\Psi}_{\beta} = 
S_{\Lambda_{1}}\mathrm{D}_{\Lambda_{1}\tilde{\mathrm{Q}} }\bar{\psi}_{\beta, \tilde{\mathrm{Q}} } \\
& \varphi = C_{\Lambda_{1}}\text{T}_{\Lambda_{1}\tilde{\mathrm{Q}} } \Phi_{\tilde{\mathrm{Q}} }.
\end{aligned}
\end{equation}
Next let
\begin{equation}
\psi_{\alpha, \Lambda_{1}} \rightarrow \psi_{\alpha, \Lambda_{1}} - \Psi_{\alpha} \hspace{1 cm}
\bar{\psi}_{\beta, \Lambda_{1}} \rightarrow \bar{\psi}_{\beta, \Lambda_{1}} - \bar{\Psi}_{\beta} \hspace{1 cm}
\Phi_{\Lambda_{1}} \rightarrow \Phi_{\Lambda_{1}} - \varphi \nonumber
\end{equation}
 and rewrite (2.11) as
\begin{equation}
\begin{aligned}
\mathrm{Z} [\text{J}, \bar{\text{J}}]  &= \Big(\frac{2\pi}{e_{0}}\Big)^{-|\mathbb{T}| + 1} \sum_{\mathrm{Q}\subset \mathbb{T}} 
\sum_{v: dv=0} \int\mathcal{D}\Phi_{\tilde{\mathrm{Q}}} 
   \int \mathcal{D}\bar{\psi}_{\tilde{\mathrm{Q}}} \mathcal{D}\psi_{\tilde{\mathrm{Q}}} \hspace{0.1 cm}
  \zeta_{\mathrm{Q}}(\Phi) \chi_{\Lambda_{0} - \Lambda_{1}}(\Phi) \\ &
e^{-\frac{1}{2}\langle\Phi, (\text{T}_{\tilde{\mathrm{Q}}} - 
\text{T}_{\tilde{\mathrm{Q}}\Lambda_{1}}C_{\Lambda_{1}}\text{T}_{\Lambda_{1}\tilde{\mathrm{Q}}}) \Phi\rangle} \\
& e^{- \langle\bar{\psi}_{\beta}, (\mathrm{D} _{\tilde{\mathrm{Q}}} - 
\mathrm{D}_{\tilde{\mathrm{Q}}\Lambda_{1}}S_{\Lambda_{1}}\mathrm{D} _{\Lambda_{1}\tilde{\mathrm{Q}}}) \psi_{\alpha} \rangle}
 e^{ - V(\tilde{\mathrm{Q}}, \psi, \bar{\psi}, \Phi)} e^{- V_{s}(\tilde{\mathrm{Q}})}   \\
& \int\mathcal{D}\Phi_{\Lambda_{1}}  \int \mathcal{D}\bar{\psi}_{\Lambda_{1}}  \mathcal{D}\psi_{\Lambda_{1}}    \hspace{0.1 cm} 
 \chi_{\Lambda_{1}}(\Phi_{\Lambda_{1}} - \varphi) \hspace{0.1 cm}
e^{-\frac{1}{2}\langle\Phi, \text{T}_{\Lambda_{1}}\Phi\rangle
  - \langle\bar{\psi}_{\beta}, \mathrm{D} _{\Lambda_{1}}\psi_{\alpha} \rangle} \\ &
  e^{- V(\Lambda_{1}, \psi -  \Psi, \bar{\psi} -  \bar{\Psi}, \Phi -  \varphi)} e^{-V_{s}(\Lambda_{1})}.
\end{aligned}
\end{equation} 
This is known as conditioning the small field integral on $\Lambda_{1}$ on values in the large field region $\tilde{\mathrm{Q}}$.

\item The integral over $\psi_{\alpha,\Lambda_{1}}, \bar{\psi}_{\beta, \Lambda_{1}}$ is Gaussian with 
covariance $S_{\Lambda_{1}}$ while the integral over $\Phi_{\Lambda_{1}}$ 
is Gaussian with covariance $C_{\Lambda_{1}}$.
The covariance operators $S_{\Lambda_{1}}$ and $C_{\Lambda_{1}} $ couple sites everywhere in $\Lambda_{1}$.
Therefore, we construct localized covariance operators. Define the kernel of operator $S^{\text{loc}}$ as 
\begin{equation}
S^{\text{loc}}(x,y) = \begin{cases}
    S (x,y)  & \text{if} \hspace{0.4 cm} |x-y| < r_{\lambda} \\
     0 & \text{otherwise}
\end{cases}
\end{equation}
and the kernel of operator $C^{\text{loc}}$ as 
\begin{equation}
C^{\text{loc}}(x,y) = \begin{cases}
    C (x,y)  & \text{if} \hspace{0.4 cm} |x-y| < r_{\lambda} \\
     0 & \text{otherwise}.
\end{cases}
\end{equation}
Define $\delta S (x,y) = (S - S^{\text{loc}}) (x,y)$ as
\begin{equation}
\delta S (x,y) = \begin{cases}
    0  & \text{if} \hspace{0.4 cm} |x-y| < r_{\lambda} \\
     S(x,y) & \text{otherwise}
\end{cases}
\end{equation}
and $\delta C^{\frac{1}{2}} (x,y) = (C^{\frac{1}{2}} - C^{\frac{1}{2},\text{loc}}) (x,y)$ as
\begin{equation}
\delta C^{\frac{1}{2}}  (x,y) = \begin{cases}
    0  & \text{if} \hspace{0.4 cm} |x-y| < r_{\lambda} \\
     C^{\frac{1}{2}}(x,y) & \text{otherwise}.
\end{cases}
\end{equation}
The representation we use for  
$C^{\frac{1}{2}}$ and $C^{\frac{1}{2},\text{loc}}$ is discussed in  \cite{D2},
\begin{equation}
C^{\frac{1}{2}} = \frac{1}{\pi}\int_{0}^{\infty}\frac{d r}{\sqrt{r}}\hspace{0.1 cm} C_{r},  \hspace{0.5 cm}
C^{\frac{1}{2},\text{loc}} = \frac{1}{\pi}\int_{0}^{\infty}\frac{d r}{\sqrt{r}}\hspace{0.1 cm} C^{\text{loc}}_{r}, 
 \hspace{0.5 cm}   C_{r} = (\text{T} + r)^{-1}.
\end{equation}
To work in unit covariance, first note that the operators $S^{\text{loc}}_{\Lambda_{1}}$ and 
$C^{\frac{1}{2},\text{loc}}_{\Lambda_{1}}$ are small perturbation of  $S_{\Lambda_{1}}$ and $C^{\frac{1}{2}}_{\Lambda_{1}}$ 
respectively and hence invertible. Then
make the change of variables $\psi_{\alpha} \rightarrow S^{\text{loc}}_{\Lambda_{1}} \psi_{\alpha}$
\begin{equation}
\begin{aligned}
\langle\bar{\psi}_{\beta}, \mathrm{D}_{\Lambda_{1}} \psi_{\alpha}\rangle &= 
 \langle \bar{\psi}_{\beta}, \mathrm{D}_{\Lambda_{1}} S^{\text{loc}}_{\Lambda_{1}}\psi_{\alpha} \rangle  
=  \langle \bar{\psi}, \psi \rangle_{\Lambda_{1}} + V_{\epsilon} (\Lambda_{1}, \bar{\psi}_{\beta}, \psi_{\alpha}) 
\end{aligned}
\end{equation} 
This is the definition of $V_{\epsilon} (\Lambda_{1}, \bar{\psi}_{\beta}, \psi_{\alpha})$. Make another change of variables
\\ $\Phi = C^{\frac{1}{2},\text{loc}}_{\Lambda_{1}} \Phi^{\prime}$,  
\begin{equation}
\langle\Phi, \text{T}_{\Lambda_{1}} \Phi\rangle =  
\langle\Phi^{\prime}, C^{\frac{1}{2},\text{loc}}_{\Lambda_{1}} \text{T}_{\Lambda_{1}} 
C^{\frac{1}{2},\text{loc}}_{\Lambda_{1}} \Phi^{\prime}\rangle 
 = ||\Phi^{\prime}||_{\Lambda_{1}}^{2} + V_{\varepsilon}(\Lambda_{1}, \Phi^{\prime}).  
\end{equation}
This is the definition of $V_{\varepsilon}(\Lambda_{1}, \Phi^{\prime})$.

\item  After the change of variables, the characteristic function,
$\chi_{\Lambda_{1}}(C^{\frac{1}{2},\text{loc}}_{\Lambda_{1}} \Phi^{\prime} - \varphi)$ becomes non local as 
 $C^{\frac{1}{2},\text{loc}}_{\Lambda_{1}} \Phi^{\prime}(u)$ spreads $\Phi^{\prime}(u)$ to a distance of $r_{\lambda}$.
Thus, we construct a new small field region.   
Let $p_{0,\lambda} = |\text{log} \hspace{0.05 cm} \lambda|^{a}$ (with $8 < a < 9$).  
 We split the region $\Lambda_{1}$ into a region $\mathrm{P}$ and $\Lambda_{1} - \mathrm{P}$ 
 by using decomposition of unity as before. Define
\begin{equation}
\hat{\chi}_{\Box}(\Phi^{\prime}) = \begin{cases}
1 & \hspace{0.1 cm} \sup_{u \in \Box} |\Phi^{\prime}(u)| < p_{0, \lambda}, \\
0 & \hspace{0.1 cm} \text{otherwise}
\end{cases}
\hspace{0.5 cm} \hat{\zeta}_{\Box}(\Phi^{\prime}) = 1- \hat{\chi}_{\Box}(\Phi^{\prime}). 
\end{equation}
Then the decomposition of unity is
 \begin{equation}
1 = \prod_{\Box \in \Lambda_{1}} (\hat{\chi}_{\Box} + \hat{\zeta}_{\Box}) = \sum_{\mathrm{P} \subset \Lambda_{1}} 
\prod_{\Box \in \Lambda_{1} - \mathrm{P}} \hat{\chi}_{\Box} \prod_{\Box \in \mathrm{P}} \hat{\zeta}_{\Box}
= \sum_{\mathrm{P} \subset \Lambda_{1}} \hat{\chi}_{\Lambda_{1} - \mathrm{P}} \hat{\zeta}_{\mathrm{P}}.
\end{equation} 
Define
\begin{equation}
\Omega = \Lambda_{1} - \mathrm{P}.   \nonumber
\end{equation}
As covariance is unity in $\Omega$, conditioning is not required along the boundary $\partial\Omega$.
The potential is localized and analytic but the characteristic function $\chi_{\Lambda_{1}}$ is a mess.  
To clean $\chi_{\Lambda_{1}}$, as a first step, contract $\Omega$ by $[r_{\lambda}]$ to get $\Omega_{0}$
and split the potential term $V(\Lambda_{1}) = V(\Lambda_{1}, S^{\text{loc}}_{\Lambda_{1}} \psi - \Psi, \bar{\psi} -  \bar{\Psi}, 
 C^{\frac{1}{2},\text{loc}}_{\Lambda_{1}} \Phi^{\prime} - \varphi)$ as
\begin{equation}
\begin{aligned}
V(\Lambda_{1}) = V(\Omega_{0}) + V(\Lambda_{1} - \Omega_{0}) 
\end{aligned}
\end{equation}
 and factorize  $\chi_{\Lambda_{1}}$ as
\begin{equation}
\begin{aligned}
\chi_{\Lambda_{1}}(C^{\frac{1}{2},\text{loc}}_{\Lambda_{1}} \Phi^{\prime} - \varphi) &= 
 \chi_{\Lambda_{1} - \Omega_{0}}(C^{\frac{1}{2},\text{loc}}_{\Lambda_{1}} \Phi^{\prime} - \varphi) \hspace{0.1 cm}
 \chi_{\Omega_{0}}(C^{\frac{1}{2},\text{loc}}_{\Lambda_{1}} \Phi^{\prime} - \varphi).
\end{aligned}
\end{equation}
For $u \in \Omega_{0}$,
\begin{equation}
|C^{\frac{1}{2},\text{loc}}_{\Lambda_{1}} \Phi^{\prime}(u)| \leqslant c\hspace{0.1 cm} ||\Phi^{\prime}||_{\infty,\Omega}
\leqslant c \hspace{0.1 cm} \text{p}_{0,\lambda} <  \frac{p_{\lambda}}{2} \nonumber
\end{equation}
(see lemma 2.2 to follow) and 
\begin{equation}
|\varphi| \leqslant
c \hspace{0.05 cm} e^{-\gamma r_{\lambda}} p_{\lambda} <  \frac{p_{\lambda}}{2} \nonumber
\end{equation}
where we have used estimate of covariance (lemma 2.1) and
$|\Phi_{\tilde{\mathrm{Q}}}| \leqslant p_{\lambda}$ since this is near $\Lambda_{1}$. Thus,
\begin{equation}
\chi_{\Omega_{0}}(C^{\frac{1}{2},\text{loc}}_{\Lambda_{1}} \Phi^{\prime} - \varphi) = 1. 
\end{equation}
Note that  $\chi_{\Lambda_{1} - \Omega_{0}}(C^{\frac{1}{2},\text{loc}}_{\Lambda_{1}} \Phi^{\prime} - \varphi) $ is non local which
 is why we cannot integrate over the region $\Omega_{0}$. Thus, we 
shrink the region $\Omega_{0}$ by $[r_{\lambda}]$ to get a new small field region $\Omega_{1}$,
that is, $d(\Omega_{0}^{c}, \Omega_{1}) =  [r_{\lambda}]$. As mentioned earlier,  
 $C^{\frac{1}{2},\text{loc}}_{\Lambda_{1}} \Phi^{\prime}(u)$ only spreads $\Phi^{\prime}(u)$ to a distance of $r_{\lambda}$,
 therefore, $\chi_{\Lambda_{1} - \Omega_{0}}(C^{\frac{1}{2},\text{loc}}_{\Lambda_{1}} \Phi^{\prime} - \varphi)$ and   
 $V(\Lambda_{1} - \Omega_{0}) $ do not 
  depend on $\Phi^{\prime}_{\Omega_{1}}$ and thus can be taken out of the integral over $\Phi^{\prime}_{\Omega_{1}}$.
  The generating functional is given by 
 \begin{equation}
\begin{aligned}
\mathrm{Z} [\text{J}, \bar{\text{J}}]  &= \Big(\frac{2\pi}{e_{0}}\Big)^{-|\mathbb{T}| + 1} \sum_{\mathrm{Q}\subset \mathbb{T}} 
\sum_{\mathrm{P}\subset \Lambda_{1}} \sum_{v: dv=0} 
\frac{\text{det}\hspace{0.1 cm}C^{\frac{1}{2},\text{loc}}_{\Lambda_{1}}}
{\text{det}\hspace{0.1 cm}S^{\text{loc}}_{\Lambda_{1}}} \\
& \int\mathcal{D}\Phi_{\tilde{\mathrm{Q}}} \int \mathcal{D}\bar{\psi}_{\tilde{\mathrm{Q}}} \mathcal{D}\psi_{\tilde{\mathrm{Q}}}
 \hspace{0.1 cm}  \zeta_{\mathrm{Q}}(\Phi) \chi_{\Lambda_{0} - \Lambda_{1}}(\Phi) \hspace{0.1 cm}
e^{-\frac{1}{2}\langle\Phi, (\text{T}_{\tilde{\mathrm{Q}}} - 
\text{T}_{\tilde{\mathrm{Q}}\Lambda_{1}}C_{\Lambda_{1}}\text{T}_{\Lambda_{1}\tilde{\mathrm{Q}}}) \Phi\rangle} \\ &
e^{- \langle\bar{\psi}_{\beta}, (\mathrm{D}_{\tilde{\mathrm{Q}}} - 
\mathrm{D}_{\tilde{\mathrm{Q}}\Lambda_{1}}S_{\Lambda_{1}}\mathrm{D}_{\Lambda_{1}\tilde{\mathrm{Q}}}) \psi_{\alpha} \rangle}
 e^{ - V(\tilde{\mathrm{Q}}, \psi, \bar{\psi}, \Phi)} e^{-V_{s}(\tilde{\mathrm{Q}})} \\
& \int\mathcal{D}\Phi^{\prime}_{\Lambda_{1} - \Omega_{1}} 
\int \mathcal{D}\bar{\psi}_{\Lambda_{1} - \Omega_{1}} \mathcal{D}\psi_{\Lambda_{1} - \Omega_{1}}  \hspace{0.1 cm} 
\hat{\zeta}_{\Lambda_{1} - \Omega}(\Phi^{\prime}) \hspace{0.05 cm}
 \chi_{\Lambda_{1} - \Omega_{0}}(C^{\frac{1}{2},\text{loc}}_{\Lambda_{1}} \Phi^{\prime} - \varphi) \\
&e^{-\frac{1}{2}||\Phi^{\prime}||^{2}_{\Lambda_{1} - \Omega_{1}}
  - \langle\bar{\psi}, \psi \rangle_{\Lambda_{1} - \Omega_{1}}}  
  e^{- V(\Lambda_{1}-\Omega_{0}, S^{\text{loc}}_{\Lambda_{1}} \psi - \Psi, \bar{\psi} -  \bar{\Psi}, 
  C^{\frac{1}{2},\text{loc}}_{\Lambda_{1}} \Phi^{\prime} -  \varphi)}  \\
& \int\mathcal{D}\Phi^{\prime}_{\Omega_{1}} \int \mathcal{D}\bar{\psi}_{\Omega_{1}} \mathcal{D}\psi_{\Omega_{1}} 
\hspace{0.1 cm} \hat{\chi}_{\Omega_{1}}(\Phi^{\prime}) \hspace{0.1 cm}
 e^{-\frac{1}{2}||\Phi^{\prime}||^{2}_{\Omega_{1}} - \langle\bar{\psi}, \psi \rangle_{\Omega_{1}}} \\
&   e^{- V(\Omega_{0}, S^{\text{loc}}_{\Lambda_{1}} \psi - \Psi, \bar{\psi} -  \bar{\Psi}, 
  C^{\frac{1}{2},\text{loc}}_{\Lambda_{1}} \Phi^{\prime} -  \varphi) 
  + V_{\epsilon} (\Lambda_{1}, \bar{\psi}_{\beta}, \psi_{\alpha}, \Phi^{\prime})} e^{-V_{s}(\Lambda_{1})}.
\end{aligned}
\end{equation}  
The above equation is the required expansion of the generating functional.  
   
\end{enumerate} 

We bring the term 
$\frac{\text{det}\hspace{0.1 cm}C^{\frac{1}{2},\text{loc}}_{\Lambda_{1}}}
{\text{det}\hspace{0.1 cm}S^{\text{loc}}_{\Lambda_{1}}} $ 
outside of the summations in two steps:
\begin{enumerate}
  \item For some $W_{1}(\Lambda_{1}) = \sum_{\Box \subset \Lambda_{1}} W_{1}(\Box)$ (see Lemma 2.6) and
  $W^{\prime}_{1}(\Lambda_{1}) = \sum_{\Box \subset \Lambda_{1}} W^{\prime}_{1}(\Box)$ (see Lemma 2.7)
\begin{equation}
\begin{aligned}
\text{det}\hspace{0.1 cm}C^{\frac{1}{2},\text{loc}}_{\Lambda_{1}}  
 = (\text{det}\hspace{0.1 cm}C^{\frac{1}{2}}_{\Lambda_{1}}) \hspace{0.1 cm} 
 e^{W_{1}(\Lambda_{1})}, \hspace{0.2 cm} 
 \text{det}\hspace{0.1 cm}S^{\text{loc}}_{\Lambda_{1}}   
  = \text{det}\hspace{0.1 cm}S_{\Lambda_{1}} \hspace{0.1 cm} e^{W^{\prime}_{1}(\Lambda_{1})}.
\end{aligned}
\end{equation}

  \item For some $W_{2}(\mathbb{T}) = \sum_{\Box \subset \mathbb{T}} W_{2}(\Box)$ (see 2.57) and
 $W^{\prime}_{2}(\mathbb{T}) = \sum_{\Box \subset \mathbb{T}} W^{\prime}_{2}(\Box)$ (see 2.57)
\begin{equation}
\begin{aligned}
\text{det}\hspace{0.1 cm}C^{\frac{1}{2}}_{\Lambda_{1}}  
= (\text{det}\hspace{0.1 cm}C^{\frac{1}{2}}) \hspace{0.1 cm} e^{W_{2}(\mathbb{T})}, \hspace{0.2 cm}
\text{det}\hspace{0.1 cm}S_{\Lambda_{1}}  
 = \text{det}\hspace{0.1 cm}S \hspace{0.1 cm} e^{W^{\prime}_{2}(\mathbb{T})}
\end{aligned}
\end{equation}
and we write 
\begin{center}
$W_{2} = W_{2}(\mathrm{Q}) + W_{2}(\Lambda_{0} - \Omega_{0}) + W_{2}(\Omega_{0})$
 \\ $W^{\prime}_{2} = W^{\prime}_{2}(\mathrm{Q}) + W^{\prime}_{2}(\Lambda_{0} - \Omega_{0}) 
+ W^{\prime}_{2}(\Omega_{0})$.
\end{center}

\end{enumerate}
 
Define
\begin{equation}
\begin{aligned}
V_{1}(\Omega_{0},\bar{\psi}, \psi, \Phi^{\prime}_{\Lambda_{1}}, 
\Phi_{\tilde{\mathrm{Q}}}, \bar{\text{J}}, \text{J})
&= - (W^{\prime}_{1}(\Lambda_{1})  - W_{1}(\Lambda_{1})) - (W^{\prime}_{2}(\Omega_{0}) - W_{2}(\Omega_{0})) \\
&+ V(\Omega_{0}, S^{\text{loc}}_{\Lambda_{1}} \psi -  \Psi, \bar{\psi} -  \bar{\Psi}, 
  C^{\frac{1}{2},\text{loc}}_{\Lambda_{1}} \Phi^{\prime} - \varphi)  \\
& - e_{0} \langle \bar{\psi}, \text{J}\rangle - e_{0} \langle \bar{\text{J}}, S^{\text{loc}}_{\Lambda_{1}} \psi \rangle 
+ V_{\epsilon} (\Lambda_{1}, \bar{\psi}, \psi, \Phi^{\prime}) 
\end{aligned}
\end{equation}
 Define the normalized Gaussian measures with unit covariance as
\begin{equation}
\begin{aligned}
 d\mu_{I}(\bar{\psi}_{\Omega_{1}}, \psi_{\Omega_{1}}) &=  
 \frac{ \mathcal{D}\bar{\psi}_{\Omega_{1}} \mathcal{D}\psi_{\Omega_{1}}  \hspace{0.05 cm} 
  e^{- \langle \bar{\psi}, \psi \rangle_{\Omega_{1}}}} 
{\int \mathcal{D}\bar{\psi}_{\Omega_{1}} \mathcal{D}\psi_{\Omega_{1}}  \hspace{0.05 cm} 
e^{- \langle \bar{\psi}, \psi \rangle_{\Omega_{1}}}} \\
d\mu_{\text{I}}(\Phi^{\prime}_{\Omega_{1}})  &= \frac{\mathcal{D}\Phi^{\prime}_{\Omega_{1}} 
e^{-\frac{1}{2}||\Phi^{\prime}||^{2}_{\Omega_{1}}}}
{\int \mathcal{D}\Phi^{\prime}_{\Omega_{1}} e^{-\frac{1}{2}||\Phi^{\prime}||^{2}_{\Omega_{1}}}}.
\end{aligned}
\end{equation}
Denote
\begin{equation}
 \mathrm{Z}_{0}(\Omega_{1}) = \int  \mathcal{D}\bar{\psi}_{\Omega_{1}} \mathcal{D}\psi_{\Omega_{1}} 
\mathcal{D}\Phi^{\prime}_{\Omega_{1}}  \hspace{0.05 cm} e^{- \langle \bar{\psi}, \psi \rangle_{\Omega_{1}} 
-\frac{1}{2}||\Phi^{\prime}||^{2}_{\Omega_{1}}}
\end{equation}
Define the localized small field integral as
\begin{equation}
\Xi(\Omega_{1},\bar{\psi}, \psi, \Phi^{\prime}_{\Lambda_{1}}, \Phi_{\tilde{\mathrm{Q}}}, \bar{\text{J}}, \text{J}) =
\int d\mu_{I}(\bar{\psi}_{\Omega_{1}}, \psi_{\Omega_{1}}) d\mu_{\text{I}}(\Phi^{\prime}_{\Omega_{1}})
\hspace{0.1 cm} \hat{\chi}_{\Omega_{1}}(\Phi^{\prime})  \hspace{0.05 cm} 
e^{- V_{1}(\Omega_{0},\bar{\psi}, \psi, \Phi^{\prime}_{\Lambda_{1}}, \Phi_{\tilde{\mathrm{Q}}}, \bar{\text{J}}, \text{J})}
\end{equation}
Rewrite the generating functional
\begin{equation}
\begin{aligned}
\mathrm{Z} [\text{J}, \bar{\text{J}}]  &= \Big(\frac{2\pi}{e_{0}}\Big)^{-|\mathbb{T}| + 1}  
\frac{\text{det}\hspace{0.1 cm}C^{\frac{1}{2}}}{\text{det}\hspace{0.1 cm}S} 
 \sum_{\mathrm{Q}\subset \mathbb{T}} \sum_{\mathrm{P}\subset \Lambda_{1}} \sum_{v: dv=0}  \\ 
&  \int\mathcal{D}\Phi_{\tilde{\mathrm{Q}}} \int \mathcal{D}\bar{\psi}_{\tilde{\mathrm{Q}}} \mathcal{D}\psi_{\tilde{\mathrm{Q}}}
 \hspace{0.1 cm}  \zeta_{\mathrm{Q}}(\Phi) \chi_{\Lambda_{0} - \Lambda_{1}}(\Phi) \hspace{0.1 cm}
e^{-\frac{1}{2}\langle\Phi, (\text{T}_{\tilde{\mathrm{Q}}} - 
\text{T}_{\tilde{\mathrm{Q}}\Lambda_{1}}C_{\Lambda_{1}}\text{T}_{\Lambda_{1}\tilde{\mathrm{Q}}}) \Phi\rangle} \\ &
e^{- \langle\bar{\psi}_{\beta}, (\mathrm{D}_{\tilde{\mathrm{Q}}} - 
\mathrm{D}_{\tilde{\mathrm{Q}}\Lambda_{1}}S_{\Lambda_{1}}\mathrm{D}_{\Lambda_{1}\tilde{\mathrm{Q}}}) \psi_{\alpha} \rangle}
\hspace{0.05 cm} e^{ - V(\tilde{\mathrm{Q}}, \psi, \bar{\psi}, \Phi)} e^{-V_{s}(\tilde{\mathrm{Q}})}
 \hspace{0.05 cm} e^{W_{2}(\mathrm{Q}) - W^{\prime}_{2}(\mathrm{Q})} \\
& \int\mathcal{D}\Phi^{\prime}_{\Lambda_{1} - \Omega_{1}} 
\int \mathcal{D}\bar{\psi}_{\Lambda_{1} - \Omega_{1}} \mathcal{D}\psi_{\Lambda_{1} - \Omega_{1}}  \hspace{0.1 cm} 
 e^{-\frac{1}{2}||\Phi^{\prime}||^{2}_{\Lambda_{1} - \Omega_{1}}
  - \langle\bar{\psi}, \psi \rangle_{\Lambda_{1} - \Omega_{1}}} \\ 
& \hat{\zeta}_{\mathrm{P}}(\Phi^{\prime}) \hspace{0.05 cm}
 \chi_{\Lambda_{1} - \Omega_{0}}(C^{\frac{1}{2},\text{loc}}_{\Lambda_{1}} \Phi^{\prime} - \varphi) 
 \hspace{0.05 cm} e^{-V_{s}(\Lambda_{1}-\Omega_{1})} \\
&  e^{- V(\Lambda_{1}-\Omega_{0}, S^{\text{loc}}_{\Lambda_{1}}\psi - \Psi, \bar{\psi} -  \bar{\Psi}, 
  C^{\frac{1}{2},\text{loc}}_{\Lambda_{1}} \Phi^{\prime} - \varphi)} 
 e^{W_{2}(\Lambda_{0} - \Omega_{0}) - W^{\prime}_{2}(\Lambda_{0} - \Omega_{0})}  \\ &
\mathrm{Z}_{0}(\Omega_{1}) \hspace{0.1 cm}  
\Xi(\Omega_{1},\bar{\psi}, \psi, \Phi^{\prime}_{\Lambda_{1}}, \Phi_{\tilde{\mathrm{Q}}}, \bar{\text{J}}, \text{J}).
\end{aligned}
\end{equation}  
Now we quote and prove some lemmas.
 
\textbf{Lemma 2.1} \cite{G1} Let \textit{C} be the positive, self-adjoint operator as defined.
Then for some constant $0 < \gamma_{1} < \text{min} \big(\mu^{2}, m_{A}^{2}\big)$, 
$|C(x,y)| \leqslant c \hspace{0.05 cm} e^{-\gamma_{1} |x-y|}$. 

\textbf{Lemma 2.2} \cite{G1} For $\Phi : \mathbb{T} \rightarrow  \mathbb{R}$
\begin{enumerate}
  \item Let $C^{\frac{1}{2},\text{loc}}$ and $\delta C^{\frac{1}{2}}$ be the operators as defined. Then
  \begin{equation}
|(C^{\frac{1}{2},\text{loc}} \hspace{0.05 cm}\Phi) (u)| \leqslant  c \hspace{0.05 cm}  ||\Phi||_{\infty}, \hspace{0.5 cm}
|(\delta C^{\frac{1}{2}} \hspace{0.05 cm}\Phi) (u)| \leqslant  c \hspace{0.1 cm} e^{-\gamma^{\prime} r_{\lambda}} ||\Phi||_{\infty}.
\end{equation}

 \item $C^{\frac{1}{2}}$ and $C^{\frac{1}{2},\text{loc}}$ are invertible and 
 \begin{equation}
|(C^{-\frac{1}{2}}\hspace{0.05 cm} \Phi) (u)|, |(C^{-\frac{1}{2},\text{loc}}\hspace{0.05 cm}\Phi) (u)|
\leqslant c \hspace{0.1 cm} ||\Phi||_{\infty}.
\end{equation}
\end{enumerate}
 
Before we prove the decay of fermionic covariance operator $S = (\gamma \cdot \nabla - \frac{1}{2} \Delta + m_{f})^{-1}$
note that in momentum space the eigenvalues of operators $\partial_{\mu}, \partial_{\mu}^{T}, \nabla_{\mu}, \Delta$ are
\begin{equation}
\begin{aligned}
\partial_{\mu} e^{i p\cdot x} &= e^{i p\cdot (x+e_{\mu})} -  e^{i p\cdot (x)} = (e^{i p_{\mu}} - 1)\hspace{0.05 cm} e^{i p\cdot x} \\
\partial_{\mu}^{T} e^{i p\cdot x} &= e^{i p\cdot (x-e_{\mu})} -  e^{i p\cdot (x)} = (e^{-i p_{\mu}} - 1)\hspace{0.05 cm} e^{i p\cdot x} \\
\nabla_{\mu} e^{i p\cdot x} &= \frac{1}{2}(\partial_{\mu} - \partial_{\mu}^{T}) e^{i p\cdot x}
= (i \hspace{0.05 cm} \text{sin} \hspace{0.05 cm} p_{\mu})\hspace{0.05 cm} e^{i p\cdot x} \\
\Delta \hspace{0.05 cm} e^{i p\cdot x} &= -\partial_{\mu}^{T} \partial_{\mu} e^{i p\cdot x}
= -2 \sum_{\mu} (1 - \text{cos} \hspace{0.05 cm} p_{\mu}) \hspace{0.05 cm} e^{i p\cdot x}.
\end{aligned}
\end{equation}
 
\textbf{Lemma 2.3} For periodic boundary conditions there exists a positive constant
$\gamma_{2} < \mathcal{O}(1) m_{f}$ such that $|S(x,y)| \leqslant c \hspace{0.05 cm} e^{-\gamma_{2} |x-y|}$.

\textit{Proof} We consider the infinite lattice $\mathbb{Z}^{4}$ since toroidal case can be obtained by
periodizing. In momentum space representation 
\begin{equation}
S(x,y) = \frac{1}{(2\pi)^{4}} \int_{|p_{\mu}| < \pi}
e^{i p\cdot (x-y)} \frac{M(p) - i \gamma \cdot s(p)}{M(p)^{2} + |s(p)|^{2}} \hspace{0.1 cm} d^{4} p
\end{equation}
where
\begin{equation}
M(p) = m_{f} + \sum_{\mu} (1 - \text{cos} \hspace{0.05 cm} p_{\mu}) \hspace{1 cm}
s_{\mu}(p) = \text{sin} \hspace{0.05 cm} p_{\mu}
\end{equation}
Let 
\begin{equation}
G(x,y) = \frac{1}{(2\pi)^{4}} \int_{|p_{\mu}| < \pi} 
e^{i p\cdot (x-y)} \frac{1}{M(p)^{2} + |s(p)|^{2}} \hspace{0.1 cm} d^{4} p.
\end{equation}
Then since $\nabla_{\mu} e^{i px} = i s_{\mu}(p) e^{i px}$ we can also write
\begin{equation}
S(x,y) = (-\gamma \cdot \nabla - \frac{1}{2} \Delta + m_{f}) G(x,y).
\end{equation}
Thus, getting estimates on $G(x,y)$ is enough.
We prove the decay for only $\mu=1$ direction, since by symmetry it holds for $\mu = 2, 3, 4$ as well.
Make a shift $p_{1} \rightarrow p_{1} + i \delta$ such that $\delta > 0$ if $x-y > 0$ and $\delta < 0$ if $x-y < 0$. 
Define a contour $\Gamma$ as a finite strip in  
the complex plane of $p_{1}$ as $\Gamma = \{|\text{Im}\hspace{0.05 cm} p_{1}| < \delta, 
\text{Re}\hspace{0.05 cm} p_{1}, \cdots, p_{4} \in (-\pi, \pi]^{4} \}$. Then
\begin{equation}
G(x,y) = \frac{1}{2\pi} \int_{\Gamma} 
e^{i p_{1} (x-y)} \frac{1}{M(p_{1} + i \delta)^{2} + |s(p_{1} + i \delta)|^{2}} e^{- \delta (x-y)} \hspace{0.1 cm} d p_{1}.
\end{equation}
Next we find $\delta$ such that the above integrand is bounded and analytical in $\Gamma$. 
 \begin{equation}
 \begin{aligned}
&s(p_{1}) =  \text{sin}\hspace{0.05 cm} p_{1},  \hspace{2 cm} M(p_{1})  =  m_{f} + 1 - \text{cos} \hspace{0.05 cm} p_{1}   \\
&s(p_{1} + i \delta) =  \text{sin}\hspace{0.05 cm} (p_{1}+i \delta),  \hspace{0.5 cm} 
M(p_{1} + i \delta)  =  m_{f} + 1 - \text{cos} \hspace{0.05 cm} (p_{1} + i \delta)
\end{aligned}
\end{equation}
Note that for $p_{1} \in (-\pi, \pi]$,
\begin{equation}
\begin{aligned}
&\text{sin}\hspace{0.05 cm} (p_{1}+i \delta) =  \text{sin}\hspace{0.05 cm} p_{1} + \mathcal{O}(\delta), \hspace{0.5 cm}
\text{sin}^{2}\hspace{0.05 cm} (p_{1}+i \delta) - \text{sin}^{2}\hspace{0.05 cm} p_{1} = \mathcal{O}(\delta) \\
&(1 - \text{cos} \hspace{0.05 cm} (p_{1} + i \delta))^{2} - (1 - \text{cos} \hspace{0.05 cm} p_{1})^{2} =  \mathcal{O}(\delta).
\end{aligned}
\end{equation}
Thus,
\begin{equation}
\begin{aligned}
M(p_{1}+i \delta)^{2} + |s(p_{1}+i \delta)|^{2}  &=  (1 - \text{cos} \hspace{0.05 cm} p_{1} + m_{f})^{2} + 
\text{sin}^{2}\hspace{0.05 cm} p_{1} +   \mathcal{O}(\delta) \\
|M(p_{1}+i \delta)^{2} + |s(p_{1}+i \delta)|^{2}| &\geqslant |\text{sin}^{2}\hspace{0.05 cm} p_{1} + 
(1 - \text{cos} \hspace{0.05 cm} p_{1} + m_{f})^{2}| -  \mathcal{O}(\delta) \\
&\geqslant m_{f} -  \mathcal{O}(\delta) \\
|M(p_{1}+i \delta)^{2} + |s(p_{1}+i \delta)|^{2}|^{-1} &\leqslant \frac{1}{m_{f} -  \mathcal{O}(\delta)}.
\end{aligned}
\end{equation}
Thus, $\delta < \mathcal{O}(1) m_{f}$.
\begin{equation}
|G(x,y)| \leqslant \frac{1}{2\pi} e^{- \delta (x-y)} \abs{ \int_{\Gamma} 
e^{i p_{1} (x-y)} \frac{1}{M(p_{1} + i \delta)^{2} + |s(p_{1} + i \delta)|^{2}} \hspace{0.1 cm} d p_{1}}.
\end{equation}
Set $|\delta| = \gamma_{2}$. Therefore, for some constant $0 < \gamma_{2} < \mathcal{O}(1) m_{f}$, 
$|S(x,y)| \leqslant c \hspace{0.05 cm} e^{-\gamma_{2} |x-y|}$. This completes the proof of Lemma 2.3.

Recall that $\mathbb{T} = (\mathbb{Z}/L \mathbb{Z})^{4}$ and $L \gg [r_{\lambda}]$. Thus,
\begin{equation}
G_{[-\frac{L}{2}, \frac{L}{2}]}(x) =  G(x) + \sum_{n \in \mathbb{Z}, n \neq 0} G(x + n L).  \nonumber
\end{equation}
Note that for $n \neq 0$, $|x+nL| \geqslant \frac{L}{2} \geqslant |x|$, therefore,
\begin{equation}
\begin{aligned}
|G_{[-\frac{L}{2}, \frac{L}{2}]}(x)| &\leqslant |G(x)| + \sum_{n \in \mathbb{Z}, n \neq 0} |G(x + n L)| \\
&\leqslant \mathcal{O}(e^{-\gamma_{2} |x|}) + \sum_{n \in \mathbb{Z}, n \neq 0} \mathcal{O}(e^{-\gamma_{2} |x+nL|}) \\
&\leqslant \mathcal{O}(e^{-\gamma_{2} |x|}) + \mathcal{O}(e^{-\frac{\gamma_{2}}{2} |x|})
\sum_{n \in \mathbb{Z}, n \neq 0} \mathcal{O}(e^{-\frac{\gamma_{2}}{2} |x+nL|}) \\
&\leqslant \mathcal{O}(e^{-\frac{\gamma_{2}}{2} |x|}).  \nonumber
\end{aligned}
\end{equation}

\textbf{Lemma 2.4} \cite{G1} For $\Phi : \mathbb{T} \rightarrow  \mathbb{R}$,
with change of variables, $\Phi = C^{\frac{1}{2},\text{loc}}_{\Lambda_{1}} \Phi^{\prime}$, we have
\begin{equation}
\langle\Phi, C^{-1}_{\Lambda_{1}} \Phi\rangle =  
\langle\Phi^{\prime}, C^{\frac{1}{2},\text{loc}}_{\Lambda_{1}} C^{-1}_{\Lambda_{1}} C^{\frac{1}{2},\text{loc}}_{\Lambda_{1}} \Phi^{\prime}\rangle 
  = ||\Phi^{\prime}||_{\Lambda_{1}}^{2} + V_{\varepsilon}(\Lambda_{1}, \Phi^{\prime}). 
\end{equation}
Then $V_{\varepsilon}(\Lambda_{1}, \Phi^{\prime})$ has a local expansion in $\Box$
\begin{equation}
V_{\varepsilon} = \sum_{\Box \subset \Lambda_{1}} V_{\varepsilon} (\Box) \hspace{1 cm} \text{with}  \hspace{1 cm}
|V_{\varepsilon}(\Box)| \leqslant c \hspace{0.05 cm} [r_{\lambda}]^{4} \hspace{0.05 cm} e^{- \gamma^{\prime} r_{\lambda}}
||\Phi^{\prime}||_{\infty}^{2}.
\end{equation} 
 
\textbf{Lemma 2.5} Under a transformation $\psi \rightarrow S^{\text{loc}} \psi$
\begin{equation}
\langle\bar{\psi},  \mathrm{D}\psi\rangle =  \langle \bar{\psi}, \psi \rangle -  \langle \bar{\psi},  \mathrm{D} \delta S \psi \rangle
=  \langle \bar{\psi}, \psi \rangle + V_{\epsilon} (\bar{\psi}, \psi).
 \end{equation}
Then $V_{\epsilon} (\bar{\psi}, \psi)$ has a local expansion in $\Box$
\begin{equation}
V_{\epsilon} (\bar{\psi}, \psi) = \sum_{\Box} V_{\epsilon} (\bar{\psi}, \psi, \Box) \hspace{1 cm} \text{with}  \hspace{1 cm}
||V_{\epsilon}(\bar{\psi}, \psi, \Box)||_{h} \leqslant c \hspace{0.05 cm} [r_{\lambda}]^{4} \hspace{0.05 cm} 
e^{-\frac{\gamma_{2}}{2} r_{\lambda}} h^{2}.
\end{equation}
\textit{Proof} Using the identity $1 = \sum_{\Box} 1_{\Box}$, where,
\begin{equation}
\begin{aligned}
V_{\varepsilon}(\bar{\psi}, \psi, \Box) = \langle \bar{\psi}, 1_{\Box}  \mathrm{D} \delta S \psi \rangle 
=  \sum_{y : x \in \Box} ( \mathrm{D} \delta S)(x,y) \psi(x) \bar{\psi}(y).
\end{aligned}
\end{equation}
Then
\begin{equation}
||V_{\epsilon}(\bar{\psi}, \psi, \Box)||_{h} = [r_{\lambda}]^{4} \hspace{0.05 cm} 
\sum_{y} |( \mathrm{D} \delta S)(x,y)| \hspace{0.05 cm} h^{2}
\end{equation}
Note that $||\mathrm{D}|| \sim \mathcal{O}(m_{f})$ which is bounded by some constant $c$ and $\delta S$ is non zero only
when $|x-y| > r_{\lambda}$,
\begin{equation}
\begin{aligned}
||V_{\epsilon}(\bar{\psi}, \psi, \Box)||_{h} &\leqslant c \hspace{0.05 cm} [r_{\lambda}]^{4} \hspace{0.05 cm} 
\sum_{y} e^{-\gamma_{2} |x-y|} \hspace{0.05 cm} h^{2} \\
&\leqslant c \hspace{0.05 cm} [r_{\lambda}]^{4} \hspace{0.05 cm} 
e^{-\frac{\gamma_{2}}{2} r_{\lambda}} h^{2} \sum_{y} e^{-\frac{\gamma_{2}}{2} |x-y|} \\
&\leqslant c \hspace{0.05 cm} [r_{\lambda}]^{4} \hspace{0.05 cm} e^{-\frac{\gamma_{2}}{2} r_{\lambda}} h^{2}.
\end{aligned}
\end{equation} 
 
\textbf{Lemma 2.6} \cite{G1} Let $W_{1}(\Lambda_{1}) = -\sum_{n=1}^{\infty} \frac{1}{n} \text{Tr} 
\big[(C^{-\frac{1}{2}}_{\Lambda_{1}}\delta C^{\frac{1}{2}}_{\Lambda_{1}})^{n}\big]$. Then $W_{1}$ has a local expansion in $\Box$, 
\begin{equation}
\begin{aligned}
W_{1} = \sum_{\Box \subset \Lambda_{1}} W_{1}(\Box), \hspace{1 cm} \text{with} \hspace{1 cm}
 |W_{1}(\Box)| \leqslant c \hspace{0.05 cm}[r_{\lambda}]^{4} e^{-\gamma^{\prime} r_{\lambda}}
\end{aligned}
\end{equation}

\textbf{Lemma 2.7} Let $W_{1}^{\prime}(\Lambda_{1}) = -\sum_{n=1}^{\infty} \frac{1}{n} 
\text{Tr} \big[(\mathrm{D}_{\Lambda_{1}}\delta S_{\Lambda_{1}})^{n}\big]$. Then $W_{1}^{\prime}$ has a local expansion in $\Box$, 
\begin{equation}
\begin{aligned}
W_{1}^{\prime} = \sum_{\Box \subset \Lambda_{1}} W_{1}^{\prime}(\Box), \hspace{1 cm} \text{with} \hspace{1 cm}
 |W_{1}^{\prime}(\Box)| \leqslant c \hspace{0.05 cm}[r_{\lambda}]^{4} e^{-\gamma_{2} r_{\lambda}}
\end{aligned}
\end{equation}
\textit{Proof} From the definition of $W_{1}^{\prime}$,
\begin{equation}
\begin{aligned}
W_{1}^{\prime}(\Box) = -\sum_{n=1}^{\infty} \frac{1}{n} \text{Tr} \big[1_{\Box}(\mathrm{D}_{\Lambda_{1}}\delta S_{\Lambda_{1}})^{n}\big]
= -\sum_{n=1}^{\infty} \frac{1}{n} \sum_{x \in \Box} \big[(\mathrm{D}_{\Lambda_{1}} \delta S_{\Lambda_{1}})^{n} \delta_{x} \big] (x).
\end{aligned}
\end{equation}
Note that $||\mathrm{D}|| \sim \mathcal{O}(m_{f})$ and $|\delta S(x,y)| < c e^{-\gamma_{2}|x-y|}$ with $|x-y| > r_{\lambda}$,  
\begin{equation}
\left| \big[(\mathrm{D}_{\Lambda_{1}} \delta S_{\Lambda_{1}})^{n} \delta_{x} \big] (x)\right|
\leqslant (c \hspace{0.05 cm} e^{-\gamma_{2} r_{\lambda}})^{n} ||\delta_{x}||
\leqslant (c \hspace{0.05 cm} e^{-\gamma_{2} r_{\lambda}})^{n}
\end{equation}
and so $|W_{1}^{\prime}(\Box)| \leqslant \sum_{n=1}^{\infty} \frac{1}{n} \sum_{x} (c \hspace{0.05 cm} e^{-\gamma_{2} r_{\lambda}})^{n}
= [r_{\lambda}]^{4} \sum_{n=1}^{\infty} \frac{1}{n} (c \hspace{0.05 cm} e^{-\gamma_{2} r_{\lambda}})^{n}$ and summing over
n gives the result.

Define
\begin{equation}
\begin{aligned}
W_{2} &=   \text{Tr}\hspace{0.6 mm}\text{log}\hspace{0.5 mm}(C^{\frac{1}{2}}_{\Lambda_{1}}) - 
  \text{Tr}\hspace{0.6 mm}\text{log} \hspace{0.5 mm} (C^{\frac{1}{2}}) 
=  - \frac{1}{2} \text{Tr}\hspace{0.6 mm}\text{log}\hspace{0.5 mm}(\text{T}_{\Lambda_{1}}) + 
\frac{1}{2}  \text{Tr}\hspace{0.6 mm}\text{log} \hspace{0.5 mm} (\text{T}) \\
W_{2}^{\prime} &=   \text{Tr}\hspace{0.6 mm}\text{log}\hspace{0.5 mm}(S_{\Lambda_{1}}) - 
  \text{Tr}\hspace{0.6 mm}\text{log} \hspace{0.5 mm} (S) 
=  - \text{Tr}\hspace{0.6 mm}\text{log}\hspace{0.5 mm}(\mathrm{D}_{\Lambda_{1}}) + 
  \text{Tr}\hspace{0.6 mm}\text{log} \hspace{0.5 mm} (\mathrm{D}).
\end{aligned}
\end{equation}

\textbf{Lemma 2.8} Determinant identity. 
\begin{enumerate}
  \item \cite{B} Let \textit{K} be a positive self-adjoint matrix. Then $\text{det} \hspace{0.1 cm} K= 
e^{\text{Tr\hspace{0.02 cm} log} \hspace{0.1 cm}K}$ where for 
any $R_{0} > 0$
\begin{equation}
 \text{log} \hspace{0.1 cm}K= K\int_{R_{0}}^{\infty} \frac{dx}{x} (K+x)^{-1} - \int_{0}^{R_{0}}dx (K+x)^{-1} +  \text{log} \hspace{0.1 cm} R_{0}.
\end{equation}
\item \cite{D3} Let \textit{M} be an invertible self-adjoint matrix. Then $\text{det} \hspace{0.1 cm} M = 
e^{\text{Tr\hspace{0.02 cm} log} \hspace{0.1 cm}M}$ where for any $R_{0} > 0$
\begin{equation}
 \text{log} \hspace{0.1 cm}M= M\int_{R_{0}}^{\infty} \frac{dy}{y} (M+i y)^{-1} - i \int_{0}^{R_{0}}dy (M+i y)^{-1} 
 +  \text{log} \hspace{0.1 cm} R_{0} + i \frac{\pi}{2}.
\end{equation}
\end{enumerate}

\textbf{Lemma 2.9} \cite{G1} Let 
\begin{equation}
\begin{aligned}
\mathcal{A}(\Box) = \text{Tr\hspace{0.02 cm}} (1_{\Box}\hspace{0.05 cm} \mathcal{A} \hspace{0.05 cm} 1_{\Box}) &= 
\text{Tr\hspace{0.02 cm}} (1_{\Box}(\text{T}_{\Lambda_{1}}+ r)^{-1} - (\text{T}+ r)^{-1} 1_{\Box}) \\
\mathcal{B}(\Box) = \text{Tr\hspace{0.02 cm}} (1_{\Box}\hspace{0.05 cm} \mathcal{B} \hspace{0.05 cm} 1_{\Box}) &= 
\text{Tr\hspace{0.02 cm}} (1_{\Box} \text{T} (\text{T}+ r)^{-1} - \text{T}_{\Lambda_{1}} (\text{T}_{\Lambda_{1}}+ r)^{-1} 1_{\Box}).
\end{aligned}
\end{equation}
Let $m = \text{min} (\mu^{2}, m_{A}^{2})$. For $\Box \subset \Omega_{1}$ and $\frac{1}{m} < \frac{1}{32}$
\begin{equation}
\begin{aligned}
|\mathcal{A}(\Box)| \leqslant \mathcal{O}(m^{-2[r_{\lambda}]})  \hspace{0.5 cm} \text{and} \hspace{0.5 cm} 
|\mathcal{B}(\Box)| \leqslant   \mathcal{O}(m^{-2[r_{\lambda}]}).
\end{aligned}
\end{equation}

\textbf{Lemma 2.10} Let 
\begin{equation}
\begin{aligned}
\mathcal{A}^{\prime}(\Box) = \text{Tr\hspace{0.02 cm}} (1_{\Box}\hspace{0.05 cm} \mathcal{A}^{\prime} \hspace{0.05 cm} 1_{\Box}) 
&= \text{Tr\hspace{0.02 cm}} (1_{\Box}(\mathrm{D}_{\Lambda_{1}}+ r)^{-1} - (\mathrm{D}+ r)^{-1} 1_{\Box}) \\
\mathcal{B}^{\prime}(\Box) = \text{Tr\hspace{0.02 cm}} (1_{\Box}\hspace{0.05 cm} \mathcal{B}^{\prime} \hspace{0.05 cm} 1_{\Box}) 
&= \text{Tr\hspace{0.02 cm}} (1_{\Box} \mathrm{D} (\mathrm{D}+ r)^{-1} 
- \mathrm{D}_{\Lambda_{1}} (\mathrm{D}_{\Lambda_{1}}+ r)^{-1} 1_{\Box}).
\end{aligned}
\end{equation}
For $\Box \subset \Omega_{1}$ and $\frac{1}{m_{f}} < \frac{1}{12}$
\begin{equation}
\begin{aligned}
|\mathcal{A}^{\prime}(\Box)| \leqslant \mathcal{O}(m_{f}^{-2[r_{\lambda}]})  \hspace{0.5 cm} \text{and} \hspace{0.5 cm} 
|\mathcal{B}^{\prime}(\Box)| \leqslant   \mathcal{O}(m_{f}^{-2[r_{\lambda}]}).
\end{aligned}
\end{equation}
\textit{Proof} is similar to Lemma 2.9 using $||\mathfrak{D}|| \leqslant 12$ which follows from (1.9).

\textbf{Lemma 2.11} \cite{G1}  $W_{2}$ has a local expansion in $\Box$, 
\begin{equation}
\begin{aligned}
W_{2} = \sum_{\Box \subset \mathbb{T}} W_{2}(\Box), \hspace{0.7 cm} \text{with} \hspace{0.7 cm}
 |W_{2}(\Box)| \leqslant  \begin{cases}
\mathcal{O}(m^{-2[r_{\lambda}]}) & \hspace{0.1 cm}  \text{if}  \hspace{0.2 cm} \Box \subset \Omega_{1}, \\
\mathcal{O}([r_{\lambda}]^{4}) & \hspace{0.1 cm} \text{if}  \hspace{0.2 cm} \Box \subset \Omega_{1}^{c}.
\end{cases}
\end{aligned}
\end{equation}

\textbf{Lemma 2.12}  $W_{2}^{\prime}$ has a local expansion in $\Box$, 
\begin{equation}
\begin{aligned}
W_{2}^{\prime} = \sum_{\Box \subset \mathbb{T}} W_{2}(\Box), \hspace{0.7 cm} \text{with} \hspace{0.7 cm}
 |W_{2}^{\prime}(\Box)| \leqslant  \begin{cases}
\mathcal{O}(m_{f}^{-2[r_{\lambda}]}) & \hspace{0.1 cm}  \text{if}  \hspace{0.2 cm} \Box \subset \Omega_{1}, \\
\mathcal{O}([r_{\lambda}]^{4}) & \hspace{0.1 cm} \text{if}  \hspace{0.2 cm} \Box \subset \Omega_{1}^{c}.
\end{cases}
\end{aligned}
\end{equation}
\textit{Proof} is similar to Lemma 2.11.

\section{Power series representation}
Let $X \subset \mathbb{T}$ and $x, y \in X$. Every site has real valued bosonic fields $\rho(x)$ and 
$A_{\mu}(x, x+e_{\mu})$ (an oriented bond $(x, x + e_{\mu})$) that are scalar functions.
Let $\text{J}(y)$ and $\text{J}(y, y + e_{\mu})$ be the scalar external bosonic source fields defined on $y$.  
Every site also has fermionic fields $\psi_{\alpha}(x), \bar{\psi}_{\alpha}(x)$ 
and external fermionic source fields $\eta_{\alpha}(y), \bar{\eta}_{\alpha}(y)$
where $\alpha$ is spinor index.  Let $V$ be a vector space over $\mathbb{C}$ with basis $\psi_{\alpha}(x)$ 
and $\bar{\psi}_{\alpha}(x)$ and let $V^{\prime}$
be a vector space over $\mathbb{C}$ with basis $\eta_{\alpha}(y)$ and $\bar{\eta}_{\alpha}(y)$.
$\psi_{\alpha}(x), \bar{\psi}_{\alpha}(x)$ and $\eta_{\alpha}(y), \bar{\eta}_{\alpha}(y)$ are taken to be
anticommuting Grassmann variables.  
Let $\xi$ stands for $(x, \alpha, \omega)$ and $z$ stands for $(y, \alpha, \omega)$ with $\omega = (0,1)$.
Define
\begin{equation}
 \Phi(u) = \begin{cases}
  \rho(x)      & \text{if} \hspace{0.4 cm} u = x  \\
 A_{\mu}(x)   & \text{if} \hspace{0.4 cm} u = (x, x+e_{\mu})
\end{cases}
\end{equation} 
\begin{equation}
 \Theta(s) = \begin{cases}
  \text{J}(y)      & \text{if} \hspace{0.4 cm} s = y  \\
 \text{J}_{\mu}(y)   & \text{if} \hspace{0.4 cm} s = (y, y+e_{\mu})
\end{cases}
\end{equation} 
\begin{equation}
 \psi(\xi) = \begin{cases}
 \psi_{\alpha}(x)      & \text{if} \hspace{0.4 cm} \xi = (x, \alpha, 0) \\
\bar{\psi}_{\alpha}(x)   & \text{if} \hspace{0.4 cm} \xi = (x, \alpha, 1)
\end{cases}
\end{equation} 
\begin{equation}
 \eta(z) = \begin{cases}
 \eta_{\alpha}(y)      & \text{if} \hspace{0.4 cm} z = (y, \alpha, 0) \\
\bar{\eta}_{\alpha}(y)   & \text{if} \hspace{0.4 cm} z = (y, \alpha, 1)
\end{cases}
\end{equation} 
Then Grassmann algebra is generated by
\begin{equation}
\begin{aligned}
 \{\psi, \eta\} = 0, \hspace{0.6 cm} \{\psi, \psi\} = 0, \hspace{0.6 cm} \{\eta, \eta\} = 0
\end{aligned}
\end{equation}
where $\{,\}$ denotes the anticommutator. 

\textbf{Power series.}  A power series of bosonic and fermionic fields is written as
\begin{equation}
\begin{aligned}
f(\Phi, \Theta, \psi, \eta) = & \sum_{n,m \geqslant 0}  \sum_{\substack{\xi_{1},\cdots, \xi_{n} \in X \\ z_{1},\cdots, z_{m} \in X}}
a(\Phi, \Theta, \xi_{1},\cdots, \xi_{n},  z_{1},\cdots, z_{m}) \hspace{0.05 cm}
 \psi(\xi_{1}) \cdots \psi(\xi_{n}) \hspace{0.05 cm} \eta(z_{1}) \cdots \eta(z_{m}) \\
= & \sum_{k,l,n,m \geqslant 0} 
 \sum_{\substack{u_{1}, \cdots, u_{k} \in X \\ s_{1}, \cdots, s_{l} \in X \\ \xi_{1},\cdots, \xi_{n} \in X \\ z_{1},\cdots, z_{m} \in X}}
 a(u_{1}, \cdots, u_{k}, s_{1}, \cdots, s_{l}, \xi_{1},\cdots, \xi_{n}, z_{1},\cdots, z_{m}) \\  
& \Phi(u_{1}) \cdots \Phi(u_{k})\hspace{0.05 cm} \Theta(s_{1}) \cdots \Theta(s_{l}) \hspace{0.05 cm}
 \psi(\xi_{1}) \cdots \psi(\xi_{n}) \hspace{0.05 cm} \eta(z_{1}) \cdots \eta(z_{m})
\end{aligned}
\end{equation}
where the coefficients $a(u_{1}, \cdots, u_{k}, s_{1}, \cdots, s_{l}, \xi_{1},\cdots, \xi_{n}, z_{1},\cdots, z_{m})
\in \mathbb{C}$.  The coefficients are not assumed to be symmetric or antisymmetric. 

\textit{Notation}. Define a n-component vector $\vec{u}$ as  
\begin{equation}
\vec{u} = \{u_{1}, u_{2}, \cdots, u_{n} \}.
\end{equation}
Define concatenation of two vectors $\vec{u}$ and $\vec{z}$ as
\begin{equation}
\begin{aligned}
\vec{u} \circ \vec{z} &= \{u_{1}, \cdots, u_{n}, z_{1}, \cdots, z_{m}\} \\
(\vec{u}_{1},\cdots, \vec{u}_{s}) &\circ (\vec{z}_{1},\cdots, \vec{z}_{s}) =
(\vec{u}_{1} \circ \vec{z}_{1}, \cdots, \vec{u}_{s} \circ \vec{z}_{s}).
\end{aligned}
\end{equation}
Then for a n-component vector $\vec{u}$, write
\begin{equation}
\Phi (\vec{u}) = \Phi (u_{1}) \cdots \Phi (u_{n}).
\end{equation}
Rewrite the power series representation as 
\begin{equation}
\begin{aligned}
f(\Phi, \Theta, \psi, \eta) &= \sum_{n,m \geqslant 0} 
\sum_{\vec{\xi}, \vec{z} \subset X} 
a(\Phi, \Theta, \vec{\xi}, \vec{z}) \hspace{0.05 cm} \psi (\vec{\xi})\hspace{0.05 cm} \eta(\vec{z}) \\
&= \sum_{k,l,n,m \geqslant 0} \sum_{\vec{u}, \vec{s}, \vec{\xi}, \vec{z} \subset X} a(\vec{u}, \vec{s},\vec{\xi}, \vec{z}) \hspace{0.05 cm} 
\Phi(\vec{u}) \hspace{0.05 cm} \Theta(\vec{s})\hspace{0.05 cm} \psi (\vec{\xi})\hspace{0.05 cm} \eta(\vec{z})
\end{aligned}
\end{equation}
Here, we assume that the total number fermion fields is even, that is, $n+m = $ even.
Note that any change in the order of fermionic fields $\psi, \eta$ is accompanied by a negative sign.
Let
\begin{equation}
\begin{aligned}
a_{\Theta \eta}(\vec{u}, \vec{\xi}) &= \sum_{\vec{s}, \vec{z} \subset X}  a(\vec{u}, \vec{s},\vec{\xi}, \vec{z})
 \hspace{0.05 cm}  \Theta(\vec{s})\hspace{0.05 cm}  \eta(\vec{z})
\end{aligned}
\end{equation}
and rewrite (3.10) as
\begin{equation}
f(\Phi, \Theta, \psi, \eta) = \sum_{\vec{u}, \vec{\xi} \subset X} a_{\Theta \eta}(\vec{u}, \vec{\xi}) 
\hspace{0.05 cm} \Phi(\vec{u}) \hspace{0.05 cm} \psi (\vec{\xi}). 
\end{equation}
The power series representation of a functional; $e^{f(\Phi, \Theta, \psi, \eta)}$ 
is written as
\begin{equation}
\begin{aligned}
& e^{f(\Phi, \Theta, \psi, \eta)} = \sum_{l=0}^{\infty} \frac{1}{l!} f(\Phi, \Theta, \psi, \eta)^{l} \\
&= 1 + \sum_{l=1}^{\infty}  \frac{1}{l!}
\sum_{\substack{\vec{u}_{1},\cdots, \vec{u}_{l} \subset X \\ \vec{\xi}_{1},\cdots, \vec{\xi}_{l} \subset X}} 
(-1)^{\#} a_{\Theta \eta}(\vec{u}_{1}, \vec{\xi}_{1})  \cdots  a_{\Theta \eta}(\vec{u}_{l}, \vec{\xi}_{l})  
 \hspace{0.05 cm} \Phi(\vec{u}_{1}) \cdots \Phi(\vec{u}_{l}) \hspace{0.05 cm} \psi (\vec{\xi}_{1}) \cdots  \psi (\vec{\xi}_{l})
\end{aligned}
\end{equation}
where $\#$ denotes the total number of interchanges of fermionic fields $\psi, \eta$. 
The normalized Gaussian measures with unit covariance are
\begin{equation}
\begin{aligned}
 d\mu_{I}(\bar{\psi}, \psi) &=  
 \frac{ \mathcal{D}\bar{\psi} \mathcal{D}\psi  \hspace{0.05 cm} 
  e^{- \langle \bar{\psi}, \psi \rangle}} 
{\int \mathcal{D}\bar{\psi} \mathcal{D}\psi  \hspace{0.05 cm} 
e^{- \langle \bar{\psi}, \psi \rangle}} \\
d\mu_{\text{I}}(\Phi)  &= \frac{\mathcal{D}\Phi e^{-\frac{1}{2}||\Phi||^{2}}}
{\int \mathcal{D}\Phi e^{-\frac{1}{2}||\Phi||^{2}}}.
\end{aligned}
\end{equation}
Let $\hat{\chi}(\Phi)$ be some characteristic function which is zero if $\Phi$ is unbounded.
Denote
\begin{equation}
\begin{aligned}
\Xi(\Theta, \eta) &=  \int d\mu_{I}(\psi) \hspace{0.05 cm} d\mu_{I}(\Phi) \hat{\chi}(\Phi) \hspace{0.05 cm} e^{f(\Phi, \Theta, \psi, \eta)} \\
\Xi(0, 0) &=  \int d\mu_{I}(\psi) \hspace{0.05 cm} d\mu_{I}(\Phi) \hat{\chi}(\Phi) \hspace{0.05 cm} e^{f(\Phi, 0, \psi, 0)}
\end{aligned}
\end{equation}
We would like to show that a power series expansion of 
$\text{log}\hspace{0.1 cm} \frac{\Xi(\Theta, \eta)}{\Xi(0, 0)}$ exists and is given by
\begin{equation}
\begin{aligned}
\text{log}\hspace{0.1 cm} \frac{\Xi(\Theta, \eta)}{\Xi(0, 0)} &=  b(s_{1}) \Theta(s_{1}) + 
 \eta(z_{1}) \hspace{0.05 cm} b(z_{1}, z_{2}) \hspace{0.05 cm}  \eta(z_{2}) + 
  \eta(z_{1}) \hspace{0.05 cm} b(s_{1}, z_{1}, z_{2}) \hspace{0.05 cm}  \eta(z_{2}) \Theta(s_{1}) \\
& + \cdots +   \eta(z_{1}) \hspace{0.05 cm} b(s_{1},\cdots,s_{l}, z_{1}, z_{2}) \hspace{0.05 cm}  \eta(z_{2})
 \Theta(s_{1}) \cdots \Theta(s_{l}) + \\
 & b(s_{1},s_{2},s_{3})  \Theta(s_{1}) \Theta(s_{2}) \Theta(s_{3}) +  \eta(z_{1}) \eta(z_{2}) \hspace{0.05 cm}
  b(z_{1}, z_{2}, z_{3}, z_{4}) \hspace{0.05 cm} \eta(z_{3}) \eta(z_{4}) + \cdots \\
 &=   \sum_{l + m \geqslant 1} \eta(z_{1}) \cdots \eta(z_{m}) \hspace{0.05 cm} b(s_{1}, \cdots, s_{l}, z_{1}, \cdots, z_{2m}) 
\hspace{0.05 cm}  \eta(z_{m+1}) \cdots \eta(z_{2m})  \Theta(s_{1}) \cdots \Theta(s_{l}) \\
&= \sum_{l + m \geqslant 1}\eta(\vec{z}_{m}) \hspace{0.05 cm} b(\vec{s}, \vec{z})
 \hspace{0.05 cm} \eta(\vec{z}_{2m}) \Theta(\vec{s})
\end{aligned}
\end{equation}
where the coefficient system $ b(s_{1}, \cdots, s_{l}, z_{1}, \cdots, z_{2m})$ is also not assumed to be 
symmetric or antisymmetric. 

\textbf{Weight system.} For a n-component vector $\vec{x}$ a weight system $w(\vec{x})$ is a function
which assigns a positive number $w(x_{1}, \cdots, x_{n})$ to $\vec{x}$ and satisfies
the following two properties:
\begin{enumerate}
  \item $w(x_{1}, \cdots, x_{n})$ is invariant under the permutations of the components
  of $\vec{x}$ and
  \item for any two vectors $\vec{x}$ and $\vec{z}$ with 
  $\text{supp}(\vec{x}) \cap  \text{supp}(\vec{z}) \neq \oslash $
  \begin{equation}
w(\vec{x}, \vec{z}) = w(\vec{x} \circ \vec{z}) \leqslant w(\vec{x}) \hspace{0.05 cm} w(\vec{z}).      \nonumber
\end{equation}
\end{enumerate}

\textbf{Norm.}
Let $w_{p_{1}, h_{1}, p_{2}, h_{2}}(\vec{u}, \vec{s},\vec{\xi}, \vec{z}) =  
e^{\kappa t(\text{supp}(\vec{u}, \vec{s},\vec{\xi}, \vec{z}))} p_{1}^{k} h_{1}^{l} p_{2}^{n} h_{2}^{m}$ 
be the weight system with mass $\kappa$ giving weight at least $p_{1}$ to $\Phi, h_{1}$ to $\Theta, p_{2}$ to $\psi$ and 
$h_{2}$ to $\eta$. Let the coefficients $a$ be the symmetric and antisymmetric realization of the coefficient system 
$a(\vec{u}, \vec{s},\vec{\xi}, \vec{z})$ in bosonic and fermionic variables respectively. Define  
\begin{equation}
\begin{aligned}
 |a|_{w_{p_{1}, h_{1}, p_{2}, h_{2}}} = \sum_{k, l, n, m \geqslant 0} &
\max\limits_{x \in X}  \hspace{0.05 cm} \max\limits_{\substack{1\leqslant i \leqslant k \\ 1\leqslant j \leqslant l \\
1\leqslant i^{\prime} \leqslant n \\ 1\leqslant j^{\prime} \leqslant m}} 
\sum_{\substack{u_{1}, \cdots, u_{k}, s_{1}, \cdots, s_{l} \in X \\
\xi_{1}, \cdots, \xi_{n}, z_{1}, \cdots, z_{m} \in X \\ r_{i}   \hspace{0.05 cm} \text{or} s_{j}  \hspace{0.05 cm} \text{or} \hspace{0.05 cm}
\xi_{i^{\prime}} \hspace{0.05 cm} \text{or} \hspace{0.05 cm} z_{j^{\prime}} = x}} 
e^{\kappa t(\text{supp}(\vec{u}, \vec{s},\vec{\xi}, \vec{z}))} \\ 
& p_{1}^{k} h_{1}^{l} p_{2}^{n} h_{2}^{m} \hspace{0.05 cm}  
|a(u_{1}, \cdots, u_{k}, s_{1}, \cdots, s_{l}, \xi_{1},\cdots, \xi_{n}, z_{1},\cdots, z_{m})|.
\end{aligned}
\end{equation}
Then the norm is defined as  $||f||_{w_{p_{1}, h_{1}, p_{2}, h_{2}}} = |a|_{w_{p_{1}, h_{1}, p_{2}, h_{2}}}$. 
Here, $\max\limits_{x \in X}$ breaks translation invariance.

The norm for $\text{log}\hspace{0.1 cm} \frac{\Xi(\Theta, \eta)}{\Xi(0, 0)}$ is defined in the same manner. 
Let $w_{h_{1},h_{2}}(\vec{s}, \vec{z}) = e^{\kappa t(\text{supp}(\vec{s},\vec{z}))} h_{1}^{n(\vec{s})} h_{2}^{n(\vec{z})}$ 
be the weight system of mass $\kappa$ (where $n(\vec{s})$ and $n(\vec{z})$ denote the $\#$ of sites in 
$\vec{s}$ and $\vec{z}$ respectively) giving weight $h_{1}$ to the field $\Theta$ and $h_{2}$ to the field $\eta$. Define
\begin{equation}
|b|_{w_{h_{1},h_{2}}} = \sum_{l+m \geqslant 1}   
\max\limits_{x \in X} \max\limits_{\substack{1\leqslant i \leqslant l \\  1 \leqslant j \leqslant 2m}} 
\sum_{\substack{s_{1}, \cdots, s_{l}, z_{1}, \cdots, z_{2m} \in X 
 \\ s_{i} \hspace{0.05 cm} \text{or} \hspace{0.05 cm} z_{j} = x}} w_{h_{1},h_{2}}(\vec{s}, \vec{z}) 
\hspace{0.05 cm} |b(s_{1}, \cdots, s_{l}, z_{1}, \cdots, z_{2m})|
\end{equation}
where coefficients $b$ is the symmetric and antisymmetric realization of the coefficient system 
$b(\vec{s}, \vec{z})$ in $\vec{s}$ and $\vec{z}$ respectively.
Then the norm  is
\begin{equation}
||\text{log}\hspace{0.1 cm} \frac{\Xi(\Theta, \eta)}{\Xi(0, 0)}||_{w_{h_{1},h_{2}}} = |b|_{w_{h_{1},h_{2}}}. \nonumber 
\end{equation}

\textbf{Theorem 3.1} Let $f(\Phi, \Theta, \psi, \eta)$ has a power series representation with 
coefficient system $a(\vec{u}, \vec{s},\vec{\xi}, \vec{z})$ and $||f||_{w_{4p_{1}, h_{1}, 4p_{2}, h_{2}}} < \frac{1}{16}$.
 Then there exists a coefficient system $b(\vec{s}, \vec{z})$ having the form (3.16) such that
$\text{log}\hspace{0.1 cm} \frac{\Xi(\Theta, \eta)}{\Xi(0, 0)} = 
\sum_{\vec{s}, \vec{z}} b(\vec{s}, \vec{z}) \hspace{0.05 cm} \Theta(\vec{s})\hspace{0.05 cm} \eta(\vec{z})$ and
\begin{equation}
|b|_{w_{h_{1},h_{2}}} \leqslant 
\frac{||f||_{w_{4p_{1}, h_{1}, 4p_{2}, h_{2}}}}{1 - 16 ||f||_{w_{4p_{1}, h_{1}, 4p_{2}, h_{2}}}} .
\end{equation}

\subsection{Proof}
The proof of the theorem builds upon the proof of the similar theorems for purely bosonic functional
integrals due to Balaban, Feldman, Kn$\ddot{\text{o}}$rrer and Trubowitz \cite{BFKT} 
and purely fermionic functional integrals in our previous work \cite{G2}. The proof consists of two
parts; (i) we show the existence of a desired coefficient system $b(\vec{s}, \vec{z})$ and 
(ii) then we prove the bound as written in the theorem.

\subsubsection{}
\textbf{Definition.} Let $X \subset \mathbb{T}$ and let $\vec{x}_{1}, \cdots, \vec{x}_{l}$ be a collection of vectors 
in $X$. Denote $X_{i} = \text{supp}(\vec{x}_{i})$ for $i = 1\hspace{0.1 cm} \text{to} \hspace{0.1 cm}l$.
Consider the set $\{X_{1}, \cdots, X_{l}\}$. Define \textit{incidence graph} $G(X_{1}, \cdots, X_{l})$ 
to be a graph with vertices $\{1, \cdots, l\}$ and edges $(i,j)$ whenever $X_{i} \cap X_{j} \neq \oslash$.
The set $\{\vec{x}_{1}, \cdots, \vec{x}_{l}\}$ is called \textit{connected} if the incidence graph  $G(X_{1}, \cdots, X_{l})$ 
is connected. For any set $Z \subset X$ define a \textit{connected cover} $\mathcal{C}(Z)$ 
to be the set of all ordered connected \textit{l}-tuples $\{\vec{x}_{1}, \cdots, \vec{x}_{l}\}$ for which
$Z = X_{1} \cup \cdots \cup X_{l}$.

\textbf{Proposition 3.1} Let $a(\vec{u}, \vec{s},\vec{\xi}, \vec{z})$ be a coefficient system  
representing a power series of $f(\Phi, \Theta, \psi, \eta)$.   
Let $Z = \text{supp} (\vec{u}_{1}, \cdots, \vec{u}_{j}, \vec{\xi}_{1}, \cdots, \vec{\xi}_{j})$.  
Let $\mathcal{C}(Z)$ denote a connected cover of $Z$. Then there exists a function $K_{0}(Z, \Theta, \eta)$ given by
\begin{equation}
\begin{aligned}
K_{0}(Z, \Theta, \eta) = (\pm 1) \sum_{k=1}^{\infty} \frac{1}{k!}  
 \sum_{(\vec{u}_{1},\vec{\xi}_{1}; \cdots;  \vec{u}_{k}, \vec{\xi}_{k}) \in  \mathcal{C}(Z)}
& a_{\Theta \eta}(\vec{u}_{1}, \vec{\xi}_{1}) \cdots  a_{\Theta \eta}(\vec{u}_{k}, \vec{\xi}_{k}) \\
& \int d\mu_{I}(\Phi) \hat{\chi}(\Phi)   \Phi(\vec{u}_{1}) \cdots \Phi(\vec{u}_{k}) 
\end{aligned}
\end{equation}
such that
\begin{equation}
\int d\mu_{I}(\psi) \hspace{0.05 cm} d\mu_{I}(\Phi) \hat{\chi}(\Phi) \hspace{0.05 cm} e^{f(\Phi, \Theta, \psi, \eta)}
= 1 +  \sum_{n=1}^{\infty} \frac{1}{n!} 
\sum_{\substack{Z_{1}, \cdots, Z_{n} \subset X \\ Z_{i} \cap Z_{j} = \oslash}} (-1)^{\#} \prod_{j=1}^{n} K_{0}(Z_{j}, \Theta, \eta).
\end{equation}

\textit{Proof} Let $a(-, \vec{s}, -, \vec{z})$ denote the $k = n = 0$ terms in (3.10). Consider 
\begin{equation}
 \frac{\Xi(\Theta, \eta)}{\Xi(0, 0)} =  
 \frac{\int d\mu_{I}(\psi) \hspace{0.05 cm} d\mu_{I}(\Phi) \hat{\chi}(\Phi) \hspace{0.05 cm} e^{f(\Phi, \Theta, \psi, \eta)}}
 {\int d\mu_{I}(\psi) \hspace{0.05 cm} d\mu_{I}(\Phi) \hat{\chi}(\Phi) \hspace{0.05 cm} e^{f(\Phi, 0, \psi, 0)}}.   \nonumber
\end{equation}
We factor out $e^{f(0, \Theta, 0, \eta)}$ from the integral in the numerator and $e^{f(0, 0, 0, 0)}$ from the integral in the denominator
and assume that $f(0, \Theta, 0, \eta) = 0$ such that $a(-, \vec{s}, -, \vec{z}) = 0$ for all $\vec{s}, \vec{z} \in X$.
We would like to arrange the power series of functional $e^{f(\Phi, \Theta, \psi, \eta)}$  
\begin{equation}
\begin{aligned}
& e^{f(\Phi, \Theta, \psi, \eta)} = \sum_{l=0}^{\infty} \frac{1}{l!} f(\Phi, \Theta, \psi, \eta)^{l} \\
&= 1 + \sum_{l=1}^{\infty}  \frac{1}{l!}\sum_{\substack{\vec{u}_{1},\cdots, \vec{u}_{l} \in X \\ \vec{\xi}_{1},\cdots, \vec{\xi}_{l} \in X}} 
(-1)^{\#} a_{\Theta \eta}(\vec{u}_{1}, \vec{\xi}_{1})  \cdots  a_{\Theta \eta}(\vec{u}_{l}, \vec{\xi}_{l})  
 \hspace{0.05 cm} \Phi(\vec{u}_{1}) \cdots \Phi(\vec{u}_{l}) \hspace{0.05 cm} \psi (\vec{\xi}_{1}) \cdots  \psi (\vec{\xi}_{l})
\end{aligned} \nonumber
\end{equation}
such that the $\text{supp} (\vec{u}_{i}, \vec{\xi}_{i})$ should overlap with the $\text{supp} (\vec{u}_{j}, \vec{\xi}_{j})$
for $i \neq j$. This is required so that the decay properties of the coefficient system $a(\vec{u}, \vec{s},\vec{\xi}, \vec{z})$
also manifest themselves in the resulting coefficient system $b(\vec{s}, \vec{z})$. 
Recall that $\xi = (x, \alpha, a, \omega)$. Here, \textit{overlap} refers to the overlap of sites and the bonds only.  
We identify the sites of $\vec{u}_{i}, \vec{\xi}_{i}$ as $\vec{x}_{i}$ for $i = 1\hspace{0.1 cm} \text{to} \hspace{0.1 cm}l$.
Let $X_{i} = \text{supp}(\vec{u}_{i}, \vec{\xi}_{i})$ and $Z = X_{1} \cup \cdots \cup X_{l}$.
   
Next decompose $Z$ into pairwise disjoint subsets $\{Z_{1}, \cdots, Z_{n}\}$ by
dividing $\{1, \cdots, l\}$ into pairwise disjoint subsets $I_{1}, \cdots, I_{n}$
such that for each $1 \leqslant j \leqslant n$, $(\vec{u}_{i}, \vec{\xi}_{i}), i \in I_{j}$ is $\mathcal{C}(Z_{j})$. 
Therefore, the power series is \cite{BFKT}  
\begin{equation}
\begin{aligned}
e^{f(\Phi, \Theta, \psi, \eta)} &= 1 + \sum_{l=1}^{\infty}  \frac{1}{l!} \sum_{n=1}^{l} \frac{1}{n!} 
\sum_{\substack{Z_{1}, \cdots, Z_{n} \subset X \\ Z_{i} \cap Z_{j} = \oslash}}
\sum_{\substack{I_{1} \cup \cdots \cup I_{n} = {1, \cdots, l} \\ I_{1}, \cdots, I_{n} \text{pairwise disjoint}}}
\sum_{\substack{(\vec{u}_{1},\vec{\xi}_{1}); \cdots; (\vec{u}_{l},\vec{\xi}_{l}) \in X \\ (\vec{u}_{i}, \vec{\xi}_{i}, i \in I_{j}) \in  \mathcal{C}(Z_{j})}}
\\ &
(-1)^{\#} a_{\Theta \eta}(\vec{u}_{1}, \vec{\xi}_{1})  \cdots  a_{\Theta \eta}(\vec{u}_{l}, \vec{\xi}_{l})  
 \hspace{0.05 cm} \Phi(\vec{u}_{1}) \cdots \Phi(\vec{u}_{l}) \hspace{0.05 cm} \psi (\vec{\xi}_{1}) \cdots  \psi (\vec{\xi}_{l}). 
\end{aligned}
\end{equation}
Following the combinatorics in \cite{BFKT}   
\begin{equation}
\begin{aligned}
e^{f(\Phi, \Theta, \psi, \eta)} &= 1 +  \sum_{n=1}^{\infty} \frac{1}{n!} 
\sum_{\substack{Z_{1}, \cdots, Z_{n} \subset X \\ Z_{i} \cap Z_{j} = \oslash}}  \sum_{k_{1}, \cdots, k_{n} \geqslant 1}  
(-1)^{\#}   
 \prod_{j=1}^{n}  \frac{1}{k_{j}!}  \sum_{(\vec{u}_{1,j},\vec{\xi}_{1,j}); \cdots; (\vec{u}_{k_{j},j},\vec{\xi}_{k_{j},j}) \in  \mathcal{C}(Z_{j})}
\\ &
 a_{\Theta \eta}(\vec{u}_{1,j},\vec{\xi}_{1,j}) \cdots  a_{\Theta \eta}(\vec{u}_{k_{j},j}, \vec{\xi}_{k_{j},j})
 \hspace{0.05 cm} \Phi(\vec{u}_{1,l}) \cdots \Phi(\vec{u}_{k_{l},l}) \hspace{0.05 cm} 
\psi (\vec{\xi}_{1,l}) \cdots \psi (\vec{\xi}_{k_{l},l}).
\end{aligned}
\end{equation}
As $Z_{i} \cap Z_{j} = \oslash$, the measure factorizes
\begin{equation}
\begin{aligned}
 \int d\mu_{I}(\psi) \hspace{0.05 cm}& d\mu_{I}(\Phi) \hat{\chi}(\Phi) \prod_{j=1}^{n}  
 \Phi(\vec{u}_{1,j}) \cdots \Phi(\vec{u}_{k_{j},j}) \hspace{0.05 cm} \psi (\vec{\xi}_{1,j}) \cdots  \psi (\vec{\xi}_{k_{j},j}) \\
 &= \prod_{l=1}^{n} \int d\mu_{I}(\psi) \hspace{0.05 cm} d\mu_{I}(\Phi) \hat{\chi}(\Phi) 
 \Phi(\vec{u}_{1,j}) \cdots \Phi(\vec{u}_{k_{j},j}) \hspace{0.05 cm} \psi (\vec{\xi}_{1,j}) \cdots  \psi (\vec{\xi}_{k_{j},j}).
\end{aligned}
\end{equation}
For every $\vec{\xi}_{i,j}$ (assuming there are \textit{n} sites in $\vec{\xi}_{i,j}$)
\begin{equation}
\begin{aligned}
\int d\mu_{I}(\psi) \hspace{0.05 cm} & \bar{\psi}_{\beta_{1},b_{1}}(x_{1}) \cdots \bar{\psi}_{\beta_{n},b_{n}}(x_{n})
\psi_{\alpha_{1},a_{1}}(y_{1}) \cdots \psi_{\alpha_{m},a_{m}}(y_{m}) = \\ 
&\begin{cases}
\pm 1    & \text{if} \hspace{0.4 cm} n =m \hspace{0.2 cm} \text{and} \hspace{0.2 cm} 
\{(\beta_{i}, x_{i}) \} = \{(\alpha_{i}, y_{i})\} \\
0   & \text{otherwise}.
\end{cases}
\end{aligned}
\end{equation}
Therefore, for every $Z \in \{Z_{1},\cdots, Z_{n}\}$, we identify $K_{0}(Z, \Theta, \eta)$ 
such that
\begin{equation}
\int d\mu_{I}(\psi) \hspace{0.05 cm} d\mu_{I}(\Phi) \hat{\chi}(\Phi) \hspace{0.05 cm} e^{f(\Phi, \Theta, \psi, \eta)}
= 1 +  \sum_{n=1}^{\infty} \frac{1}{n!} 
\sum_{\substack{Z_{1}, \cdots, Z_{n} \subset X \\ Z_{i} \cap Z_{j} = \oslash}} (-1)^{\#} \prod_{j=1}^{n} K_{0}(Z_{j}, \Theta, \eta).
\end{equation}
This completes the proof of the proposition 3.1. 

Next we take care of the pairwise disjoint condition of $\{Z_{j}\}$, by following the standard procedure,
\cite{BFKT}, and using the standard argument \cite{Sal},
\begin{equation}
\begin{aligned}
 \text{log}\hspace{0.1 cm} \Xi(\Theta, \eta) &= \text{log}\hspace{0.1 cm} \int d\mu_{I}(\psi) \hspace{0.05 cm} 
 d\mu_{I}(\Phi) \hat{\chi}(\Phi) \hspace{0.05 cm} e^{f(\Phi, \Theta, \psi, \eta)} \\ 
&= \sum_{n=1}^{\infty} \frac{1}{n!} \sum_{Z_{1}, \cdots, Z_{n} \subset X } \rho^{T}(Z_{1}, \cdots, Z_{n}) 
\hspace{0.05 cm} (-1)^{\#} \prod_{j=1}^{n} K_{0}(Z_{j}, \Theta, \eta)
\end{aligned}
\end{equation}
where $\rho^{T}(Z_{1}, \cdots, Z_{n}) = 0$ if $\{Z_{j}\}$ has any disjoint pair.

The standard argument for taking the logarithms which is often used for scalar functions
works in our case as well which contains Grassmann variables along with scalar functions.
As we have assumed that the total number of fermionic fields in the series expansion of 
$f(\Phi, \Theta, \psi, \eta)$ (3.10) is even. This property is preserved by each term 
$a_{\Theta \eta}(\vec{u}_{1,j},\vec{\xi}_{1,j}) \cdots  a_{\Theta \eta}(\vec{u}_{k_{j},j}, \vec{\xi}_{k_{j},j})
 \hspace{0.05 cm} \Phi(\vec{u}_{1,l}) \cdots \Phi(\vec{u}_{k_{l},l}) \hspace{0.05 cm} 
\psi (\vec{\xi}_{1,l}) \cdots \psi (\vec{\xi}_{k_{l},l})$ in (3.23). Then (3.25) integrates out even number of $\psi$ fields
leaving behind an even number of $\eta$ fields. Thus, every $K_{0}(Z_{j}, \Theta, \eta)$ is even in fermionic fields
and therefore, commutes with every other $K_{0}$.

\textbf{Proposition 3.2} Let $a(\vec{u}, \vec{s},\vec{\xi}, \vec{z})$ be a coefficient system and 
$K_{0}(Z, \Theta, \eta)$ be the function as defined. Then there exists a coefficient system
$b(\vec{s}, \vec{z})$  such that $\text{log}\hspace{0.1 cm} \Xi(\Theta, \eta) = \sum_{\vec{s}, \vec{z}}
 b(\vec{s}, \vec{z}) \hspace{0.05 cm} \Theta(\vec{s})\hspace{0.05 cm} \eta(\vec{z})$.

\textit{Proof} Using $a_{\Theta \eta}(\vec{u}, \vec{\xi}) = \sum_{\vec{s}, \vec{z} \in X}  a(\vec{u}, \vec{s},\vec{\xi}, \vec{z})
 \hspace{0.05 cm}  \Theta(\vec{s}) \hspace{0.05 cm}  \eta(\vec{z})$
rewrite $K_{0}(Z, \Theta, \eta)$ as
\begin{equation}
\begin{aligned}
K_{0}(Z, \Theta, \eta) &= (\pm 1) \sum_{k=1}^{\infty} \frac{1}{k!} 
\sum_{\substack{(\vec{u}_{1}, \cdots, \vec{u}_{k},\vec{\xi}_{1}, \cdots, \vec{\xi}_{k}) \in  
\mathcal{C}(\text{supp} \hspace{0.5 mm} \vec{u},\vec{\xi}) \\ \vec{u}_{1}\circ \cdots \circ \vec{u}_{k} = \vec{u}
\\ \vec{\xi}_{1}\circ \cdots \circ \vec{\xi}_{k} = \vec{\xi}}}  
\sum_{\substack{(\vec{s}_{1}, \cdots, \vec{s}_{k},\vec{z}_{1}, \cdots, \vec{z}_{k}) \\
\vec{s}_{1}\circ \cdots \circ \vec{s}_{k} = \vec{s} \\ \vec{z}_{1}\circ \cdots \circ \vec{z}_{k} = \vec{z}}} \\
& a(\vec{u}_{1}, \vec{s}_{1},\vec{\xi}_{1}, \vec{z}_{1}) \cdots a(\vec{u}_{k}, \vec{s}_{k},\vec{\xi}_{k}, \vec{z}_{k})
\hspace{0.1 cm}\Theta(\vec{s})\hspace{0.05 cm} \eta(\vec{z}) 
\int d\mu_{I}(\Phi) \hat{\chi}(\Phi)   \Phi(\vec{u}_{1}) \cdots \Phi(\vec{u}_{k}) \\
&= \sum_{\substack{\vec{u}, \vec{s}, \vec{\xi}, \vec{z} \in X \\ \text{supp} \vec{u}, \vec{\xi} = Z}} 
(\pm 1) \sum_{k=1}^{\infty} \frac{1}{k!} 
\sum_{\substack{(\vec{u}_{1}, \cdots, \vec{u}_{k},\vec{\xi}_{1}, \cdots, \vec{\xi}_{k}) \in  
\mathcal{C}(\text{supp} \hspace{0.5 mm} \vec{u},\vec{\xi}) \\ \vec{u}_{1}\circ \cdots \circ \vec{u}_{k} = \vec{u}
\\ \vec{\xi}_{1}\circ \cdots \circ \vec{\xi}_{k} = \vec{\xi}} } 
\sum_{\substack{(\vec{s}_{1}, \cdots, \vec{s}_{k},\vec{z}_{1}, \cdots, \vec{z}_{k}) \\
\vec{s}_{1}\circ \cdots \circ \vec{s}_{k} = \vec{s} \\ \vec{z}_{1}\circ \cdots \circ \vec{z}_{k} = \vec{z}}} \\
& a(\vec{u}_{1}, \vec{s}_{1},\vec{\xi}_{1}, \vec{z}_{1}) \cdots a(\vec{u}_{k}, \vec{s}_{k},\vec{\xi}_{k}, \vec{z}_{k}) 
\hspace{0.1 cm}\Theta(\vec{s})\hspace{0.05 cm} \eta(\vec{z})
\int d\mu_{I}(\Phi) \hat{\chi}(\Phi)   \Phi(\vec{u}_{1}) \cdots \Phi(\vec{u}_{k}) \\
&=  \sum_{\substack{\vec{u}, \vec{s}, \vec{\xi}, \vec{z} \in X \\ \text{supp} \vec{u}, \vec{\xi} = Z}} 
\tilde{a}(\vec{u}, \vec{s}, \vec{\xi}, \vec{z}) \hspace{0.05 cm}\Theta(\vec{s}) \hspace{0.05 cm} \eta(\vec{z})
\end{aligned}
\end{equation}
where in the second line we have grouped vectors together as concatenation and then 
in the third line we sum over concatenations. Note that
\begin{equation}
|\int d\mu_{I}(\Phi) \hat{\chi}(\Phi)   \Phi(\vec{u}_{1}) \cdots \Phi(\vec{u}_{k})| \leqslant 
p_{1}^{\sum_{i=1}^{k} n(\vec{u}_{i})}. \nonumber
\end{equation}
where $n(\vec{u}_{i})$ denotes $\#$ sites in $\vec{u}_{i}$.
Following (3.27) and (3.28) define
\begin{equation}
\begin{aligned}
b(\vec{s}, \vec{z}) =  \sum_{n=1}^{\infty} \frac{1}{n!} &
\sum_{\substack{(\vec{s}_{1}, \cdots, \vec{s}_{n},\vec{z}_{1}, \cdots, \vec{z}_{n}) \\
\vec{s}_{1}\circ \cdots \circ \vec{s}_{n} = \vec{s} \\ \vec{z}_{1}\circ \cdots \circ \vec{z}_{n} = \vec{z}}}
 \sum_{\vec{u}_{1}, \cdots, \vec{u}_{n}, \vec{\xi}_{1}, \cdots, \vec{\xi}_{n} \in X}
(-1)^{\#} \\  & \rho^{T}(\text{supp} \vec{u}_{1},\vec{\xi}_{1}; \cdots; \text{supp}\vec{u}_{n}, \vec{\xi}_{n})
\prod_{l=1}^{n} \tilde{a}(\vec{u}_{l}, \vec{s}_{l}, \vec{\xi}_{l}, \vec{z}_{l}) 
\end{aligned}
\end{equation}
and rewrite (3.27) as
\begin{equation}
 \text{log}\hspace{0.1 cm} \Xi(\Theta, \eta) = \sum_{\vec{s},\vec{z} \in X} 
 b(\vec{s}, \vec{z})  \hspace{0.05 cm}\Theta(\vec{s}) \hspace{0.05 cm} \eta(\vec{z}). \nonumber
\end{equation}

\subsubsection{Bound on norms} 
First rewrite  coefficient systems  as follows \cite{BFKT},
\begin{equation}
 b(\vec{s}, \vec{z})  =  \sum_{n=1}^{\infty} \frac{1}{n!} \sum_{\substack{T, \text{labelled tree with} \\ \text{vertices} 1,\cdots, n}}
\sum_{\vec{u}, \vec{\xi} \in X} \tilde{a}_{T} (\vec{u}, \vec{s}, \vec{\xi}, \vec{z})
\end{equation}
where
\begin{equation}
 \tilde{a}_{T} (\vec{u}, \vec{s}, \vec{\xi}, \vec{z}) = 
 \sum_{\substack{\vec{u}_{1}, \cdots, \vec{u}_{n}, \vec{\xi}_{1}, \cdots, \vec{\xi}_{n} \in  X \\
\vec{u}_{1}\circ \cdots \circ \vec{u}_{n} = \vec{u}\\ \vec{\xi}_{1}\circ \cdots \circ \vec{\xi}_{n} = \vec{\xi} \\ 
T \subset G(\vec{u}_{1}, \cdots, \vec{u}_{n}, \vec{\xi}_{1}, \cdots, \vec{\xi}_{n})}}
\sum_{\substack{(\vec{s}_{1}, \cdots, \vec{s}_{n},\vec{z}_{1}, \cdots, \vec{z}_{n}) \\
\vec{s}_{1}\circ \cdots \circ \vec{s}_{n} = \vec{s} \\ \vec{z}_{1}\circ \cdots \circ \vec{z}_{n} = \vec{z}}}
 \tilde{a} (\vec{u}_{1}, \vec{s}_{1}, \vec{\xi}_{1}, \vec{z}_{1}) \cdots  \tilde{a} (\vec{u}_{n}, \vec{s}_{n}, \vec{\xi}_{n}, \vec{z}_{n})
\end{equation}
and
\begin{equation}
\tilde{a} (\vec{u}, \vec{s}, \vec{\xi}, \vec{z}) = (\pm 1)  \sum_{k=1}^{\infty} \frac{1}{k!}
 \sum_{\substack{T, \text{labelled tree with} \\ \text{vertices} 1,\cdots, k}} a_{T}(\vec{u}, \vec{s}, \vec{\xi}, \vec{z})
 \int d\mu_{I}(\Phi) \hat{\chi}(\Phi)   \Phi(\vec{u}_{1}) \cdots \Phi(\vec{u}_{k})
\end{equation}
where  
\begin{equation}
a_{T} (\vec{u}, \vec{s}, \vec{\xi}, \vec{z}) =  \sum_{\substack{\vec{u}_{1}, \cdots, \vec{u}_{k}, \vec{\xi}_{1}, \cdots, \vec{\xi}_{k} \in  X \\
\vec{u}_{1}\circ \cdots \circ \vec{u}_{k} = \vec{u}\\ \vec{\xi}_{1}\circ \cdots \circ \vec{\xi}_{k} = \vec{\xi} \\ 
T \subset G(\vec{u}_{1}, \cdots, \vec{u}_{k}, \vec{\xi}_{1}, \cdots, \vec{\xi}_{k})}}
\sum_{\substack{(\vec{s}_{1}, \cdots, \vec{s}_{k},\vec{z}_{1}, \cdots, \vec{z}_{k}) \\
\vec{s}_{1}\circ \cdots \circ \vec{s}_{k} = \vec{s} \\ \vec{z}_{1}\circ \cdots \circ \vec{z}_{k} = \vec{z}}}
a (\vec{u}_{1}, \vec{s}_{1}, \vec{\xi}_{1}, \vec{z}_{1}) \cdots  a(\vec{u}_{k}, \vec{s}_{k}, \vec{\xi}_{k}, \vec{z}_{k})
\end{equation} 
and then use the following two lemmas.  

\textbf{Lemma 3.1}\cite{BFKT} Let $w_{p_{1}, h_{1}, p_{2}, h_{2}}(\vec{u}, \vec{s},\vec{\xi}, \vec{z}) =  
e^{\kappa t(\text{supp}(\vec{u}, \vec{s},\vec{\xi}, \vec{z}))} p_{1}^{k} h_{1}^{l} p_{2}^{n} h_{2}^{m}$ 
be the weight system with mass $\kappa$ giving weight at least $p_{1}$ to $\Phi, h_{1}$ to $\Theta, p_{2}$ to $\psi$ and 
$h_{2}$ to $\eta$.
Let \textit{T} be a labelled tree with vertices $1, \cdots, n$ and coordination numbers $d_{1}, \cdots, d_{n}$.
Let $a^{\prime}(\vec{u}, \vec{s}, \vec{\xi}, \vec{z})$ be a coefficient system. Define a new
coefficient system $a^{\prime}_{T}$ by
\begin{equation}
a^{\prime}_{T}(\vec{u}, \vec{s}, \vec{\xi}, \vec{z}) = 
\sum_{\substack{\vec{u}_{1}, \cdots, \vec{u}_{n}, \vec{\xi}_{1}, \cdots, \vec{\xi}_{n} \in  X \\
\vec{u}_{1}\circ \cdots \circ \vec{u}_{n} = \vec{u}\\ \vec{\xi}_{1}\circ \cdots \circ \vec{\xi}_{n} = \vec{\xi} \\ 
T \subset G(\vec{u}_{1}, \cdots, \vec{u}_{n}, \vec{\xi}_{1}, \cdots, \vec{\xi}_{n})}}
\sum_{\substack{(\vec{s}_{1}, \cdots, \vec{s}_{n},\vec{z}_{1}, \cdots, \vec{z}_{n}) \\
\vec{s}_{1}\circ \cdots \circ \vec{s}_{n} = \vec{s} \\ \vec{z}_{1}\circ \cdots \circ \vec{z}_{n} = \vec{z}}}
a^{\prime}(\vec{u}_{1}, \vec{s}_{1}, \vec{\xi}_{1}, \vec{z}_{1}) \cdots 
a^{\prime}(\vec{u}_{1}, \vec{s}_{1}, \vec{\xi}_{n}, \vec{z}_{n})
\end{equation}
Then
\begin{equation}
|a^{\prime}_{T}|_{w_{p_{1}, h_{1}, p_{2}, h_{2}}} \leqslant d_{1}! \cdots d_{n}! \hspace{0.05 cm} 
|a^{\prime}|_{w_{2p_{1}, h_{1}, 2p_{2}, h_{2}}}^{n}.
\end{equation}

Next assuming that all the norms are less than $\frac{1}{8}$, Lemma 3.2 provides estimates of the sums arising in Lemma 3.1. 

\textbf{Lemma 3.2}\cite{BFKT} Let $0 < \epsilon < \frac{1}{8}$. Then
\begin{equation}
\epsilon + \sum_{n=2}^{\infty} \frac{1}{(n-1)!} \sum_{\substack{d_{1}, \cdots, d_{n} \\ d_{1} + \cdots + d_{n} = 2(n-1)}}
 \sum_{\substack{T, \text{labelled tree} \\ \text{with coordination} \\ \text{numbers} \hspace{0.05 cm} d_{1}, \cdots, d_{n} }}
d_{1}! \cdots d_{n}! \hspace{0.05 cm} \epsilon^{n} \leqslant \frac{\epsilon}{1-8\epsilon}. 
\end{equation} 

Norm is estimated in two steps. First rewrite (3.30) 
 \begin{equation}
|b(\vec{s},\vec{z})|_{w_{h_{1},h_{2}}} \leqslant 
 \sum_{n=1}^{\infty} \frac{1}{n!} \sum_{\substack{T, \text{labelled tree with} \\ \text{vertices} 1,\cdots, n}}
\sum_{\vec{u}, \vec{\xi} \in X} |\tilde{a}_{T}|_{w_{p_{1},h_{1},p_{2},h_{2}}}  \nonumber
\end{equation}
Note that from (3.31) and Lemma 3.1
\begin{equation}
|\tilde{a}_{T}|_{w_{p_{1},h_{1},p_{2},h_{2}}} \leqslant d_{1}! \cdots d_{n}! \hspace{0.05 cm} 
|\tilde{a}|_{w_{2 p_{1},h_{1},2 p_{2},h_{2}}}^{n}. \nonumber
\end{equation}
Now using Lemma 3.2 with $\epsilon = |\tilde{a}|_{w_{2 p_{1},h_{1},2 p_{2},h_{2}}}$ 
\begin{equation}
|b(\vec{s},\vec{z})|_{w_{h_{1},h_{2}}} \leqslant 
\frac{|\tilde{a}|_{w_{2 p_{1},h_{1},2 p_{2},h_{2}}}}{1 - 8|\tilde{a}|_{w_{2 p_{1},h_{1},2 p_{2},h_{2}}}}.
\end{equation}
Next denote $n(\vec{u})$ as the number of sites in $\vec{u}$ and rewrite (3.32)
\begin{equation}
\begin{aligned}
|\tilde{a}|_{w_{2 p_{1},h_{1},2 p_{2},h_{2}}} &\leqslant 
\sum_{k=1}^{\infty} \frac{1}{k!} \sum_{\substack{T, \text{labelled tree with} \\ \text{vertices} 1,\cdots, k}}
| \int d\mu_{I}(\Phi) \hat{\chi}(\Phi)   \Phi(\vec{u}_{1}) \cdots \Phi(\vec{u}_{k})| \hspace{0.05 cm}
|a_{T}|_{w_{2 ,h_{1},2 p_{2},h_{2}}} \\ 
&\leqslant \sum_{k=1}^{\infty} \frac{1}{k!} \sum_{\substack{T, \text{labelled tree with} \\ \text{vertices} 1,\cdots, k}}   
p_{1}^{n(\vec{u})} |a_{T}|_{w_{2 ,h_{1},2 p_{2},h_{2}}} \\  
&\leqslant \sum_{k=1}^{\infty} \frac{1}{k!} \sum_{\substack{T, \text{labelled tree with} \\ \text{vertices} 1,\cdots, k}}   
 |a_{T}|_{w_{2 p_{1},h_{1},2 p_{2},h_{2}}}   
\end{aligned}
\end{equation}
Note that from (3.33) and Lemma 3.1
\begin{equation}
|a_{T}|_{w_{2 p_{1},h_{1},2 p_{2},h_{2}}}  \leqslant d_{1}! \cdots d_{k}! 
\hspace{0.05 cm} |a|_{w_{4 p_{1},h_{1},4 p_{2},h_{2}}}^{k}.       \nonumber
\end{equation}
Now using Lemma 3.2 with $\epsilon = |a|_{w_{4 p_{1},h_{1},4 p_{2},h_{2}}}$
\begin{equation}
|\tilde{a}|_{w_{2 p_{1},h_{1},2 p_{2},h_{2}}} \leqslant  
\frac{|a|_{w_{4 p_{1},h_{1},4 p_{2},h_{2}}}}{1 - 8|a|_{w_{4 p_{1},h_{1},4 p_{2},h_{2}}}}.
\end{equation}
Substituting $|\tilde{a}|_{w_{2 p_{1},h_{1},2 p_{2},h_{2}}}$ from (3.39) in (3.37)
\begin{equation}
|b|_{w_{h_{1},h_{2}}}  
\leqslant \frac{|a|_{w_{4 p_{1},h_{1},4 p_{2},h_{2}}}}{1 - 16|a|_{w_{4 p_{1},h_{1},4 p_{2},h_{2}}}}.
\end{equation}
Setting $p_{2} = h_{1} = h_{2} = 1$ we get
\begin{equation}
|b|_{w_{1,1}} \leqslant \frac{|a|_{w_{4 p_{1},1,4,1}}}{1 - 16|a|_{w_{4 p_{1},1,4,1}}}. \nonumber
\end{equation}

\subsection{Application to U(1) Higgs-Yukawa model}  The potential is given by
\begin{equation}
\begin{aligned}
&V_{1}(\Omega_{0},\bar{\psi}, \psi, \Phi^{\prime}_{\Lambda_{1}}, 
\Phi_{\Lambda^{c}_{1}}, \bar{\text{J}}, \text{J})
= - W_{1}(\Lambda_{1}) - W_{1}^{\prime}(\Lambda_{1}) - W_{2}(\Omega_{0})  - W_{2}^{\prime}(\Omega_{0})  \\
&+ V(\Omega_{0}, S^{\text{loc}}_{\Lambda_{1}} \psi - \Psi,
  \bar{\psi} -  \bar{\Psi}, C^{\frac{1}{2},\text{loc}}_{\Lambda_{1}} \Phi^{\prime}- \varphi)  
+ V_{\epsilon} (\Lambda_{1}, \bar{\psi}, \psi, \Phi^{\prime}) -
\langle \bar{\psi}, \text{J}\rangle - \langle \bar{\text{J}}, S^{\text{loc}}_{\Lambda_{1}} \psi \rangle \nonumber
\end{aligned}
\end{equation}
and the localized small field integral is
\begin{equation}
\begin{aligned}
\Xi(\Omega_{1},\bar{\psi}_{\Omega_{1}^{c}}, \psi_{\Omega_{1}^{c}}, 
& \Phi^{\prime}_{\Lambda_{1} - \Omega_{1}}, \Phi_{\Lambda^{c}_{1}}, \bar{\text{J}}, \text{J}) =
\\ & \int d\mu_{I}(\bar{\psi}_{\Omega_{1}}, \psi_{\Omega_{1}}) d\mu_{\text{I}}(\Phi^{\prime}_{\Omega_{1}})
\hspace{0.1 cm} \hat{\chi}_{\Omega_{1}}(\Phi^{\prime})  \hspace{0.05 cm}   \nonumber
e^{- V_{1}(\Omega_{0},\bar{\psi}, \psi, \Phi^{\prime}_{\Lambda_{1}}, \Phi_{\Lambda^{c}_{1}}, \bar{\text{J}}, \text{J})}
\end{aligned}
\end{equation}
Recall that $\Omega \subseteq \Lambda_{1}$, $\Omega_{0} = \Omega - [r_{\lambda}]$ and 
$\Omega_{1} = \Omega_{0} - [r_{\lambda}]$.
Set $\Phi_{\Lambda_{1}^{c}} = 0, \psi_{\Lambda_{1}^{c}} = 0$. Note that
\begin{equation}
\begin{aligned}
C^{\frac{1}{2},\text{loc}}_{\Lambda_{1}} \Phi^{\prime} &= C^{\frac{1}{2},\text{loc}}_{\Lambda_{1}}
(\Phi_{\Lambda_{1} - \Omega_{1}}^{\prime}, \Phi_{\Omega_{1}}^{\prime}) \\
S^{\text{loc}}_{\Lambda_{1}} \psi &= S^{\text{loc}}_{\Lambda_{1}} (\psi_{\Lambda_{1} - \Omega_{1}}, \psi_{\Omega_{1}}).
\end{aligned}
\end{equation}
Denote
\begin{equation}
\begin{aligned}
&\Phi = \Phi_{\Omega_{1}}^{\prime}, \hspace{0.5 cm} \Theta = \Phi_{\Lambda_{1}-\Omega_{1}}^{\prime},  \hspace{0.5 cm}
\psi = \psi_{\Omega_{1}}, \hspace{0.5 cm} \eta = (\eta_{1}, \eta_{2}) = (\psi_{\Lambda_{1}-\Omega_{1}}, \text{J}) \\
&\Xi(\Omega_{1},\bar{\psi}_{\Omega_{1}^{c}}, \psi_{\Omega_{1}^{c}}, \Phi^{\prime}_{\Lambda_{1} - \Omega_{1}}, 
\Phi_{\Lambda^{c}_{1}}, \bar{\text{J}}, \text{J}) = \Xi(\Omega_{1}, \Theta, \eta).
\end{aligned}
\end{equation}
Here, $\Theta = \Phi_{\Lambda_{1}-\Omega_{1}}^{\prime}$ is a spectator field and not an external source.
They are just present in the region but are not the integration variables. They are treated as a part of the region 
$\mathrm{P} = \Lambda_{1} - \Omega$. 

\textbf{Theorem 3.2} There exists a coefficient system $ a(\vec{u}, \vec{s},\vec{\xi}, \vec{z})$ 
and a positive constant, \\ $\kappa^{\prime} \leqslant \text{min} (\gamma_{1}, \gamma_{2})$ such that 
\begin{equation}
V_{1}(\Omega_{0},\bar{\psi}, \psi, \Phi^{\prime}_{\Lambda_{1}}, 
\Phi_{\Lambda^{c}_{1}}, \bar{\text{J}}, \text{J}) = \sum_{k,l,n,m \geqslant 0} 
\sum_{\vec{u}, \vec{s}, \vec{\xi}, \vec{z} \subset \Omega_{0}} a(\vec{u}, \vec{s},\vec{\xi}, \vec{z}) \hspace{0.05 cm} 
\Phi(\vec{u}) \hspace{0.05 cm} \Theta(\vec{s})\hspace{0.05 cm} \psi (\vec{\xi})\hspace{0.05 cm} \eta(\vec{z})
\end{equation}
and if $g \sim \mathcal{O}(e_{0})$ then 
\begin{equation}
|a(\vec{u}, \vec{s},\vec{\xi}, \vec{z})|  \leqslant c \hspace{0.05 cm} e_{0}
\hspace{0.05 cm} e^{- \kappa^{\prime}t(u_{1},\cdots,u_{k},s_{1},\cdots,s_{l},\xi_{1}, \cdots,  \xi_{n}, z_{1}, \cdots, z_{m})}
\end{equation}
where $t(u_{1},\cdots,u_{k},s_{1},\cdots,s_{l},\xi_{1}, \cdots,  \xi_{n}, z_{1}, \cdots, z_{m})$ is the length of minimal tree
and
\begin{equation}
||V_{1}||_{w_{4p_{0,\lambda},1,4,1}} \leqslant c \hspace{0.05 cm} e_{0}^{\frac{1}{2} - \epsilon}. \nonumber
\end{equation}

\textit{Proof}
First consider the Yukawa term 
\begin{equation}
\begin{aligned}
V_{Y}(\Omega_{0}) &=  \sum_{x,\alpha} g\hspace{0.05 cm} 
\bar{\psi}_{L,\alpha}(x)\hspace{0.05 cm}  C^{\frac{1}{2}, \text{loc}}_{\Lambda_{1}} \Phi^{\prime} (x) 
\hspace{0.05 cm} (S^{\text{loc}}_{\Lambda_{1}}\psi_{R})_{\alpha}(x) \\
 &= \sum_{x,\alpha} g\hspace{0.05 cm} \bar{\psi}_{L,\alpha}(x)\hspace{0.05 cm}
[C^{\frac{1}{2}, \text{loc}}_{\Lambda_{1}} \Phi (x) + C^{\frac{1}{2}, \text{loc}}_{\Lambda_{1}} \Theta (x)] 
[(S^{\text{loc}}_{\Lambda_{1}}\psi_{R})_{\alpha}(x) + (S^{\text{loc}}_{\Lambda_{1}}\eta_{1,R})_{\alpha}(x)] \\
&= \sum_{x,\alpha} g\hspace{0.05 cm}\bar{\psi}_{L,\alpha}(x)\hspace{0.05 cm}
C^{\frac{1}{2}, \text{loc}}_{\Lambda_{1}} \Phi (x) (S^{\text{loc}}_{\Lambda_{1}}\psi_{R})_{\alpha}(x) +
 \sum_{x,\alpha} g\hspace{0.05 cm}\bar{\psi}_{L,\alpha}(x)\hspace{0.05 cm}
C^{\frac{1}{2}, \text{loc}}_{\Lambda_{1}} \Phi (x) (S^{\text{loc}}_{\Lambda_{1}}\eta_{1,R})_{\alpha}(x) \\
&+  \sum_{x,\alpha} g\hspace{0.05 cm}\bar{\psi}_{L,\alpha}(x)\hspace{0.05 cm}
C^{\frac{1}{2}, \text{loc}}_{\Lambda_{1}} \Theta (x) (S^{\text{loc}}_{\Lambda_{1}}\psi_{R})_{\alpha}(x) +
 \sum_{x,\alpha} g\hspace{0.05 cm}\bar{\psi}_{L,\alpha}(x)\hspace{0.05 cm}
C^{\frac{1}{2}, \text{loc}}_{\Lambda_{1}} \Theta (x)(S^{\text{loc}}_{\Lambda_{1}}\eta_{1,R})_{\alpha}(x)
\end{aligned}
\end{equation}
We can extract the series coefficients by considering any one of the above terms.
For example, rewrite
\begin{equation}
\begin{aligned} 
 \sum_{x,\alpha} g\hspace{0.05 cm}\bar{\psi}_{L,\alpha}(x)\hspace{0.05 cm}
& C^{\frac{1}{2}, \text{loc}}_{\Lambda_{1}} \Phi (x) (S^{\text{loc}}_{\Lambda_{1}}\eta_{1,R})_{\alpha}(x) \\
&= \sum_{x,x_{1},x_{2}} \sum_{\alpha,\beta} g\hspace{0.05 cm}\bar{\psi}_{L,\alpha}(x)\hspace{0.05 cm}
C^{\frac{1}{2}, \text{loc}}_{\Lambda_{1}}(x,x_{1}) \Phi (x_{1})
S^{\text{loc}}_{\Lambda_{1},\alpha\beta}(x,x_{2}) \eta_{1,R,\beta} (x_{2}) \\
&= \sum_{x,x_{1},x_{2}} \sum_{\alpha,\beta} a_{\alpha\beta}(x,x_{1},x_{2}) \bar{\psi}_{L,\alpha}(x) \hspace{0.05 cm}
\Phi (x_{1})  \hspace{0.05 cm} \eta_{1,R,\beta} (x_{2})
\end{aligned}
\end{equation}
where $a_{\alpha\beta}(x,x_{1},x_{2}) =  g\hspace{0.05 cm}C^{\frac{1}{2}, \text{loc}}_{\Lambda_{1}}(x,x_{1}) 
S^{\text{loc}}_{\Lambda_{1},\alpha\beta}(x,x_{2})$ and $x_{1}, x_{2} \in \Lambda_{1}$. Note that \\
$|C^{\frac{1}{2}, \text{loc}}_{\Lambda_{1}} (x,x_{1})| \leqslant c \hspace{0.05 cm} e^{-\gamma_{1} |x-x_{1}|}$
and $|S^{\text{loc}}_{\Lambda_{1},\alpha\beta}(x,x_{2})| \leqslant c \hspace{0.05 cm} e^{-\gamma_{2} |x-x_{2}|}$.
Let $\kappa^{\prime} = \text{min} \hspace{0.05 cm}(\gamma_{1}, \gamma_{2}),$ and
$t(x,x_{1},x_{2})$ be the length of the
shortest tree joining $x, x_{1}, x_{2}$ and $g \sim \mathcal{O}(e_{0})$. Then
\begin{equation}
|a_{\alpha\beta}(x,x_{1},x_{2})|  \leqslant c \hspace{0.05 cm} e_{0}\hspace{0.05 cm}e^{-\gamma_{1} |x-x_{1}|}
\hspace{0.05 cm} e^{-\gamma_{2} |x-x_{2}|} 
\leqslant  c \hspace{0.05 cm} e_{0}\hspace{0.05 cm}e^{-\kappa^{\prime} t(x,x_{1},x_{2})}.
\end{equation}
\textbf{Norm of $V_{Y}$}. In (3.17) set $p_{1} = p_{0,\lambda}$ and $h_{1} = p_{2} = h_{2} = 1$
with $\kappa < \kappa^{\prime}$ and  $k=n=m=1$ and $l = 0$, 
\begin{equation}
\begin{aligned}
||V_{Y}||_{w_{4p_{0,\lambda},1,4,1}} &\leqslant 
\max\limits_{y \in  \Omega_{0}}  \hspace{0.05 cm} \sum_{\substack{ x,x_{1},x_{2} \in  \Omega_{0} \\
 x \hspace{0.05 cm} \text{or}\hspace{0.05 cm} x_{1}  \hspace{0.05 cm} \text{or} \hspace{0.05 cm} x_{2} = y}}  
\sum_{\alpha,\beta} 4 \hspace{0.05 cm} e^{\kappa t(x,x_{1},x_{2})} \hspace{0.05 cm} (4)  \hspace{0.05 cm}
4p_{0,\lambda}  \hspace{0.05 cm} c \hspace{0.05 cm} e_{0}\hspace{0.05 cm}e^{-\kappa^{\prime} t(x,x_{1},x_{2})} \\
&\leqslant \max\limits_{y \in  \Omega_{0}}  \hspace{0.05 cm} \sum_{\substack{ x,x_{1},x_{2} \in  \Omega_{0} \\
 x \hspace{0.05 cm} \text{or}\hspace{0.05 cm} x_{1}  \hspace{0.05 cm} \text{or} \hspace{0.05 cm} x_{2} = y}} 
4 \hspace{0.05 cm} c \hspace{0.05 cm} 4^{2} \hspace{0.05 cm}4  \hspace{0.05 cm}  e_{0}^{-\epsilon}\hspace{0.05 cm}
e_{0}\hspace{0.05 cm}  e^{-(\kappa^{\prime} - \kappa)t(x,x_{1},x_{2})} \\
&\leqslant  c \hspace{0.05 cm} 256 \hspace{0.05 cm} e_{0}^{1- \epsilon}
\end{aligned}
\end{equation}
where we have assumed that $4 \hspace{0.05 cm}  p_{0,\lambda} \leqslant e_{0}^{-\epsilon}$.

Next consider the term 
\begin{equation}
\begin{aligned}
V^{\prime}_{\epsilon}(\Omega_{0}) &= \sum_{x}
\bar{\psi}_{L}(x)[\mathfrak{D}_{A} - \mathfrak{D}] (S^{\text{loc}}_{\Lambda_{1}}\psi_{L})(x) \\
&= \sum_{x} \bar{\psi}(x) [\gamma \cdot \nabla_{A} - \frac{1}{2} \Delta_{A} - 
(\gamma \cdot \nabla - \frac{1}{2} \Delta)] (S^{\text{loc}}_{\Lambda_{1}}\psi)(x) \\
&= - \sum_{x} \sum_{\mu} \Big[ \bar{\psi}(x) \Big(\frac{1-\gamma_{\mu}}{2}\Big)
(e^{i e_{0} (C^{\frac{1}{2},\text{loc}}_{\Lambda_{1}}A_{\mu})(x)} - 1)(S^{\text{loc}}_{\Lambda_{1}}\psi)(x+e_{\mu}) \\
& \hspace{0.7 cm} + \bar{\psi}(x) \Big(\frac{1+ \gamma_{\mu}}{2}\Big) (e^{-i e_{0} (C^{\frac{1}{2},\text{loc}}_{\Lambda_{1}}A_{\mu})(x)} - 1)
(S^{\text{loc}}_{\Lambda_{1}}\psi)(x-e_{\mu})\Big] \\
&= - \sum_{x,y,z} \sum_{\mu} \Big[ \bar{\psi}(x) \Big(\frac{1-\gamma_{\mu}}{2}\Big)
(e^{i e_{0} (C^{\frac{1}{2},\text{loc}}_{\Lambda_{1}}A_{\mu})(x)} - 1) S^{\text{loc}}_{\Lambda_{1}}(x+e_{\mu},y) \psi(y) \\
& \hspace{0.7 cm} + \bar{\psi}(x) \Big(\frac{1+ \gamma_{\mu}}{2}\Big) (e^{-i e_{0} (C^{\frac{1}{2},\text{loc}}_{\Lambda_{1}}A_{\mu})(x)} - 1)
S^{\text{loc}}_{\Lambda_{1}}(x-e_{\mu},z) \psi(z)\Big] \\
&= \sum_{x,y,z} [a(x,y) \bar{\psi}(x) \psi(y) + a(x,z) \bar{\psi}(x) \psi(z)]
\end{aligned}
\end{equation}
where for notation purposes we have suppressed the spinor indices and
\begin{equation}
\begin{aligned}
a(x,y) &= -\sum_{\mu} \Big(\frac{1-\gamma_{\mu}}{2}\Big) 
(e^{i e_{0} (C^{\frac{1}{2},\text{loc}}_{\Lambda_{1}}A_{\mu})(x)} - 1) S^{\text{loc}}_{\Lambda_{1}}(x+e_{\mu},y) \\
a(x,z) &= -\sum_{\mu} \Big(\frac{1+ \gamma_{\mu}}{2}\Big) 
(e^{-i e_{0} (C^{\frac{1}{2},\text{loc}}_{\Lambda_{1}}A_{\mu})(x)} - 1) S^{\text{loc}}_{\Lambda_{1}}(x-e_{\mu},z).
\end{aligned}
\end{equation}
Note that since $|A_{\mu}(x)| \leqslant p_{0,\lambda}$ therefore, from Lemma 2.2
$|(C^{\frac{1}{2},\text{loc}}_{\Lambda_{1}}A_{\mu})(x)| \leqslant c \hspace{0.05 cm} p_{0,\lambda}$ and 
$|e^{i e_{0}(C^{\frac{1}{2},\text{loc}}_{\Lambda_{1}}A_{\mu})(x)} - 1| \leqslant 
c \hspace{0.05 cm} e_{0} p_{0,\lambda} \leqslant c \hspace{0.05 cm} e_{0}^{1 - \epsilon}$. Thus, 
$|a(x,y)| \leqslant c \hspace{0.05 cm} e_{0}^{1 - \epsilon} \hspace{0.05 cm} \sum_{\mu} e^{-\gamma_{2} |x+e_{\mu}-y|}$ and similarly
$|a(x,z)| \leqslant c \hspace{0.05 cm} e_{0}^{1 - \epsilon} \hspace{0.05 cm} \sum_{\mu} e^{-\gamma_{2} |x-e_{\mu}-z|}$.
Let $\kappa < \gamma_{2}$, then
\begin{equation}
\begin{aligned}
||V^{\prime}_{\epsilon}||_{w_{4p_{0,\lambda},1,4,1}} &\leqslant 
\max\limits_{\xi \in \Omega_{0}} \hspace{0.05 cm} \sum_{\substack{ x,y \in \Omega_{0} \\
 x \hspace{0.05 cm} \text{or} \hspace{0.05 cm} y = \xi}}  
\sum_{\alpha,\beta,\delta} \sum_{\mu} 2 \hspace{0.05 cm} e^{\kappa t(x,y)} \hspace{0.05 cm} 4^{2}  \hspace{0.05 cm}
 c \hspace{0.05 cm} e_{0}^{1-\epsilon} \hspace{0.05 cm}e^{- \gamma_{2} t(x,y)} \\
&\leqslant \max\limits_{\xi \in \Omega_{0}} \hspace{0.05 cm} \sum_{\substack{ x,y \in \Omega_{0} \\
 x \hspace{0.05 cm} \text{or} \hspace{0.05 cm} y = \xi}}  (32) \hspace{0.05 cm} 256  \hspace{0.05 cm}
 c \hspace{0.05 cm} e_{0}^{1-\epsilon} \hspace{0.05 cm}e^{- (\gamma_{2} - \kappa) t(x,y)} \\
 &\leqslant c \hspace{0.05 cm} 8192 \hspace{0.05 cm} e_{0}^{1-\epsilon}. 
\end{aligned}
\end{equation}

Next consider the source term 
\begin{equation}
\begin{aligned}
e_{0} \langle \bar{\text{J}}, S^{\text{loc}}_{\Lambda_{1}} \psi \rangle &= 
e_{0} \sum_{x,\alpha}  \eta_{2,\alpha}(x) (S^{\text{loc}}_{\Lambda_{1}} \psi)_{\alpha}(x) \\
&= e_{0} \sum_{x,\alpha}  \eta_{2,\alpha}(x) [(S^{\text{loc}}_{\Lambda_{1}} \psi)_{\alpha}(x) 
+ (S^{\text{loc}}_{\Lambda_{1}} \eta_{1})_{\alpha}(x)] \\
&=  e_{0} \sum_{x, x_{1}, \alpha,\beta} \eta_{2,\alpha}(x)S^{\text{loc}}_{\Lambda_{1},\alpha,\beta}(x,x_{1}) \psi_{\beta}(x_{1})
+ e_{0} \sum_{x,x_{1},\alpha,\beta} \eta_{2,\alpha}(x)S^{\text{loc}}_{\Lambda_{1},\alpha,\beta}(x,x_{1})  \eta_{1,\beta} (x_{1}) \\
&= \sum_{x,x_{1},\alpha,\beta} a_{\alpha\beta}(x,x_{1}) \eta_{2,\alpha}(x)  \psi_{\beta}(x_{1}) + 
\sum_{x,x_{1},\alpha,\beta} a_{\alpha\beta}(x,x_{1}) \eta_{2,\alpha}(x)  \eta_{1,\beta} (x_{1})
\end{aligned}
\end{equation}
where $a_{\alpha\beta}(x,x_{1}) = e_{0} \hspace{0.05 cm} S^{\text{loc}}_{\Lambda_{1},\alpha,\beta}(x,x_{1})$
with $|a_{\alpha\beta}(x,x_{1})| \leqslant c \hspace{0.05 cm} e_{0} \hspace{0.05 cm}e^{-\gamma_{2} |x-x_{1}|}$
and $x_{1} \in \Lambda_{1}$. To estimate the norm  set $p_{1} = p_{0,\lambda}$ and $h_{1} = p_{2} = h_{2} = 1$
with $\kappa < \gamma_{2}$ and  $k=m=1$ and $l=n= 0$ in (3.17), 
\begin{equation}
\begin{aligned}
||e_{0} \langle \bar{\text{J}}, S^{\text{loc}}_{\Lambda_{1}} \psi \rangle||_{w_{4p_{0,\lambda},1,4,1}}
&\leqslant \max\limits_{y \in X}  \hspace{0.05 cm} \sum_{\substack{ x,x_{1} \in X \\
 x \hspace{0.05 cm} \text{or}\hspace{0.05 cm} x_{1}  = y}}  
\sum_{\alpha,\beta} 2 \hspace{0.05 cm} e^{\kappa |x-x_{1}|}   \hspace{0.05 cm} 4  \hspace{0.05 cm}
 c \hspace{0.05 cm} e_{0} \hspace{0.05 cm}e^{-\gamma_{2} |x-x_{1}|} \\
&\leqslant \max\limits_{y \in X}  \hspace{0.05 cm} \sum_{\substack{ x,x_{1} \in X \\
 x \hspace{0.05 cm} \text{or}\hspace{0.05 cm} x_{1}  = y}}  c \hspace{0.05 cm} (8) \hspace{0.05 cm} 4^{2} \hspace{0.05 cm}
  e_{0} \hspace{0.05 cm} e^{-(\gamma_{2} - \kappa)|x-x_{1}|} \\
&\leqslant  c \hspace{0.05 cm} 128 \hspace{0.05 cm} e_{0}.
\end{aligned}
\end{equation}
Recall that from Lemma 2.5,
$||V_{\epsilon}(\bar{\psi}, \psi, \Box)||_{h} \leqslant c \hspace{0.05 cm} [r_{\lambda}]^{4} \hspace{0.05 cm} 
e^{-\frac{\gamma_{2}}{2} r_{\lambda}} h^{2}$. 
Setting $h=1$, using $e^{-\frac{\gamma_{2}}{2} r_{\lambda}} \sim \mathcal{O}(e_{0}^{2})$ and summing over
spinor indices, 
\begin{equation}
||V_{\epsilon}(\bar{\psi}, \psi, \Box)||_{1} \leqslant c \hspace{0.05 cm} 16 [r_{\lambda}]^{4} \hspace{0.05 cm} e_{0}^{2}.
\end{equation}
For the bosonic terms $V_{B}$ in the potential we quote our result from the U(1) Higgs model \cite{G1}
\begin{equation}
||V_{B}||_{w_{4p_{0,\lambda},1,4,1}} \leqslant c \hspace{0.05 cm} (c_{0} \hspace{0.05 cm} e_{0}^{1-2 \epsilon})^{\frac{1}{2}}.
\end{equation}
Then the norm of $V_{1}$ is
\begin{equation}
\begin{aligned}
||V_{1}||_{w_{4p_{0,\lambda},1,4,1}} &\leqslant ||V_{B}||_{w_{4p_{0,\lambda},1,4,1}} +
2  \hspace{0.05 cm} ||V_{Y}||_{w_{4p_{0,\lambda},1,4,1}} + ||V^{\prime}_{\epsilon}||_{w_{4p_{0,\lambda},1,4,1}} +
 ||V_{s}||_{w_{4p_{0,\lambda},1,4,1}} \\
&\leqslant c \hspace{0.05 cm} (c_{0} \hspace{0.05 cm} e_{0}^{1-2 \epsilon})^{\frac{1}{2}} +
c \hspace{0.05 cm} 512 \hspace{0.05 cm} e_{0}^{1- \epsilon} + c \hspace{0.05 cm} 8192 \hspace{0.05 cm} e_{0}^{1-\epsilon}
+  c \hspace{0.05 cm} 256 \hspace{0.05 cm} e_{0} \\
&\leqslant c \hspace{0.05 cm} e_{0}^{\frac{1}{2} - \epsilon}.      \nonumber
\end{aligned} 
\end{equation}
This completes the proof of Theorem 3.2.

If $||V_{1}||_{w_{4p_{0,\lambda},1,4,1}} < \frac{1}{16}$, then from the Theorem 3.1
\begin{equation}
||\text{log}\hspace{0.1 cm} \Xi(\Omega_{1}, \Theta, \eta) -  \text{log}\hspace{0.1 cm} \Xi(\Omega_{1}, 0, 0)||_{w_{1, 1}} 
\leqslant \frac{||V_{1}||_{w_{4p_{0,\lambda},1,4,1}}}{1 - 16 ||V_{1}||_{w_{4p_{0,\lambda},1,4,1}}} 
\leqslant  c \hspace{0.05 cm} e_{0}^{\frac{1}{2} - \epsilon}.
\end{equation}
Define a function $H$
\begin{equation}
\begin{aligned}
H(\Theta, \eta) &= \sum_{Y} \sum_{\substack{\vec{s}, \vec{z} \\ \text{supp} (\vec{s},\vec{z}) = Y}}
 b(\vec{s}, \vec{z}) \hspace{0.05 cm} \Theta(\vec{s})\hspace{0.05 cm} \eta(\vec{z})
= \sum_{Y} H(Y, \Theta, \eta)  \\
 ||H - H(0)||_{w_{1, 1}} &= 
 ||\text{log}\hspace{0.1 cm} \Xi(\Omega_{1}, \Theta, \eta) -  \text{log}\hspace{0.1 cm} \Xi(\Omega_{1}, 0, 0)||_{w_{1, 1}} .
\end{aligned}
\end{equation}
\textbf{Shift.} Denote
\begin{equation}
\begin{aligned}
&\Upsilon = 
S_{\Lambda_{1}}\mathrm{D}_{\Lambda_{1}\tilde{\mathrm{Q}} }\eta_{\tilde{\mathrm{Q}} } \hspace{1 cm}
\varphi = C_{\Lambda_{1}}\text{T}_{\Lambda_{1}\tilde{\mathrm{Q}} } \Phi_{\tilde{\mathrm{Q}} }.
\end{aligned}
\end{equation}
Define $H^{\#}$ by
 \begin{equation}
H^{\#}(\Theta, \phi, \eta, \Psi) = H(\Theta - \varphi, \eta - \Upsilon)  \nonumber
\end{equation}
Let $w_{h_{1}, h_{1}^{\prime}, h_{2}, h_{2}^{\prime}}(\vec{s}, \vec{z}) =  
e^{\kappa t(\text{supp}(\vec{s}, \vec{z}))} h_{1}^{l_{1}} h_{1}^{\prime l_{2}} h_{2}^{m_{1}} h_{2}^{\prime m_{2}}$ 
be the weight system with mass $\kappa$ giving weight at least $h_{1}$ to $\Theta$, $h_{1}^{\prime}$ to 
$\phi$, $h_{2}$ to $\eta$ and $h_{2}^{\prime}$ to $\Psi$. Then 
\begin{equation}
\begin{aligned}
||H^{\#}||_{w_{h_{1}, h_{1}^{\prime}, h_{2}, h_{2}^{\prime}}} &\leqslant
||H||_{w_{h_{1}+h_{1}^{\prime}, h_{2}+h_{2}^{\prime} }}   
\end{aligned}
\end{equation}
The above bound follows from \cite{BFKT} where $H$ and $H^{\#}$ are functions of bosonic variables and
thus, $||H|| = ||H^{\#}||$. In our case due to antisymmetric nature of the fermionic variables we have an inequality instead. 

\textbf{Norm}. Let $w_{h_{1}, h_{2}}(\vec{s}, \vec{z}) = e^{\kappa t(\text{supp}(\vec{s},\vec{z}))} h_{1}^{n(\vec{s})} h_{2}^{n(\vec{z})}$ 
be the weight system of mass $\kappa$ (where $n(\vec{s})$ and $n(\vec{z})$ denote the $\#$ of sites in 
$\vec{s}$ and $\vec{z}$ respectively) giving weight $h_{1}$ to the field $\Theta$ and $h_{2}$ to the field $\eta$. 
Let $Y = \text{supp} (\vec{s}, \vec{z})$. Define
\begin{equation}
||H^{\#}||_{w_{h_{1}, h_{2}}} = \sup_{Y} e^{\kappa t(Y)} ||H^{\#}(Y)||_{w_{h_{1}, h_{2}}}. 
\end{equation}
In our case field $\Theta$ (external bosonic field) is absent. We take support of $\Theta$ to be empty but the weight to
be non zero. Set $h_{1} = h_{2} = 1$. We have shown that
\begin{equation}
||H^{\#}||_{w_{1,1}} \leqslant c \hspace{0.05 cm} e_{0}^{\frac{1}{2} - \epsilon}
\end{equation}
set $\kappa = \kappa_{2} < \text{min} (\gamma_{1}, \gamma_{2})$ therefore,
\begin{equation}
||H^{\#}(Y)||_{w_{1,1}} \leqslant  c \hspace{0.05 cm} e_{0}^{\frac{1}{2} - \epsilon} \hspace{0.05 cm} e^{-\kappa_{2} t(Y)}.
\end{equation}

\subsubsection{From $\vec{s}, \vec{z}$ to polymer}
 We now make a transition from vector notation to polymers. 
A \textit{polymer} is a union of $[r_{\lambda}]$ blocks $\Box_{i}$ such that any two blocks centered on nearest $[r_{\lambda}]$  
neighbor sites, have a common hypersurface, face, edge or a site; \textit{connected} union of blocks.  
For a polymer X, let $|\text{X}|$ denote the $\#$ of $\Box$ in X.
 It is assumed that a series expansion of type $H^{\#}(Y, \Theta, \eta)$ is contained in
 each polymer.  Define the set
\begin{equation}
\mathrm{X}_{Y} = \{\text{all}\hspace{0.1 cm} [r_{\lambda}] \hspace{0.1 cm} \text{blocks}, 
\Box\hspace{0.1 cm} \text{containing}\hspace{0.1 cm} Y  : \hspace{0.1 cm} \mathrm{X}_{Y} \cap \Omega_{1} \neq \oslash\}. 
\end{equation}
Let $\mathrm{X} =\{\mathrm{X}_{i}\} \subset \Omega_{0}$ be a collection of polymers.
Rewrite
\begin{equation}
\text{log}\hspace{0.1 cm} \Xi(\Omega_{1}, \Theta, \eta) = 
\sum_{Y} H^{\#}(Y, \Theta, \eta) = \sum_{\mathrm{X}} \sum_{Y : \mathrm{X}_{Y} = \mathrm{X}} H^{\#} (Y, \Theta, \eta) 
= \sum_{\mathrm{X}} H^{\#}(\mathrm{X}, \Theta, \eta).
\end{equation}
Rewrite $\Xi(\Omega_{1}, \Theta, \eta)$ using Mayer expansion as
\begin{equation}
\begin{aligned}
\Xi(\Omega_{1}, \Theta, \eta) = e^{\sum_{\mathrm{X}}H^{\#}(\mathrm{X}, \Theta, \eta)} &=
\prod_{\mathrm{X}}\Big[e^{H^{\#}(\mathrm{X}, \Theta, \eta)} -1 + 1\Big] \\
&= \sum_{\{\mathrm{X}_{i}\} \rightarrow \mathrm{Z}}
\prod_{i}\Big[e^{H^{\#}(\mathrm{X}_{i}, \Theta, \eta)} -1\Big] \\
&= \sum_{\substack{\{\mathrm{Z}_{i}\}, \\ \text{disjoint}}} \prod_{i} K(\mathrm{Z}_{i}, \Theta, \eta)
\end{aligned}
\end{equation}
where $\{\mathrm{Z}_{i}\}$ are polymers and for a particular $\mathrm{Z}_{i}$,
\begin{equation}
K(\mathrm{Z}, \Theta, \eta) = \sum_{\cup \mathrm{X}_{i} = \mathrm{Z}} \prod_{i} \Big[e^{H^{\#}(\mathrm{X}_{i}, \Theta, \eta)} - 1\Big].
\end{equation}
For a polymer X, define
\begin{center}
$\tau$(X) : $[r_{\lambda}]^{-1}$ (length of the shortest tree in X joining $[r_{\lambda}] - \Box$ in X).
\end{center}
Note that for $Y = \text{supp} (\vec{s}, \vec{z})$ and 
$||H^{\#}(Y)||_{w_{1,1}} \leqslant c  \hspace{0.05 cm} e_{0}^{\frac{1}{2} - \epsilon} e^{- \kappa_{2} t(Y)}$,
\begin{equation}
\begin{aligned}
t(Y) &\geqslant [r_{\lambda}] \hspace{0.05 cm} \tau(\text{X}) \\ 
||H^{\#}(\text{X})||_{w_{1,1}} &\leqslant 
c  \hspace{0.05 cm} e_{0}^{\frac{1}{2} - \epsilon} e^{- \kappa_{2} [r_{\lambda}] \hspace{0.05 cm} \tau(\text{X})} \nonumber
\end{aligned}
\end{equation}
then for some $\kappa_{3}$ sufficiently large
\begin{equation}
||H^{\#}(\text{X})||_{w_{1,1}} \leqslant c \hspace{0.05 cm} e_{0}^{\frac{1}{2} - \epsilon} e^{- \kappa_{3} \tau(\text{X})}.
\end{equation}

\textbf{Lemma 3.3} Let $\mathrm{Z} = \{\mathrm{Z}_{i}\}$ be a disjoint collection of polymers and 
$K(\mathrm{Z}_{i}, \Theta, \eta)$ be the function as defined above. Then for a new constant $\kappa_{4} = \frac{\kappa_{3}}{2}$, 
$||K(\mathrm{Z})||_{w_{1,1}} \leqslant  c \hspace{0.05 cm} e_{0}^{\frac{1}{2} - \epsilon}  
\hspace{0.1 cm}  e^{-\kappa_{4} \tau(\mathrm{Z})}$.

\textit{Proof}  Rewrite the unordered collection $\{\mathrm{X}_{i}\}$ as ordered sets 
$(\mathrm{X}_{1}, \mathrm{X}_{2}, \cdots, \mathrm{X}_{n})$.
Then for a particular $\mathrm{Z}_{i}$ with $\cup \mathrm{X}_{i} = \mathrm{Z}_{i}$,
\begin{equation}
K(\mathrm{Z}, \Theta, \eta) = \sum_{n=1}^{\infty}\frac{1}{n!} \sum_{\cup \mathrm{X}_{i} = \mathrm{Z}}
\prod_{i} \Big[e^{H^{\#}(\mathrm{X}_{i}, \Theta, \eta)} - 1\Big].
\end{equation}
Now using  (3.67), 
\begin{equation}
\begin{aligned}
|e^{H^{\#}(\mathrm{X}_{i})} - 1| = |\int_{0}^{1}\frac{d}{ds} e^{s H^{\#}(\mathrm{X}_{i})} ds|
&= |\int_{0}^{1} H^{\#}(\mathrm{X}_{i}) e^{s H^{\#}(\mathrm{X}_{i})} ds| \\
&\leqslant c \hspace{0.05 cm} \mathcal{O}(1) \hspace{0.05 cm} e_{0}^{\frac{1}{2} - \epsilon}  
\hspace{0.1 cm} e^{-\kappa_{3} \tau(\mathrm{X}_{i})}
\end{aligned}
\end{equation}
and
\begin{equation}
\begin{aligned}
||K(\mathrm{Z})||_{w_{1,1}} &\leqslant  \sum_{n=1}^{\infty}\frac{1}{n!} \sum_{\cup \mathrm{X}_{i} = \mathrm{Z}}
\prod_{i=1}^{n} c \hspace{0.05 cm} \mathcal{O}(1) \hspace{0.05 cm} e_{0}^{\frac{1}{2} - \epsilon}
 \hspace{0.1 cm} e^{-\kappa_{3} \tau(\mathrm{X}_{i})} \\ 
&\leqslant\mathcal{O}(1) \hspace{0.05 cm} e_{0}^{\frac{1}{2} - \epsilon} \hspace{0.1 cm} e^{-\frac{\kappa_{3}}{2} \tau(\mathrm{Z})}
  \sum_{n=1}^{\infty}\frac{1}{n!} \sum_{(\mathrm{X}_{1},\cdots, \mathrm{X}_{n}) \subset \mathrm{Z}}
 \prod_{i=1}^{n} c  \hspace{0.1 cm} (\mathcal{O}(1) \hspace{0.05 cm} e_{0}^{\frac{1}{2} - \epsilon})^{\frac{1}{2}} \hspace{0.1 cm} 
 e^{-\frac{\kappa_{3}}{2} \tau(\mathrm{X}_{i})}  \\  
&\leqslant \mathcal{O}(1) \hspace{0.05 cm} e_{0}^{\frac{1}{2} - \epsilon} \hspace{0.1 cm} 
e^{-\frac{\kappa_{3}}{2} \tau(\mathrm{Z})} \sum_{n=1}^{\infty}\frac{1}{n!} 
\Big[\sum_{\mathrm{X} \subset \mathrm{Z}} c  \hspace{0.1 cm}
(\mathcal{O}(1) \hspace{0.05 cm} e_{0}^{\frac{1}{2} - \epsilon})^{\frac{1}{2}} \hspace{0.1 cm} 
e^{-\frac{\kappa_{3}}{2} \tau(\mathrm{X})}\Big]^{n} \\
\end{aligned}
\end{equation}
In second line, since $\{\mathrm{X}_{i}\}$ overlap, $\tau(\mathrm{X}_{1}) + \cdots + \tau(\mathrm{X}_{n}) \geqslant \tau(\mathrm{Z})$ 
and we also extracted a factor of $\mathcal{O}(1) \hspace{0.05 cm} e_{0}^{\frac{1}{2} - \epsilon}$ from 
$\mathcal{O}(1) \hspace{0.05 cm} (e_{0}^{\frac{1}{2} - \epsilon})^{n}$, leaving behind a factor of 
$(\mathcal{O}(1) \hspace{0.05 cm} e_{0}^{\frac{1}{2} - \epsilon})^{\frac{n}{2}}$. This inequality holds for $n\geqslant 2$.
Next we use lemma 25, \cite{D1} (there exists a constant $b$, such that $\sum_{X \cap Z \neq \oslash} e^{-a \tau(\mathrm{X})} 
\leqslant b$) and rewrite the bracketed term as

\begin{equation}
\sum_{\mathrm{X} \subset \mathrm{Z}} c  \hspace{0.1 cm}(\mathcal{O}(1) \hspace{0.05 cm} e_{0}^{\frac{1}{2} - \epsilon})^{\frac{1}{2}}
 \hspace{0.1 cm} e^{-\frac{\kappa_{3}}{2} \tau(\mathrm{X})} 
 \leqslant c  \hspace{0.1 cm}(\mathcal{O}(1) \hspace{0.05 cm} e_{0}^{\frac{1}{2} - \epsilon})^{\frac{1}{2}}
\hspace{0.1 cm} b. \nonumber
\end{equation}
Thus,
\begin{equation}
\begin{aligned}
||K(\mathrm{Z})||_{w_{1,1}} &\leqslant c \hspace{0.05 cm} e_{0}^{\frac{1}{2} - \epsilon} \hspace{0.1 cm} 
e^{-\frac{\kappa_{3}}{2} \tau(\mathrm{Z})} \sum_{n=1}^{\infty}\frac{1}{n!} 
( c \hspace{0.1 cm}(\mathcal{O}(1) \hspace{0.05 cm} e_{0}^{\frac{1}{2} - \epsilon})^{\frac{1}{2}}
\hspace{0.1 cm} b)^{n} \\
&\leqslant  \mathcal{O}(1) \hspace{0.05 cm} e_{0}^{\frac{1}{2} - \epsilon} \hspace{0.1 cm} 
e^{-\frac{\kappa_{3}}{2}\tau(\mathrm{Z})}  e^{c  \hspace{0.1 cm} 
(\mathcal{O}(1) \hspace{0.05 cm} e_{0}^{\frac{1}{2} - \epsilon})^{\frac{1}{2}} b} \\
&\leqslant  c \hspace{0.05 cm} e_{0}^{\frac{1}{2} - \epsilon} \hspace{0.1 cm}   e^{-\kappa_{4} \tau(\mathrm{Z})}.
\end{aligned}
\end{equation}

\section{Large field region} 
There are two large field regions
\begin{enumerate}
 \item  intermediate large field region $\mathrm{P}$ (recall the definition $\Omega = \Lambda_{1} - \mathrm{P}$)
  such that $\forall \hspace{0.05 cm} \Box \in \mathrm{P},
 \exists$ at least one $u \in \Box,$ with $p_{0,\lambda} < |\Phi (u)|$ and $\forall u \in \Box,  |\Phi (u)| < p_{\lambda}$,
  \item  large field region $\mathrm{Q}$ such that $\forall \hspace{0.05 cm} \Box \in \mathrm{Q}, \exists$
 at least one $u \in \Box,$ with  $|\Phi (u)| \geqslant p_{\lambda}$ and $v \neq 0$.
 \end{enumerate}
Recall that $\tilde{\mathrm{Q}} = \Lambda_{1}^{c}$, that is, $\tilde{\mathrm{Q}}$ is an enlargement of $\mathrm{Q}$
by $[r_{\lambda}]^{4}$. Each component $\tilde{\mathrm{Q}}_{k}$ of $\tilde{\mathrm{Q}}$ is a union of connected
components of $\mathrm{Q}$ written $\tilde{\mathrm{Q}}_{k} \supset \cup_{i} \mathrm{Q}_{k, i}$. Then 
$\mathrm{Q} = \cup_{k,i} \mathrm{Q}_{k, i} \subset \tilde{\mathrm{Q}}$. Denote 
\begin{equation}
\tilde{\mathrm{P}} \equiv \Lambda_{0} - \Omega_{1} 
\end{equation}
such that $\tilde{\mathrm{P}}$ is an enlargement of $\mathrm{P}$ by $16 [r_{\lambda}]^{4}$.
Each component $\tilde{\mathrm{P}}_{l}$ of $\tilde{\mathrm{P}}$ is a union of connected
components of $\mathrm{P}$ written $\tilde{\mathrm{P}}_{l} \supset \cup_{i} \mathrm{P}_{l, i}$. Then 
$\mathrm{P} = \cup_{l,i} \mathrm{P}_{l, i} \subset \tilde{\mathrm{P}}$. 

Rewrite the large field quadratic part in the generating functional 
\begin{equation}
\begin{aligned}
 -\frac{1}{2} & \langle\Phi, \text{T}_{\tilde{\mathrm{Q}}} \Phi\rangle + \frac{1}{2}
\langle \Phi, \text{T}_{\tilde{\mathrm{Q}}\Lambda_{1}} C_{\Lambda_{1}} \text{T}_{\Lambda_{1}\tilde{\mathrm{Q}}}
\Phi \rangle - \langle\bar{\psi}_{\beta}, \mathrm{D}_{\tilde{\mathrm{Q}}} \psi_{\alpha}\rangle +
\langle \bar{\psi}_{\beta}, 
\mathrm{D}_{\tilde{\mathrm{Q}}\Lambda_{1}} S_{\Lambda_{1}} \mathrm{D}_{\Lambda_{1}\tilde{\mathrm{Q}}} \psi_{\alpha} \rangle \\
&= -\frac{1}{2} \sum_{i} \langle\Phi, (\text{T}_{\tilde{\mathrm{Q}}_{i}} - \text{T}_{\tilde{\mathrm{Q}}_{i}\Lambda_{1}}
C_{\Lambda_{1}} \text{T}_{\Lambda_{1}\tilde{\mathrm{Q}}_{i}})\Phi\rangle + 
\sum_{i} \langle \bar{\psi}_{\beta}, (\mathrm{D}_{\tilde{\mathrm{Q}}_{i}} - \mathrm{D}_{\tilde{\mathrm{Q}}_{i}\Lambda_{1}}
S_{\Lambda_{1}} \mathrm{D}_{\Lambda_{1}\tilde{\mathrm{Q}}_{i}}) \psi_{\alpha}\rangle  \\
& + \frac{1}{2} 
\sum_{\substack{i,j \\ i\neq j}} \langle \Phi, \text{T}_{\tilde{\mathrm{Q}}_{i}\Lambda_{1}}
C_{\Lambda_{1}} \text{T}_{\Lambda_{1}\tilde{\mathrm{Q}}_{j}} \Phi\rangle +
\sum_{\substack{i,j \\ i\neq j}} \langle \bar{\psi}_{\beta}, \mathrm{D}_{\tilde{\mathrm{Q}}_{i}\Lambda_{1}}
S_{\Lambda_{1}} \mathrm{D}_{\Lambda_{1}\tilde{\mathrm{Q}}_{j}}  \psi_{\alpha}\rangle
\end{aligned}
\end{equation}
For the terms involving different polymers $\tilde{\mathrm{Q}}_{i}$ and $\tilde{\mathrm{Q}}_{j}$
we construct polymers connecting them. 
Consider two disjoint blocks $\Box$ and $\Box^{\prime}$ such that
$d(\Box, \Box^{\prime}) \geqslant [r_{\lambda}]$. Let X be a polymer connecting $\Box$ and $\Box^{\prime}$. For example,
  \begin{center}
$\text{X} (\Box, \Box^{\prime}) : \text{all 4-dimensional rectangular paths connecting}\hspace{0.1 cm} \Box, \Box^{\prime}$.
\end{center}
(By rectangular paths we mean starting at $\Box$ and selecting coordinate axis one at a time and traveling along it until the
coordinate of $\Box^{\prime}$ is reached.)
 Let $1_{\Box}$ and $1_{\Box^{\prime}}$ denote the characteristic functions restricting operators to 
  $\Box$ and $\Box^{\prime}$ respectively. Then
  \begin{equation}
 \begin{aligned}
\frac{1}{2} & \sum_{\substack{i,j \\ i\neq j}} \langle \Phi, \text{T}_{\tilde{\mathrm{Q}}_{i}\Lambda_{1}}
C_{\Lambda_{1}} \text{T}_{\Lambda_{1}\tilde{\mathrm{Q}}_{j}} \Phi\rangle 
+ \sum_{\substack{i,j \\ i\neq j}} \langle \bar{\psi}_{\beta}, \mathrm{D}_{\tilde{\mathrm{Q}}_{i}\Lambda_{1}}
S_{\Lambda_{1}} \mathrm{D}_{\Lambda_{1}\tilde{\mathrm{Q}}_{j}}  \psi_{\alpha}\rangle \\
&= \frac{1}{2}\sum_{\Box, \Box^{\prime}} \sum_{i,j}
\langle \Phi, 1_{\Box} \text{T}_{\tilde{\mathrm{Q}}_{i}\Lambda_{1}} 
C_{\Lambda_{1}} \text{T}_{\Lambda_{1}\tilde{\mathrm{Q}}_{j}} 1_{\Box^{\prime}} \Phi\rangle 
+ \sum_{\Box, \Box^{\prime}} \sum_{i,j}
\langle \bar{\psi}_{\beta}, 1_{\Box} \mathrm{D}_{\tilde{\mathrm{Q}}_{i}\Lambda_{1}} 
S_{\Lambda_{1}} \mathrm{D}_{\Lambda_{1}\tilde{\mathrm{Q}}_{j}} 1_{\Box^{\prime}} \psi_{\alpha} \rangle \\  
&= \sum_{\text{X}} \sum_{\Box, \Box^{\prime} \rightarrow \text{X}} \sum_{i,j} 
\Big[\frac{1}{2} \langle \Phi, 1_{\Box} \text{T}_{\tilde{\mathrm{Q}}_{i}\Lambda_{1}} C_{\Lambda_{1}} \text{T}_{\Lambda_{1}\tilde{\mathrm{Q}}_{j}} 1_{\Box^{\prime}} \Phi\rangle + \langle  \bar{\psi}_{\beta}, 1_{\Box} \mathrm{D}_{\tilde{\mathrm{Q}}_{i}\Lambda_{1}} S_{\Lambda_{1}} 
\mathrm{D}_{\Lambda_{1}\tilde{\mathrm{Q}}_{j}} 1_{\Box^{\prime}} \psi_{\alpha} \rangle \Big] \\
&= \sum_{\text{X}} \sigma (\text{X})
\end{aligned}
\end{equation}
where only disjoint $\Box, \Box^{\prime}$ contribute and
\begin{equation}
\sigma (\text{X}) = \sum_{\Box, \Box^{\prime} \rightarrow \text{X}} \sum_{i,j} 
\Big[\frac{1}{2} \langle \Phi, 1_{\Box} \text{T}_{\tilde{\mathrm{Q}}_{i}\Lambda_{1}} C_{\Lambda_{1}} \text{T}_{\Lambda_{1}\tilde{\mathrm{Q}}_{j}} 1_{\Box^{\prime}} \Phi\rangle + \langle  \bar{\psi}_{\beta}, 1_{\Box} \mathrm{D}_{\tilde{\mathrm{Q}}_{i}\Lambda_{1}} S_{\Lambda_{1}} 
\mathrm{D}_{\Lambda_{1}\tilde{\mathrm{Q}}_{j}} 1_{\Box^{\prime}} \psi_{\alpha} \rangle \Big].
\end{equation} 
Thus, $e^{\sum_{i \neq j} [\frac{1}{2} \langle \Phi, \text{T}_{\tilde{\mathrm{Q}}_{i}\Lambda_{1}}
C_{\Lambda_{1}} \text{T}_{\Lambda_{1}\tilde{\mathrm{Q}}_{j}} \Phi\rangle + 
\langle \bar{\psi}_{\beta}, \mathrm{D}_{\tilde{\mathrm{Q}}_{i}\Lambda_{1}}
S_{\Lambda_{1}} \mathrm{D}_{\Lambda_{1}\tilde{\mathrm{Q}}_{j}}  \psi_{\alpha}\rangle]} =  e^{\sum_{\text{X}} \sigma (\text{X})}$
and rewrite it using Mayer expansion as 
\begin{equation}
 e^{\sum_{\text{X}} \sigma (\text{X})} = \prod_{\text{X}} \Big[ (e^{\sigma(\text{X})} - 1) + 1\Big]
= \sum_{\substack{\{\text{X}_{i}\} \\ \text{distinct}}} \prod_{i} \Big[ e^{\sigma(\text{X}_{i})} - 1\Big] = 
\sum_{\substack{\{\text{X}_{i}\} \\ \text{disjoint}}} \prod_{i} f(\text{X}_{i})
\end{equation}
where in the last step we group the collection $\{\text{X}_{i}\}$ into disjoint connected sets and define for X connected
\begin{equation}
f(\text{X}) = \sum_{\cup \text{X}_{i} = \text{X}} \prod_{i} \Big[ e^{\sigma(\text{X}_{i})} - 1\Big].
\end{equation}
Denote $\mathcal{D}\bar{\psi} \mathcal{D}\psi = \prod_{x, \alpha, \beta} d\bar{\psi}_{\beta}(x) d\psi_{\alpha}(x)$. \\
Recall that $ \mathrm{Z}_{0}(\Omega_{1}) = \int \mathcal{D}\bar{\psi}_{\Omega_{1}} \mathcal{D}\psi_{\Omega_{1}}
\mathcal{D}\Phi^{\prime}_{\Omega_{1}}  \hspace{0.05 cm} e^{- \langle \bar{\psi}_{\beta}, \psi_{\alpha} \rangle_{\Omega_{1}} 
-\frac{1}{2}||\Phi^{\prime}||^{2}_{\Omega_{1}}}$ and rewrite 
\begin{equation}
\begin{aligned}
\mathrm{Z}_{0}(\Omega_{1}) =  
\mathrm{Z}_{0}(\mathbb{T}) \hspace{0.05 cm} \mathrm{Z}_{0}(\Omega_{1}^{c})^{-1}  
= \mathrm{Z}_{0}(\mathbb{T}) \hspace{0.05 cm} \mathrm{Z}_{0}(\mathrm{Q})^{-1} 
\hspace{0.05 cm} \mathrm{Z}_{0}(\tilde{\mathrm{P}})^{-1}.
\end{aligned}
\end{equation}
Then the generating functional is
\begin{equation}
\begin{aligned}
\mathrm{Z}[\eta] = & \Big(\frac{2\pi}{e_{0}}\Big)^{-|\mathbb{T}| + 1} 
\frac{\text{det}\hspace{0.1 cm}C^{\frac{1}{2}}}{\text{det}\hspace{0.1 cm}S} \hspace{0.1 cm} \mathrm{Z}_{0}(\mathbb{T}) \hspace{0.1 cm}
 \sum_{v: dv=0} \sum_{\{\tilde{\mathrm{Q}}_{k}\}} \sum_{\{\text{X}_{i}\}}
\sum_{\{\tilde{\mathrm{P}}_{l}\}} \sum_{\{\text{Z}_{j}\}} \\ &  \hspace{0.1 cm} \prod_{k} 
e^{W_{2}(\tilde{\mathrm{Q}}_{k}) - W^{\prime}_{2}(\tilde{\mathrm{Q}}_{k})}  
\int \prod_{k} \mathcal{D}\Phi_{\tilde{\mathrm{Q}}_{k}} \hspace{0.1 cm}
  \mathcal{D}\bar{\psi}_{\tilde{\mathrm{Q}}_{k}} \mathcal{D}\psi_{\tilde{\mathrm{Q}}_{k}} \hspace{0.1 cm}
 \zeta_{\mathrm{Q}_{k}} \hspace{0.05 cm} \chi_{\tilde{\mathrm{Q}}_{k} \backslash \mathrm{Q}_{k}}(\Phi) \\
& e^{-\frac{1}{2} \langle\Phi, (\text{T}_{\tilde{\mathrm{Q}}_{k}} - \text{T}_{\tilde{\mathrm{Q}}_{k}\Lambda_{1}}
C_{\Lambda_{1}} \text{T}_{\Lambda_{1}\tilde{\mathrm{Q}}_{k}})\Phi\rangle 
- \langle \bar{\psi}_{\beta}, (\mathrm{D}_{\tilde{\mathrm{Q}}_{k}} - \mathrm{D}_{\tilde{\mathrm{Q}}_{k}\Lambda_{1}}
S_{\Lambda_{1}} \mathrm{D}_{\Lambda_{1}\tilde{\mathrm{Q}}_{k}})  \psi_{\alpha}\rangle} e^{- V(\tilde{\mathrm{Q}}_{k}, \psi, \bar{\psi}, \Phi)} \\
& e^{V_{s}(\tilde{\mathrm{Q}}_{k}) - V_{\epsilon}^{\prime}(\tilde{\mathrm{Q}}_{k})} 
\prod_{k}  \mathrm{Z}_{0}(\mathrm{Q}_{k})^{-1}  \prod_{i}  f(\text{X}_{i}, \eta) \hspace{0.1 cm}    
\prod_{l} e^{W_{2}(\tilde{\mathrm{P}}_{l}) - W^{\prime}_{2}(\tilde{\mathrm{P}}_{l})} \hspace{0.1 cm} \\
&  \int \prod_{l} \mathcal{D}\Phi^{\prime}_{\tilde{\mathrm{P}}_{l}} 
 \mathcal{D}\bar{\psi}_{\tilde{\mathrm{P}}_{l}}  \mathcal{D}\psi_{\tilde{\mathrm{P}}_{l}}    
 e^{-\frac{1}{2}||\Phi^{\prime}||_{\tilde{\mathrm{P}}_{l}}^{2} - \langle\bar{\psi}_{\beta}, \psi_{\alpha} \rangle_{\tilde{\mathrm{P}}_{l}}}
e^{- V(\tilde{\mathrm{P}}_{l}, S^{\text{loc}}_{\Lambda_{1}}\psi - \Psi,
  \bar{\psi} -  \bar{\Psi}, C^{\frac{1}{2},\text{loc}}_{\Lambda_{1}} \Phi^{\prime} - \varphi)} 
  \\ & e^{V_{s}(\tilde{\mathrm{P}}_{l}) - V_{\epsilon}^{\prime}(\tilde{\mathrm{P}}_{l})} 
    \hat{\zeta}_{\mathrm{P}_{l}}(\Phi^{\prime}) \hspace{0.1 cm}  
\chi_{\tilde{\mathrm{P}}_{l}}(C^{\frac{1}{2},\text{loc}}_{\Lambda_{1}} \Phi^{\prime} -  \varphi) 
\prod_{l} \mathrm{Z}_{0}(\tilde{\mathrm{P}}_{l})^{-1}  \hspace{0.1 cm}  \prod_{j} K(\text{Z}_{j}, \Theta, \eta).
\end{aligned}
\end{equation}
where  $\zeta_{\mathrm{P}_{l}} = \prod_{i}\zeta_{\mathrm{P}_{l, i} : \cup_{i} \mathrm{P}_{l, i} \subset \tilde{\mathrm{P}}_{l}}(\Phi^{\prime})$ 
and $\zeta_{\mathrm{Q}_{k}} = \prod_{i}\zeta_{\mathrm{Q}_{k, i} : \cup_{i} \mathrm{Q}_{k, i} \subset \tilde{\mathrm{Q}}_{k}}(\Phi)$. 
Denote 
\begin{equation}
\begin{aligned}
\rho(\tilde{\mathrm{Q}}_{k}, \eta) &= e^{W_{2}(\tilde{\mathrm{Q}}_{k}) - W^{\prime}_{2}(\tilde{\mathrm{Q}}_{k})} \hspace{0.1 cm}
e^{-\frac{1}{2} \langle\Phi, (\text{T}_{\tilde{\mathrm{Q}}_{k}} - \text{T}_{\tilde{\mathrm{Q}}_{k}\Lambda_{1}}
C_{\Lambda_{1}} \text{T}_{\Lambda_{1}\tilde{\mathrm{Q}}_{k}})\Phi\rangle 
- \langle \bar{\psi}_{\beta}, (\mathrm{D}_{\tilde{\mathrm{Q}}_{k}} - \mathrm{D}_{\tilde{\mathrm{Q}}_{k}\Lambda_{1}}
S_{\Lambda_{1}} \mathrm{D}_{\Lambda_{1}\tilde{\mathrm{Q}}_{k}})  \psi_{\alpha}\rangle}
\\ &\hspace{0.4 cm}  e^{- V(\tilde{\mathrm{Q}}_{k}, \psi, \bar{\psi}, \Phi) + V_{s}(\tilde{\mathrm{Q}}_{k}) -
V_{\epsilon}^{\prime}(\tilde{\mathrm{Q}}_{k})} \zeta_{\mathrm{Q}_{k}}(\Phi)
\hspace{0.05 cm} \chi_{\tilde{\mathrm{Q}}_{k} \backslash \mathrm{Q}_{k}}(\Phi)  \hspace{0.1 cm} \mathrm{Z}_{0}(\mathrm{Q}_{k})^{-1} \\
\rho^{\prime}(\tilde{\mathrm{P}}_{l}, \eta) &= e^{W_{2}(\tilde{\mathrm{P}}_{l}) - W^{\prime}_{2}(\tilde{\mathrm{P}}_{l})}  
e^{-\frac{1}{2}||\Phi^{\prime}||_{\tilde{\mathrm{P}}_{l}}^{2} - \langle\bar{\psi}_{\beta}, \psi_{\alpha} \rangle_{\tilde{\mathrm{P}}_{l}}}
\hspace{0.2 cm} e^{- V(\tilde{\mathrm{P}}_{l}, S^{\text{loc}}_{\Lambda_{1}}\psi - \Psi,
  \bar{\psi} -  \bar{\Psi}, C^{\frac{1}{2},\text{loc}}_{\Lambda_{1}} \Phi^{\prime} - \varphi)}
 \\ &
\hspace{0.4 cm}  e^{V_{s}(\tilde{\mathrm{P}}_{l}) - V_{\epsilon}^{\prime}(\tilde{\mathrm{P}}_{l})} 
 \hat{\zeta}_{\mathrm{P}_{l}}(\Phi^{\prime})
   \chi_{\tilde{\mathrm{P}}_{l}} (C^{\frac{1}{2},\text{loc}}_{\Lambda_{1}} \Phi^{\prime} - \varphi) \hspace{0.1 cm} 
\mathrm{Z}_{0}(\tilde{\mathrm{P}}_{l})^{-1}
\end{aligned}
\end{equation}
and rewrite the generating functional as
\begin{equation}
\begin{aligned}
\mathrm{Z}[\eta] &= \Big(\frac{2\pi}{e_{0}}\Big)^{-|\mathbb{T}| + 1} 
\frac{\text{det}\hspace{0.1 cm}C^{\frac{1}{2}}}{\text{det}\hspace{0.1 cm}S}
\hspace{0.1 cm} \mathrm{Z}_{0}(\mathbb{T}) \hspace{0.1 cm}
 \sum_{v: dv=0} \sum_{\{\tilde{\mathrm{Q}}_{k}\}} \sum_{\{\text{X}_{i}\}} \sum_{\{\tilde{\mathrm{P}}_{l}\}} \sum_{\{\text{Z}_{j}\}}
\\ &  \prod_{k} \int  \mathcal{D}\Phi_{\tilde{\mathrm{Q}}_{k}}  \hspace{0.1 cm}
  \mathcal{D}\bar{\psi}_{\tilde{\mathrm{Q}}_{k}} \mathcal{D}\psi_{\tilde{\mathrm{Q}}_{k}} 
\rho(\tilde{\mathrm{Q}}_{k}, \eta) \hspace{0.1 cm} \prod_{i}  f(\text{X}_{i}, \eta) \hspace{0.1 cm}  
\\ &  \prod_{l} \int \mathcal{D}\Phi^{\prime}_{\tilde{\mathrm{P}}_{l}}
  \mathcal{D}\bar{\psi}_{\tilde{\mathrm{P}}_{l}}  \mathcal{D}\psi_{\tilde{\mathrm{P}}_{l}} 
   \rho^{\prime}(\tilde{\mathrm{P}}_{l}, \eta) \hspace{0.1 cm}
   \prod_{j} K(\text{Z}_{j}, \Theta, \eta).
\end{aligned}
\end{equation}
Let $w_{h}(\vec{\xi}) = e^{\kappa t(\text{supp} (\vec{\xi}))} h^{n(\vec{\xi})}$ be a weight system of mass $\kappa$ 
giving weight at least $h$ to $\bar{\psi}, \psi$ and $n(\vec{\xi})$ denotes the $\#$ sites in $\vec{\xi}$.

In the estimates below $\mathcal{O}(1)$ refers to a numerical value independent of the coupling constants and 
depending only on the number of dimensions.

\textbf{Lemma 4.1} There exists a constant $\kappa^{\prime} > 0$ such that
$||f(\text{X})||_{w_{1}} \leqslant 
\mathcal{O}(1) \hspace{0.05 cm} e_{0}^{2- 2\epsilon} \hspace{0.05 cm} e^{-\kappa^{\prime} \tau(\text{X})}$.
 
\textit{Proof} Let $x \in \Box$ and $x^{\prime} \in \Box^{\prime}$.
Define a metric $d(x, x^{\prime}) = \sup_{\mu} |x_{\mu} - x^{\prime}_{\mu}| $ and
$d(\Box, \Box^{\prime}) = \inf_{x, x^{\prime}} d(x, x^{\prime})$. Then
$|C_{\Lambda_{1}}(x, x^{\prime})| \leqslant c \hspace{0.05 cm} e^{-\gamma_{1} d(x, x^{\prime})} \leqslant 
c \hspace{0.05 cm} e^{-\gamma_{1} d(\Box, \Box^{\prime})}$ and \\
$|S_{\Lambda_{1}}(x, x^{\prime})| \leqslant c \hspace{0.05 cm} e^{-\gamma_{2} d(x, x^{\prime})} \leqslant 
c \hspace{0.05 cm} e^{-\gamma_{2} d(\Box, \Box^{\prime})}$.  
Let X is a polymer joining $\Box, \Box^{\prime}$. First estimate $\sigma(\text{X})$. 
Note that

\begin{equation}
\begin{aligned}
& \langle  \bar{\psi}, 1_{\Box} \text{D}_{\tilde{\mathrm{Q}}_{i}\Lambda_{1}} S_{\Lambda_{1}} 
\text{D}_{\Lambda_{1}\tilde{\mathrm{Q}}_{j}} 1_{\Box^{\prime}} \psi \rangle
= \sum_{x \in \Lambda_{1}} \sum_{\alpha} (\bar{\psi} 1_{\Box}  \text{D}_{\tilde{\mathrm{Q}}_{i}\Lambda_{1}})_{\alpha}(x)
(S_{\Lambda_{1}} \text{D}_{\Lambda_{1}\tilde{\mathrm{Q}}_{j}} 1_{\Box^{\prime}} \psi)_{\alpha}(x) \\
&=  \sum_{\substack{x \in \Lambda_{1} : (x+e_{\mu}) \in \Box \\ y \in \Lambda_{1}}} 
\sum_{\substack{\alpha,\beta \\ \mu}} \bar{\psi}_{\beta} (x+e_{\mu})
 \text{D}_{\tilde{\mathrm{Q}}_{i}\Lambda_{1}, \alpha,\beta} (x,x+e_{\mu}) S_{\Lambda_{1},\alpha\beta}(x, y) 
(\text{D}_{\Lambda_{1}\tilde{\mathrm{Q}}_{j}}1_{\Box^{\prime}} \psi)_{\beta} (y) \\
&=  \sum_{\substack{x \in \Lambda_{1} : (x+e_{\mu}) \in \Box  \\ y \in \Lambda_{1} : (y+e_{\mu}) \in \Box^{\prime}}}
 \sum_{\substack{\alpha,\beta,\beta^{\prime} \\ \mu }} \bar{\psi}_{\beta} (x+e_{\mu})
 \text{D}_{\tilde{\mathrm{Q}}_{i}\Lambda_{1}, \alpha,\beta} (x,x+e_{\mu}) S_{\Lambda_{1},\alpha\beta}(x, y) 
\text{D}_{\Lambda_{1}\tilde{\mathrm{Q}}_{j},\beta, \beta^{\prime}} (y,y+e_{\mu})\psi_{\beta^{\prime}}(y+e_{\mu}) \\
&=  \sum_{\substack{x \in \Lambda_{1} : (x+e_{\mu}) \in \Box  \\ y \in \Lambda_{1} : (y+e_{\mu}) \in \Box^{\prime}}}
 \sum_{\substack{\beta,\beta^{\prime} \\ \mu}}
a_{\beta,\beta^{\prime}} (x+e_{\mu}, y+e_{\mu})\bar{\psi}_{\beta} (x+e_{\mu}) \psi_{\beta^{\prime}}(y+e_{\mu})
\end{aligned}
\end{equation}
where 
\begin{equation}
a_{\beta,\beta^{\prime}} (x+e_{\mu}, y+e_{\mu}) =  \sum_{\alpha}
\text{D}_{\tilde{\mathrm{Q}}_{i}\Lambda_{1}, \alpha,\beta} (x,x+e_{\mu}) S_{\Lambda_{1},\alpha\beta}(x, y) 
\text{D}_{\Lambda_{1}\tilde{\mathrm{Q}}_{j},\beta, \beta^{\prime}} (y,y+e_{\mu}) 
\end{equation}
To calculate $|a_{\beta,\beta^{\prime}} (x+e_{\mu}, y+e_{\mu})|$ first note that since 
$\Lambda_{1}, \tilde{\mathrm{Q}}$ are disjoint there is no mass term in $\text{D}_{\Lambda_{1}\tilde{\mathrm{Q}}}$, 
therefore, $||\text{D}_{\Lambda_{1}\tilde{\mathrm{Q}}}||^{2} \leqslant  \mathcal{O}(1)$. Then
\begin{equation}
\begin{aligned}
|a_{\beta,\beta^{\prime}} (x+e_{\mu}, y+e_{\mu})| \leqslant \sum_{\alpha} \mathcal{O}(1) \hspace{0.05 cm} c \hspace{0.05 cm}
e^{-\gamma_{2}d(x+e_{\mu}, y+e_{\mu})}  
 \leqslant 4 \hspace{0.05 cm} \mathcal{O}(1) \hspace{0.05 cm} e^{-\gamma_{2} d(x+e_{\mu}, y+e_{\mu})}  
\end{aligned}
\end{equation}
Therefore, for $\kappa < \gamma_{2}$
\begin{equation}
\begin{aligned}
||\langle  \bar{\psi}, & 1_{\Box} \text{D}_{\tilde{\mathrm{Q}}_{i}\Lambda_{1}} S_{\Lambda_{1}} 
\text{D}_{\Lambda_{1}\tilde{\mathrm{Q}}_{j}} 1_{\Box^{\prime}} \psi \rangle||_{w_{1}}
\leqslant \max_{z} \max_{1\leqslant i \leqslant 2} \sum_{\substack{x,y \\ x \text{or} y = z}} 
\sum_{\substack{\beta,\beta^{\prime} \\ \mu}}
e^{\kappa d(x+e_{\mu}, y+e_{\mu})} |a_{\beta,\beta^{\prime}} (x+e_{\mu}, y+e_{\mu})| \\
&\leqslant \max_{z} \max_{1\leqslant i \leqslant 2} \sum_{\substack{x,y \\ x \text{or} y = z}} 
4 \hspace{0.05 cm} \mathcal{O}(1) \hspace{0.05 cm} 4^{3} e^{-(\gamma_{2} - \kappa) d (x+e_{\mu}, y+e_{\mu})} \\
&\leqslant \mathcal{O}(1) \hspace{0.05 cm} 256 \hspace{0.05 cm} e^{-\frac{3}{4}(\gamma_{2} - \kappa) d (\Box, \Box^{\prime})}
\max_{z} \max_{1\leqslant i \leqslant 2} \sum_{\substack{x,y \\ x \text{or} y = z}} 
 e^{-\frac{1}{4}(\gamma_{2} - \kappa) d (x+e_{\mu}, y+e_{\mu})} \\
 &\leqslant \mathcal{O}(1)  \hspace{0.05 cm} e_{0}^{2} \hspace{0.05 cm} 
 e^{-\frac{1}{4}(\gamma_{2} - \kappa) d (\Box, \Box^{\prime})}
\end{aligned}
\end{equation}
where the last step follows as $\Box, \Box^{\prime}$ are disjoint $d(\Box, \Box^{\prime}) \geqslant [r_{\lambda}]$
and $e^{-\frac{1}{2}(\gamma_{2} - \kappa) [r_{\lambda}]} = \mathcal{O}(e_{0}^{2})$.
Again since $\Lambda_{1}, \tilde{\mathrm{Q}}$ are disjoint, $\text{T}_{\Lambda_{1}\tilde{\mathrm{Q}}}$ 
does not have a mass term;
$||\text{T}_{\tilde{\mathrm{Q}}_{k}\Lambda_{1}} \Phi||^{2} \leqslant 2^{4+1} ||\Phi||^{2}$ (see Lemma 4.2 in \cite{G1})
and therefore, 
\begin{equation}
||\langle \Phi, 1_{\Box} \text{T}_{\tilde{\mathrm{Q}}_{i}\Lambda_{1}} C_{\Lambda_{1}} \text{T}_{\Lambda_{1}\tilde{\mathrm{Q}}_{j}} 1_{\Box^{\prime}} \Phi\rangle|| \leqslant 
 2^{4+1}\hspace{0.05 cm} e_{0}^{2} \hspace{0.05 cm} p_{\lambda}^{2} e^{-\frac{\gamma_{1}}{2}d (\Box, \Box^{\prime})}. \nonumber
\end{equation}
For a given X joining $\Box, \Box^{\prime}$, 
the number of $(\Box, \Box^{\prime})$ with $\Box \in \tilde{\mathrm{Q}}_{i}$ and $\Box^{\prime} \in \tilde{\mathrm{Q}}_{j}$
is bounded by $\mathcal{O}(1)$ depending only on the dimension.
The number of $(\tilde{\mathrm{Q}}_{i}, \tilde{\mathrm{Q}}_{j})$ is bounded by $\mathcal{O}(|\text{X}|)$. 
There are $d! = 24$ ways  to construct X. Thus, $d(\Box, \Box^{\prime}) \geqslant \tau(\text{X}) \frac{[r_{\lambda}]}{24}$. 
Also, near the boundary $\partial \Lambda_{1}$, 
$|\Phi(x)| \leqslant p_{\lambda}, \forall x \in \Lambda_{0}$. 
Use $e^{-\frac{\gamma_{1}}{2}[r_{\lambda}]} = \mathcal{O}(e_{0}^{2})$ and
denote $\gamma^{\prime} = \text{min} \hspace{0.05 cm} (\frac{\gamma_{1}}{2}, \frac{1}{4}(\gamma_{2} - \kappa))$. Then
from (4.4)
\begin{equation}
\begin{aligned}
||\sigma (\text{X})||_{w_{1}} &\leqslant  \mathcal{O}(|\text{X}|) 
 \Big[ 2^{4+1}\hspace{0.05 cm} e_{0}^{2} \hspace{0.05 cm} p_{\lambda}^{2} e^{-\frac{\gamma_{1}}{2}d (\Box, \Box^{\prime}) } + 
\mathcal{O}(1)  \hspace{0.05 cm} e_{0}^{2}  \hspace{0.05 cm} 
 e^{-\frac{1}{4}(\gamma_{2} - \kappa) d (\Box, \Box^{\prime})} \Big] \\
&\leqslant  \mathcal{O}(|\text{X}|) 
 \Big[ 32 \hspace{0.05 cm} e_{0}^{2} \hspace{0.05 cm} p_{\lambda}^{2} + 
\mathcal{O}(1)  \hspace{0.05 cm} e_{0}^{2}  \Big]  e^{- \gamma^{\prime} d (\Box, \Box^{\prime})} \\
 &\leqslant \mathcal{O}(1) \hspace{0.05 cm} \mathcal{O}(|\text{X}|)  \hspace{0.05 cm}  e_{0}^{2- 2\epsilon} \hspace{0.05 cm} 
 e^{- \gamma^{\prime} \frac{[r_{\lambda}]}{24} \tau(\text{X}) }.
\end{aligned}
\end{equation}
Following the analysis of Lemma 3.3 for some constant $\kappa^{\prime}$ sufficiently large,
\begin{center}
$||f(\text{X})||_{w_{1}} \leqslant 
\mathcal{O}(1) \hspace{0.05 cm} e_{0}^{2- 2\epsilon} \hspace{0.05 cm} e^{- \kappa^{\prime} \tau(\text{X})}$.
\end{center}

\textbf{Lemma 4.2} The following bounds hold.
\begin{enumerate}
  \item $ ||\langle  \bar{\psi}, (\text{D}_{\tilde{\mathrm{Q}}_{k}} - \text{D}_{\tilde{\mathrm{Q}}_{k}\Lambda_{1}} S_{\Lambda_{1}} 
\text{D}_{\Lambda_{1}\tilde{\mathrm{Q}}_{k}}) \psi \rangle||_{w_{1}} \leqslant \mathcal{O}(1)$.
  \item Let $m_{\text{min}} = \text{min} (\mu^{2}, m_{A}^{2})$. Then for $m_{\text{min}} > 2^{\frac{4}{2} + 1} = 8$
and constants $\gamma_{3}, \gamma_{4} > 0$, 
\begin{equation}
\begin{aligned}
\frac{1}{2}  \langle\Phi, (\text{T}_{\tilde{\mathrm{Q}}_{k}} - \text{T}_{\tilde{\mathrm{Q}}_{k}\Lambda_{1}}
C_{\Lambda_{1}} \text{T}_{\Lambda_{1}\tilde{\mathrm{Q}}_{k}})\Phi\rangle 
+ V(\tilde{\mathrm{Q}}_{k}, \psi, \bar{\psi}, \Phi)
\geqslant \gamma_{3} \langle\Phi, \text{T} \Phi\rangle_{\tilde{\mathrm{Q}}_{k}}  
+ \gamma_{4} \hspace{0.05 cm} ||\Phi||^{2}_{\tilde{\mathrm{Q}}_{k}}. \nonumber
\end{aligned}
\end{equation}
\end{enumerate}

\textit{Proof} Note that  
\begin{equation}
\begin{aligned}
\langle \bar{\psi}, \text{D}_{\tilde{\mathrm{Q}}_{k}} \psi \rangle = \sum_{x \in \tilde{\mathrm{Q}}_{k}} \sum_{\alpha}
\bar{\psi}_{\alpha}(x)  (\text{D}_{\tilde{\mathrm{Q}}_{k}} \psi)_{\alpha} (x) &= \sum_{x \in \tilde{\mathrm{Q}}_{k}} 
\sum_{\alpha, \beta} \bar{\psi}_{\alpha}(x) \text{D}_{\tilde{\mathrm{Q}}_{k},\alpha\beta}(x,x+e_{\mu}) \psi_{\beta}(x+e_{\mu}) \\
&= \sum_{x \in \tilde{\mathrm{Q}}_{k}} \sum_{\alpha, \beta} a_{\alpha,\beta}(x,x+e_{\mu}) \bar{\psi}_{\alpha}(x) \psi_{\beta}(x+e_{\mu})
\end{aligned}
\end{equation}
where $a_{\alpha,\beta}(x,x+e_{\mu}) = \text{D}_{\tilde{\mathrm{Q}}_{k},\alpha\beta}(x,x+e_{\mu})$. Let
$\kappa  < \gamma_{2}$. Then
\begin{equation}
\begin{aligned}
||\langle \bar{\psi}, \text{D}_{\tilde{\mathrm{Q}}_{k}} \psi \rangle||_{w_{1}} \leqslant \max_{x} 
 \sum_{\alpha, \beta} e^{\kappa} \hspace{0.05 cm} |a_{\alpha,\beta}(x,x+e_{\mu})| 
 \leqslant 16 \hspace{0.05 cm} e^{\kappa} \hspace{0.05 cm} \mathcal{O}(m_{f})
\end{aligned}
\end{equation}
and from Lemma 4.1 
\begin{equation} 
||\langle  \bar{\psi}, \text{D}_{\tilde{\mathrm{Q}}_{k}\Lambda_{1}} S_{\Lambda_{1}} 
\text{D}_{\Lambda_{1}\tilde{\mathrm{Q}}_{k}} \psi \rangle||_{w_{1}} \leqslant \mathcal{O}(1). 
\end{equation}
Since $m_{f} > 1$
\begin{equation}
\begin{aligned}
||\langle  \bar{\psi}, (\text{D}_{\tilde{\mathrm{Q}}_{k}} - \text{D}_{\tilde{\mathrm{Q}}_{k}\Lambda_{1}} S_{\Lambda_{1}} 
\text{D}_{\Lambda_{1}\tilde{\mathrm{Q}}_{k}}) \psi \rangle||_{w_{1}} &\leqslant
||\langle \bar{\psi}, \text{D}_{\tilde{\mathrm{Q}}_{k}} \psi \rangle||_{w_{1}} + 
||\langle  \bar{\psi}, \text{D}_{\tilde{\mathrm{Q}}_{k}\Lambda_{1}} S_{\Lambda_{1}} 
\text{D}_{\Lambda_{1}\tilde{\mathrm{Q}}_{k}} \psi \rangle||_{w_{1}} \\ 
&\leqslant  16 \hspace{0.05 cm} e^{\kappa} \hspace{0.05 cm} \mathcal{O}(m_{f}) + \mathcal{O}(1) \\
&\leqslant \mathcal{O}(1).
\end{aligned}
\end{equation}
This completes the proof of (1).

First consider the Yukawa term $V_{Y}(\tilde{\mathrm{Q}}_{k}) = \sum_{x \in \tilde{\mathrm{Q}}_{k}}  g  (\bar{\psi} \rho \psi)(x)$ 
in $V(\tilde{\mathrm{Q}}_{k}, \psi, \bar{\psi}, \Phi)$ and note that
\begin{equation}
\begin{aligned}
||V_{Y}(\tilde{\mathrm{Q}}_{k})||_{w_{1}} = \sum_{x \in \tilde{\mathrm{Q}}_{k}} \sum_{\alpha, \beta}
||g  (\bar{\psi}_{\beta} \rho \psi_{\alpha})(x) ||_{w_{1}} 
\leqslant  g \hspace{0.05 cm} 16  \hspace{0.05 cm} ||\Phi||_{\tilde{\mathrm{Q}}_{k}} \leqslant  
g \hspace{0.05 cm} 16  \hspace{0.05 cm}  ||\Phi||^{2}_{\tilde{\mathrm{Q}}_{k}}.
\end{aligned}
\end{equation}
Let $V_{B}(\tilde{\mathrm{Q}}_{k},\Phi)$ denote the bosonic terms in $V(\tilde{\mathrm{Q}}_{k}, \psi, \bar{\psi}, \Phi)$
such that $V_{B} + V_{Y} = V$.
Next from Lemma 4.2 in \cite{G1} first rewrite
\begin{equation}
\begin{aligned}
&\langle\Phi, (\text{T}_{\tilde{\mathrm{Q}}_{k}} - \text{T}_{\tilde{\mathrm{Q}}_{k}\Lambda_{1}}
C_{\Lambda_{1}} \text{T}_{\Lambda_{1}\tilde{\mathrm{Q}}_{k}})\Phi\rangle 
+ V_{B}(\tilde{\mathrm{Q}}_{k}, \Phi) \\ &= \frac{1}{2} \langle \Phi, (\mathrm{T}_{\tilde{\mathrm{Q}}_{k}} - 
2 \hspace{0.05 cm} \text{T}_{\tilde{\mathrm{Q}}_{k}\Lambda_{1}} C_{\Lambda_{1}} \text{T}_{\Lambda_{1}\tilde{\mathrm{Q}}_{k}}) \Phi\rangle 
+ \frac{1}{2} \langle\Phi, \text{T}_{\tilde{\mathrm{Q}}_{k}} \Phi\rangle + V_{B}(\tilde{\mathrm{Q}}_{k}, \Phi) \nonumber
\end{aligned}
\end{equation}
and then use
\begin{equation}
\begin{aligned}
\frac{1}{2} \langle \Phi, (\mathrm{T}_{\tilde{\mathrm{Q}}_{k}} - 
2 \hspace{0.05 cm} \text{T}_{\tilde{\mathrm{Q}}_{k}\Lambda_{1}} C_{\Lambda_{1}} \text{T}_{\Lambda_{1}\tilde{\mathrm{Q}}_{k}}) \Phi\rangle 
&\geqslant \frac{1}{2} (m_{\text{min}} -  \frac{2\hspace{0.05 cm} (2^{5})}{m_{\text{min}}}) ||\Phi||^{2}_{\tilde{\mathrm{Q}}_{k}}.
\end{aligned}
\end{equation}
The constant on the right hand side in above equations is positive if $m_{\text{min}} > 8$.
Thus, from (4.20) and (4.21) it follows that
\begin{equation}
\begin{aligned}
\frac{1}{2} \langle \Phi, (\mathrm{T}_{\tilde{\mathrm{Q}}_{k}} - 
2 \hspace{0.05 cm} \text{T}_{\tilde{\mathrm{Q}}_{k}\Lambda_{1}} C_{\Lambda_{1}} \text{T}_{\Lambda_{1}\tilde{\mathrm{Q}}_{k}}) \Phi\rangle 
+ V_{Y}(\tilde{\mathrm{Q}}_{k}) &\geqslant 
\frac{1}{2} (m_{\text{min}} -  \frac{2\hspace{0.05 cm} (2^{5})}{m_{\text{min}}}) ||\Phi||^{2}_{\tilde{\mathrm{Q}}_{k}} - 
g \hspace{0.05 cm} 16  \hspace{0.05 cm}  ||\Phi||^{2}_{\tilde{\mathrm{Q}}_{k}} \\
&\geqslant \gamma_{4} \hspace{0.05 cm}  ||\Phi||^{2}_{\tilde{\mathrm{Q}}_{k}}.
\end{aligned}
\end{equation}
And from the stability Lemma 11.1 in \cite{BBIJ} we have for some constant $\gamma_{3} > 0$ 
\begin{equation}
\frac{1}{2} \langle\Phi, \text{T}_{\tilde{\mathrm{Q}}_{k}} \Phi\rangle + V_{B}(\tilde{\mathrm{Q}}_{k},\Phi) 
 \geqslant \gamma_{3} \hspace{0.05 cm}  \langle\Phi, \text{T}_{\tilde{\mathrm{Q}}_{k}} \Phi\rangle. 
\end{equation}
Adding (4.22) and (4.23) gives the result.

Recall that  
\begin{equation}
\begin{aligned}
\rho(\tilde{\mathrm{Q}}_{k}, \eta) &= e^{W_{2}(\tilde{\mathrm{Q}}_{k}) - W^{\prime}_{2}(\tilde{\mathrm{Q}}_{k})} \hspace{0.1 cm}
e^{-\frac{1}{2} \langle\Phi, (\text{T}_{\tilde{\mathrm{Q}}_{k}} - \text{T}_{\tilde{\mathrm{Q}}_{k}\Lambda_{1}}
C_{\Lambda_{1}} \text{T}_{\Lambda_{1}\tilde{\mathrm{Q}}_{k}})\Phi\rangle 
- \langle \bar{\psi}, (\text{D}_{\tilde{\mathrm{Q}}_{k}} - \text{D}_{\tilde{\mathrm{Q}}_{k}\Lambda_{1}}
S_{\Lambda_{1}} \text{D}_{\Lambda_{1}\tilde{\mathrm{Q}}_{k}})  \psi \rangle}
\\ &\hspace{0.4 cm}  e^{- V(\tilde{\mathrm{Q}}_{k}, \psi, \bar{\psi}, \Phi) + V_{s}(\tilde{\mathrm{Q}}_{k}) -
V_{\epsilon}^{\prime}(\tilde{\mathrm{Q}}_{k})} \zeta_{\mathrm{Q}_{k}}(\Phi)  \nonumber
\hspace{0.05 cm} \chi_{\tilde{\mathrm{Q}}_{k} \backslash \mathrm{Q}_{k}}(\Phi)  \hspace{0.1 cm} \mathrm{Z}_{0}(\mathrm{Q}_{k})^{-1}. 
\end{aligned}
\end{equation}
For the source term note that
\begin{equation}
 ||V_{s}||_{w_{1,1}} = ||e_{0} \langle \bar{\psi},\eta\rangle + e_{0} \langle\bar{\eta}, \psi \rangle||_{w_{1,1}} 
 \leqslant  e_{0} \hspace{0.05 cm} 32
\end{equation} 
where summing over spinor indices gives us a factor of 16 and two terms give another factor of 2.
Thus, $||e^{V_{s}}||_{w_{1,1}} \leqslant e^{||V_{s}||_{w_{1,1}}} \leqslant \mathcal{O}(1)$.

Next consider the term 
\begin{equation}
\begin{aligned}
V^{\prime}_{\epsilon} &= \sum_{x \in \tilde{\mathrm{Q}}_{k}}
\bar{\psi}_{L}(x)[\mathfrak{D}_{A} - \mathfrak{D}] \psi_{L}(x) 
= \sum_{x \in \tilde{\mathrm{Q}}_{k}} \bar{\psi}(x) [\gamma \cdot \nabla_{A} - \frac{1}{2} \Delta_{A} - 
(\gamma \cdot \nabla - \frac{1}{2} \Delta)] \psi(x) \\
&= - \sum_{x \in \tilde{\mathrm{Q}}_{k}} \sum_{\mu} \Big[ \bar{\psi}(x) \Big(\frac{1-\gamma_{\mu}}{2}\Big)
(e^{i e_{0} A_{\mu}(x)} - 1)\psi(x+e_{\mu}) + \\
& \hspace{1 cm}  \bar{\psi}(x) \Big(\frac{1+ \gamma_{\mu}}{2}\Big) (e^{-i e_{0} A_{\mu}(x)} - 1)\psi(x-e_{\mu})\Big]
\end{aligned}
\end{equation}
Then
\begin{equation}
||V^{\prime}_{\epsilon}||_{w_{1}} \leqslant 16 \sum_{\mu} \Big[|e^{i e_{0} A_{\mu}} - 1| +| e^{-i e_{0} A_{\mu}} - 1| \Big]
\leqslant 256
\end{equation}
 where a factor of $16$ comes from the implicit sum over spinor indices and sum over $\mu$ gives another
 factor of 4.

\textbf{Proposition 4.3} For polymers $\tilde{\mathrm{Q}} = \{\tilde{\mathrm{Q}}_{k}\}$  
let the function $\rho(\tilde{\mathrm{Q}}_{k}, \eta)$ be as defined. From Lemma 4.2
$||e^{- \langle \bar{\psi}, (\text{D}_{\tilde{\mathrm{Q}}_{k}} - \text{D}_{\tilde{\mathrm{Q}}_{k}\Lambda_{1}}
S_{\Lambda_{1}} \text{D}_{\Lambda_{1}\tilde{\mathrm{Q}}_{k}})  \psi \rangle}||_{w_{1}} \leqslant
e^{||\langle \bar{\psi}, (\text{D}_{\tilde{\mathrm{Q}}_{k}} - \text{D}_{\tilde{\mathrm{Q}}_{k}\Lambda_{1}}
S_{\Lambda_{1}} \text{D}_{\Lambda_{1}\tilde{\mathrm{Q}}_{k}})  \psi \rangle||_{w_{1}}} \leqslant \mathcal{O}(1)$. Then
(4.24), (4.26) and $e^{-\frac{p_{\lambda}}{2}} = \mathcal{O}(e_{0}^{2})$ implies
\begin{equation}
\begin{aligned}
||\sum_{v \hspace{0.05 cm} \text{on} \hspace{0.05 cm} \tilde{\mathrm{Q}}_{k} : dv=0} \rho (\tilde{\mathrm{Q}}_{k}, \eta)||_{w_{1,1}}
\leqslant  \mathcal{O}(1) \hspace{0.05 cm} e_{0}^{2} \hspace{0.05 cm} 
e^{-\frac{3}{8} p_{\lambda} \tau(\mathrm{Q}_{k})} \hspace{0.1 cm}  
e^{- (m_{\text{min}} \gamma_{3} - p_{\lambda}^{-1} + c) ||\Phi||_{\tilde{\mathrm{Q}}_{k}}^{2}}.
\end{aligned}
\end{equation}

\textit{Proof} follows directly from \cite{G1}.  

\textbf{Lemma 4.4} In region $\{\tilde{\mathrm{P}}_{l}\}$, the following relation holds
\begin{equation}
||e^{- V(\tilde{\mathrm{P}}_{l}, S^{\text{loc}}_{\Lambda_{1}}\psi - \Psi,
  \bar{\psi} -  \bar{\Psi}, C^{\frac{1}{2},\text{loc}}_{\Lambda_{1}} \Phi^{\prime} - \varphi)}||_{w_{1,1}}
 \leqslant e^{\mathcal{O}(1) \hspace{0.05 cm} e_{0}^{1- 4\epsilon} |\tilde{\mathrm{P}}_{l}|}. 
\end{equation}

\textit{Proof} From Lemma 2.2,
$|C^{\frac{1}{2},\text{loc}}_{\Lambda_{1}} \Phi^{\prime}| < \mathcal{O}(1) \hspace{0.05 cm} ||\Phi^{\prime}||_{\tilde{\mathrm{P}}_{l}, \infty}
< \mathcal{O}(1) \hspace{0.05 cm} p_{\lambda}$. Also, near the boundary $\partial \Lambda_{1}$, 
$|\Phi_{\tilde{\mathrm{Q}}}| \leqslant p_{\lambda}$, therefore,
$|\varphi| \leqslant \mathcal{O}(1) \hspace{0.05 cm} e^{-r_{\lambda}} p_{\lambda} < \frac{p_{\lambda}}{2}$.
Next note that every term in $V$ carries a factor of at least $g$ or $e_{0}$. We have assumed that for some $0 < \epsilon \ll 1$,
$p_{\lambda} < e_{0}^{-\epsilon}$ and using (4.24), (4.26),
$||V(\tilde{\mathrm{P}}_{l}, S^{\text{loc}}_{\Lambda_{1}}\psi - \Psi,
  \bar{\psi} - \bar{\Psi}, C^{\frac{1}{2},\text{loc}}_{\Lambda_{1}} \Phi^{\prime} - \varphi)||_{w_{1,1}}
  \leqslant \mathcal{O}(1) \hspace{0.05 cm} e_{0}^{1- 4\epsilon} |\tilde{\mathrm{P}}_{l}|$.
The result follows using $||e^{-V}||_{w_{1,1}} \leqslant e^{||V||_{w_{1,1}}}$.

Recall that
\begin{equation}
\begin{aligned}
\rho^{\prime}(\tilde{\mathrm{P}}_{l}, \eta) &= e^{W_{2}(\tilde{\mathrm{P}}_{l}) - W^{\prime}_{2}(\tilde{\mathrm{P}}_{l})}  
e^{-\frac{1}{2}||\Phi^{\prime}||_{\tilde{\mathrm{P}}_{l}}^{2} - \langle\bar{\psi}_{\beta}, \psi_{\alpha} \rangle_{\tilde{\mathrm{P}}_{l}}}
\hspace{0.2 cm} e^{- V(\tilde{\mathrm{P}}_{l}, S^{\text{loc}}_{\Lambda_{1}}\psi - \Psi,
  \bar{\psi} -  \bar{\Psi}, C^{\frac{1}{2},\text{loc}}_{\Lambda_{1}} \Phi^{\prime} - \varphi)}
 \\ &
\hspace{0.4 cm}  e^{V_{s}(\tilde{\mathrm{P}}_{l}) - V_{\epsilon}^{\prime}(\tilde{\mathrm{P}}_{l})} \hspace{0.2 cm}
 \hat{\zeta}_{\mathrm{P}_{l}}(\Phi^{\prime})
   \chi_{\tilde{\mathrm{P}}_{l}} (C^{\frac{1}{2},\text{loc}}_{\Lambda_{1}} \Phi^{\prime} - \varphi) \hspace{0.1 cm} 
\mathrm{Z}_{0}(\tilde{\mathrm{P}}_{l})^{-1}
\end{aligned}
\end{equation}
 
\textbf{Proposition 4.5} For polymers $\tilde{\mathrm{P}} = \{\tilde{\mathrm{P}}_{l}\}$ let 
 $\rho^{\prime}(\tilde{\mathrm{P}}_{l}, \text{J})$ be the function as defined. Then from Lemma 4.4,
 (4.24), (4.26) and using $e^{-\frac{p_{\lambda}}{2}} = \mathcal{O}(e_{0}^{2})$ we have
\begin{equation}
||\rho^{\prime}(\tilde{\mathrm{P}}_{l}, \eta)||_{w_{1,1}} \leqslant \mathcal{O}(1) \hspace{0.05 cm} e_{0}^{2} \hspace{0.05 cm} 
e^{-\frac{3}{8} p_{0,\lambda} \tau(\mathrm{P}_{l})} e^{ - \langle\bar{\psi}, \psi \rangle_{\tilde{\mathrm{P}}_{l}}
- (\frac{1}{2}- p_{0, \lambda}^{-1}) ||\Phi^{\prime}||_{\tilde{\mathrm{P}}_{l}}^{2}}.
\end{equation}

\textit{Proof} follows directly from \cite{G1}.   

\section{Convergence}
Recall that $\mathbb{T}$ is a finite lattice. $\mathrm{Q} \subset \mathbb{T}$ and $\Lambda_{0} = \mathbb{T} - \mathrm{Q}$.
$\tilde{\mathrm{Q}}$ is an enlargement of $\mathrm{Q}$ by $[r_{\lambda}]^{4}$ and $\Lambda_{1} = \mathbb{T} - \tilde{\mathrm{Q}}$. 
$\mathrm{P} \subset \Lambda_{1}$ and $\Omega = \Lambda_{1} - \mathrm{P}$. $\Omega_{1}$ is contraction of
$\Omega$ by $16 [r_{\lambda}]^{4}$ and $\tilde{\mathrm{P}} = \Lambda_{0} - \Omega_{1}$ that is $\tilde{\mathrm{P}}$
is an enlargement of $\mathrm{P}$ by $16 [r_{\lambda}]^{4}$.

$\{Z_{j}\} \subseteq \Omega_{1}$,  $\{\tilde{\mathrm{P}}_{l}\} \subseteq \Lambda_{0} - \Omega_{1}$ and
$\{\mathrm{X}_{i}\} \cup \{\tilde{\mathrm{Q}}_{k}\} \subseteq \Lambda_{1}^{c}$ are disjoint polymers.
These collection of polymers overlap near the boundaries of various regions of $\mathbb{T}$. 
We combine these overlapping parts into connected components in two steps. 
\begin{enumerate}
  \item In the region $\Lambda_{1}$,
define connected components $\{\text{Y}_{m}\}$, such that any $\text{Y} \in \{\text{Y}_{m}\}$, 
\begin{equation}
\text{Y} = \cup_{j}\{Z_{j}\} \cup_{l} \{\tilde{\mathrm{P}}_{l}\}
\end{equation}
where $Z_{j}$ and $\tilde{\mathrm{P}}_{l}$ overlap and rewrite
\begin{equation}
\begin{aligned}
\sum_{\{Z_{j}\}} \sum_{\{\tilde{\mathrm{P}}_{l}\}} &
\hspace{0.05 cm} \prod_{l} \int \mathcal{D}\Phi^{\prime}_{\tilde{\mathrm{P}}_{l}} \hspace{0.05 cm} 
 \mathcal{D}\bar{\psi}_{\tilde{\mathrm{P}}_{l}}  \mathcal{D}\psi_{\tilde{\mathrm{P}}_{l}} 
   \rho^{\prime}(\tilde{\mathrm{P}}_{l}, \eta)\hspace{0.1 cm} 
 \prod_{j} K(Z_{j}, \eta)\\ &= \sum_{\{\text{Y}_{m}\}} \sum_{\cup_{j}\{Z_{j}\} \cup_{l} \{\tilde{\mathrm{P}}_{l}\} \rightarrow \text{Y}_{m}}  
 \hspace{0.05 cm} \prod_{l} \int \mathcal{D}\Phi^{\prime}_{\tilde{\mathrm{P}}_{l}} \hspace{0.05 cm} 
  \mathcal{D}\bar{\psi}_{\tilde{\mathrm{P}}_{l}}  \mathcal{D}\psi_{\tilde{\mathrm{P}}_{l}} 
   \rho^{\prime}(\tilde{\mathrm{P}}_{l}, \eta) \hspace{0.1 cm} 
 \prod_{j} K(Z_{j}, \eta) \\ &= \sum_{\{\text{Y}_{m}\}} \prod_{m} K^{\prime} (\text{Y}_{m}, \eta)
\end{aligned}
\end{equation}
where for a connected $\text{Y}$,
\begin{equation}
K^{\prime} (\text{Y}, \eta) = \sum_{\cup_{j}\{Z_{j}\} \cup_{l} \{\tilde{\mathrm{P}}_{l}\} \rightarrow \text{Y}}  
 \hspace{0.05 cm} \prod_{l} \int \mathcal{D}\Phi^{\prime}_{\tilde{\mathrm{P}}_{l}} \hspace{0.05 cm} 
 \mathcal{D}\bar{\psi}_{\tilde{\mathrm{P}}_{l}}  \mathcal{D}\psi_{\tilde{\mathrm{P}}_{l}} 
   \rho^{\prime}(\tilde{\mathrm{P}}_{l}, \eta) \hspace{0.1 cm} 
 \prod_{j} K(\text{Z}_{j}, \eta).
\end{equation}
Rewrite the generating functional
\begin{equation}
\begin{aligned}
\mathrm{Z}[\eta] &= \Big(\frac{2\pi}{e_{0}}\Big)^{-|\mathbb{T}| + 1}
\Big( \frac{\text{det}\hspace{0.1 cm}C^{\frac{1}{2}}}{\text{det}\hspace{0.1 cm}S}\Big)
\hspace{0.1 cm} \mathrm{Z}_{0}(\mathbb{T}) \hspace{0.1 cm}
 \sum_{v: dv=0} \sum_{\{\tilde{\mathrm{Q}}_{k}\}} \sum_{\{\text{X}_{i}\}}
 \sum_{\{\text{Y}_{m}\}} \\ &  \prod_{k} \int  \mathcal{D}\Phi_{\tilde{\mathrm{Q}}_{k}}\hspace{0.1 cm}
  \mathcal{D}\bar{\psi}_{\tilde{\mathrm{Q}}_{k}} \mathcal{D}\psi_{\tilde{\mathrm{Q}}_{k}} 
\rho(\tilde{\mathrm{Q}}_{k}, \eta) \hspace{0.1 cm} \prod_{i}  f(\text{X}_{i}, \eta) \hspace{0.1 cm}  
 \prod_{m} K^{\prime}(\text{Y}_{m}, \eta).
\end{aligned}
\end{equation}

 \item In the entire lattice $\mathbb{T}$, define connected components $\{\mathcal{C}_{l}\}$, 
 such that any $\mathcal{C} \in \{\mathcal{C}_{l}\}$,
\begin{equation}
\mathcal{C} =  \cup_{i}\{\text{X}_{i}\} \cup_{k}\{\tilde{\mathrm{Q}}_{k}\}\cup_{m}\{\text{Y}_{m}\}
\end{equation}
where $X_{i}$, $\tilde{\mathrm{Q}}_{k}$ and $\text{Y}_{m}$ overlap and rewrite
\begin{equation}
\begin{aligned}
& \sum_{\{\text{X}_{i}\}} \sum_{\{\tilde{\mathrm{Q}}_{k}\}} \sum_{\{\text{Y}_{m}\}} 
 \prod_{k}  \int \mathcal{D}\Phi_{\tilde{\mathrm{Q}}_{k}}  \mathcal{D}\bar{\psi}_{\tilde{\mathrm{Q}}_{k}} 
 \mathcal{D}\psi_{\tilde{\mathrm{Q}}_{k}} \\ & \hspace{1 cm}
  \sum_{v \hspace{0.05 cm} \text{on} \hspace{0.05 cm} \tilde{\mathrm{Q}}_{k} : dv=0}
 \rho(\tilde{\mathrm{Q}}_{k}, \eta)  \hspace{0.05 cm} \prod_{i} f(\text{X}_{i}, \eta) 
 \hspace{0.05 cm} \prod_{m} K^{\prime}(\text{Y}_{m}, \eta) \\  
&=  \sum_{\{\mathcal{C}_{l}\}} \sum_{\cup_{i}\{\text{X}_{i}\} \cup_{k}\{\tilde{\mathrm{Q}}_{k}\}
 \cup_{m}\{\text{Y}_{m}\} \rightarrow \{\mathcal{C}_{l}\}} 
 \prod_{k} \int \mathcal{D}\Phi_{\tilde{\mathrm{Q}}_{k}} \mathcal{D}\bar{\psi}_{\tilde{\mathrm{Q}}_{k}} \mathcal{D}\psi_{\tilde{\mathrm{Q}}_{k}} 
  \\ & \hspace{1 cm} \sum_{v \hspace{0.05 cm} \text{on} \hspace{0.05 cm} \tilde{\mathrm{Q}}_{k} : dv=0}
 \rho(\tilde{\mathrm{Q}}_{k}, \eta) \prod_{i} f(\text{X}_{i}, \eta) 
 \hspace{0.05 cm} \prod_{m} K^{\prime}(\text{Y}_{m}, \eta) \\
&= \sum_{\{\mathcal{C}_{l}\}} \prod_{l} K^{\#} (\mathcal{C}_{l}, \eta) 
\end{aligned}
\end{equation}
where for any connected $\mathcal{C}$, 
\begin{equation}
\begin{aligned}
 K^{\#} (\mathcal{C}, \eta) &= \sum_{\cup_{i}\{\text{X}_{i}\} \cup_{k}\{\tilde{\mathrm{Q}}_{k}\} \cup_{m}\{\text{Y}_{m}\}  \rightarrow \mathcal{C}} 
 \prod_{k} \int \mathcal{D}\Phi_{\mathrm{Q}_{k}}   \mathcal{D}\bar{\psi}_{\tilde{\mathrm{Q}}_{k}} \mathcal{D}\psi_{\tilde{\mathrm{Q}}_{k}} 
 \\ & \hspace{0.5 cm}  \sum_{v \hspace{0.05 cm} \text{on} \hspace{0.05 cm} \tilde{\mathrm{Q}}_{k} : dv=0}
 \rho(\tilde{\mathrm{Q}}_{k}, \eta) \prod_{i} f(\text{X}_{i}, \eta) 
 \hspace{0.05 cm} \prod_{m} K^{\prime}(\text{Y}_{m}, \eta). 
\end{aligned}
\end{equation}
Rewrite the generating functional
\begin{equation}
\mathrm{Z}[\eta] = \Big(\frac{2\pi}{e_{0}}\Big)^{-|\mathbb{T}| + 1}
\Big( \frac{\text{det}\hspace{0.1 cm}C^{\frac{1}{2}}}{\text{det}\hspace{0.1 cm}S}\Big)
\hspace{0.1 cm} \mathrm{Z}_{0}(\mathbb{T}) \hspace{0.1 cm} \sum_{\{\mathcal{C}_{l}\}} \prod_{l} K^{\#} (\mathcal{C}_{l}, \eta).
\end{equation}

\end{enumerate}

Let $w_{p_{1}, p_{2}, h_{1}, h_{2}}(\vec{u}, \vec{s}, \vec{\xi}, \vec{z}) =  
e^{\kappa t(\text{supp}(\vec{u}, \vec{\xi}, \vec{z}))} p_{1}^{n(\vec{u})} h_{1}^{n(\vec{s})} p_{2}^{n(\vec{\xi})} h_{2}^{n(\vec{z})}$ 
be the weight system with mass $\kappa$ giving weight at least $p_{1}$ to $\Phi, h_{1}$ to $\Theta, p_{2}$ to $\psi$ and 
$h_{2}$ to $\eta$ where $n(\vec{u}), n(\vec{s}), n(\vec{\xi}), n(\vec{z})$ denote the $\#$ sites in 
$\vec{u}, \vec{s}, \vec{\xi}, \vec{z}$ respectively.. 

Set $p_{1} = p_{0, \lambda},  h_{1} = 1, p_{2} = 1, h_{2} = 1$ and $n(\vec{s}) = 0$ since there are no 
external bosonic sources. Denote $w = w_{p_{0,\lambda},1,1,1}$.

\textbf{Lemma 5.1} There exists constant $\kappa_{5} > 0$ and sufficiently large such that 
\begin{equation}   
||K^{\prime}(\text{Y}, \eta)||_{w} \leqslant c \hspace{0.05 cm} e_{0}^{\frac{1}{2} - \epsilon}  
\hspace{0.05 cm} e^{-\kappa_{5} \tau(\text{Y})}. \nonumber
\end{equation}

\textit{Proof}  From the definition (5.3) 
\begin{equation}
\begin{aligned}
||K^{\prime} (\text{Y}, \eta)||_{w} &\leqslant  \sum_{\cup_{j}\{Z_{j}\} \cup_{l} \{\tilde{\mathrm{P}}_{l}\} = \text{Y}}  
 \hspace{0.05 cm} \prod_{l}|| \int \mathcal{D}\Phi^{\prime}_{\tilde{\mathrm{P}}_{l}} \hspace{0.05 cm} 
 \mathcal{D}\bar{\psi}_{\tilde{\mathrm{P}}_{l}}  \mathcal{D}\psi_{\tilde{\mathrm{P}}_{l}} 
   \rho^{\prime}(\tilde{\mathrm{P}}_{l}, \eta)||_{w} \hspace{0.1 cm} 
 \prod_{j} ||K(Z_{j}, \eta)||_{w} \\
 &\leqslant  \sum_{\cup_{j}\{Z_{j}\} \cup_{l} \{\tilde{\mathrm{P}}_{l}\} = \text{Y}}  
 \prod_{j} c \hspace{0.05 cm} e_{0}^{\frac{1}{2} - \epsilon}  e^{-\kappa_{4}\tau(Z_{j})} \prod_{l}
 c \hspace{0.05 cm} e_{0}^{2} \hspace{0.1 cm} e^{-\frac{3}{8} p_{0, \lambda} \hspace{0.05 cm} \tau(\mathrm{P}_{l})} \\
 & \hspace{1 cm} \int \mathcal{D}\Phi^{\prime}_{\tilde{\mathrm{P}}_{l}} \hspace{0.05 cm} 
 \mathcal{D}\bar{\psi}_{\tilde{\mathrm{P}}_{l}}  \mathcal{D}\psi_{\tilde{\mathrm{P}}_{l}} 
 e^{- \langle\bar{\psi}, \psi \rangle_{\tilde{\mathrm{P}}_{l}}
- (\frac{1}{2} - p_{0, \lambda}^{-1}) ||\Phi^{\prime}||_{\tilde{\mathrm{P}}_{l}}^{2}}.
\end{aligned}
\end{equation}
Since $\{Z_{j}\}$ and $\{\tilde{\mathrm{P}}_{l}\}$ overlap, $\tau(Z_{j}) + \tau(\tilde{\mathrm{P}}_{l}) \geqslant \tau(\text{Y})$,
 we can extract a factor of $e^{-\text{min}(\kappa_{4}, \frac{3}{8} p_{\lambda})\frac{\tau(\text{Y})}{2}} = e^{- \frac{\kappa_{4}}{2} \tau(\text{Y})}$. 
 We extract a factor of $e_{0}^{\frac{1}{2} - \epsilon}$ from the first term since the small field is 
 the most prominent region and also do the integration of the last term,
\begin{equation}
\begin{aligned}
||K^{\prime} (\text{Y}, \eta)||_{w} &\leqslant e_{0}^{\frac{1}{2} - \epsilon}  \hspace{0.05 cm} 
 e^{- \frac{\kappa_{4}}{2} \tau(\text{Y})} \sum_{\cup_{j}\{Z_{j}\} \cup_{l} \{\tilde{\mathrm{P}}_{l}\} \subset \text{Y}}  
\prod_{j} c \hspace{0.05 cm} e_{0}^{\frac{1}{4} - \frac{\epsilon}{2}} e^{-\frac{\kappa_{4}}{2}\tau(Z_{j})} \prod_{l}
 c \hspace{0.05 cm} e_{0}^{2} \hspace{0.1 cm} e^{-\frac{3}{16} p_{0, \lambda} \hspace{0.05 cm} \tau(\mathrm{P}_{l})} \\
 & \hspace{1 cm} \int \mathcal{D}\Phi^{\prime}_{\tilde{\mathrm{P}}_{l}} \hspace{0.05 cm} 
 \mathcal{D}\bar{\psi}_{\tilde{\mathrm{P}}_{l}}  \mathcal{D}\psi_{\tilde{\mathrm{P}}_{l}} 
 e^{- \langle\bar{\psi}, \psi \rangle_{\tilde{\mathrm{P}}_{l}}
- (\frac{1}{2} - p_{0, \lambda}^{-1}) ||\Phi^{\prime}||_{\tilde{\mathrm{P}}_{l}}^{2}} \\  
 &\leqslant  e_{0}^{\frac{1}{2} - \epsilon} \hspace{0.05 cm} 
 e^{- \frac{\kappa_{4}}{2} \tau(\text{Y})} \hspace{0.05 cm}  \Big(\frac{\pi}{1 - p_{0,\lambda}^{-1}}\Big)^{\frac{|\text{Y}|}{2}} 
   \Big[\sum_{\cup_{j}\{Z_{j}\} \subset \text{Y}} 
\prod_{j} c \hspace{0.05 cm} e_{0}^{\frac{1}{4} - \frac{\epsilon}{2}} e^{-\frac{\kappa_{4}}{2}\tau(Z_{j})} \Big]
\\ & \hspace{1 cm} \Big[\sum_{\cup_{l} \{\mathrm{P}_{l}\} \subset \text{Y}} \prod_{l} c \hspace{0.05 cm} e_{0}^{2}
\hspace{0.05 cm} e^{-\frac{3}{16} p_{0, \lambda} \hspace{0.05 cm} \tau(\mathrm{P}_{l})}\Big]. 
\end{aligned}
\end{equation}
Rewrite $\{Z_{j}\}$ and $\{\mathrm{P}_{l}\}$ as ordered collection $(Z_{1}, \cdots, Z_{n})$ and 
$(\mathrm{P}_{1}, \cdots, \mathrm{P}_{m})$ %and using $|\text{Y}| \leqslant 81 (\tau(\text{Y}) + 1)$
\begin{equation}
\begin{aligned}
||K^{\prime} (\text{Y}, \eta)||_{w} &\leqslant c \hspace{0.05 cm} e_{0}^{\frac{1}{2} - \epsilon} \hspace{0.05 cm} 
 e^{- \frac{\kappa_{4}}{2} \tau(\text{Y})}
  \Big[\sum_{n=1}^{\infty} \frac{1}{n!} \sum_{(Z_{1}, \cdots, Z_{n}) \subset \text{Y}}  
\prod_{j=1}^{n} c \hspace{0.05 cm} e_{0}^{\frac{1}{4} - \frac{\epsilon}{2}} e^{-\frac{\kappa_{4}}{2}\tau(Z_{j})} \Big] \\
& \hspace{0.5 cm}\Big[\sum_{m=1}^{\infty} \frac{1}{m!} \sum_{(\mathrm{P}_{1}, \cdots, \mathrm{P}_{m}) \subset Z} 
\prod_{l=1}^{m} c \hspace{0.05 cm} e_{0}^{2} \hspace{0.05 cm} e^{-\frac{3}{16} p_{0, \lambda} \hspace{0.05 cm} \tau(\mathrm{P}_{l})}\Big] \\
&\leqslant c \hspace{0.05 cm} e_{0}^{\frac{1}{2} - \epsilon} \hspace{0.05 cm} 
 e^{- \frac{\kappa_{4}}{2}  \tau(\text{Y})}
  \Big[\sum_{n=1}^{\infty} \frac{1}{n!} \Big( c \hspace{0.05 cm} e_{0}^{\frac{1}{4} - \frac{\epsilon}{2}}
\hspace{0.05 cm} \sum_{Z \subset \text{Y}}  e^{-\frac{\kappa_{4}}{2}\tau(Z)}\Big)^{n} \Big] \\
& \hspace{0.5 cm}\Big[\sum_{m=1}^{\infty} \frac{1}{m!} \Big( c \hspace{0.05 cm} e_{0}^{2} \hspace{0.05 cm} 
\sum_{\mathrm{P} \subset \text{Y}} e^{-\frac{3}{16} p_{0, \lambda} \hspace{0.05 cm} \tau(\mathrm{P})} \Big)^{m}\Big].
\end{aligned}
\end{equation}
Following lemma 25, \cite{D1} (there exists a constant $b$, such that  
$\sum_{X \cap Y \neq \oslash} e^{-a \tau(X)} 
\leqslant b$), the first bracketed term is bounded by $e^{c e_{0}^{\frac{1}{4} - \frac{\epsilon}{2}} b}$ and the 
second bracketed term is bounded by $e^{c \hspace{0.05 cm} e_{0}^{2} b}$. 
Set $\kappa_{5} = \frac{\kappa_{4}}{2} $,
\begin{equation}
||K^{\prime}(\text{Y}, \eta)||_{w} \leqslant c \hspace{0.05 cm} e_{0}^{\frac{1}{2} - \epsilon}  
\hspace{0.05 cm} e^{-\kappa_{5} \tau(\text{Y})}.
\end{equation}

\textbf{Lemma 5.2} There exist a constant $\kappa_{7} > 0$ and sufficiently large such that
\begin{equation}
||K^{\#} (\mathcal{C}, \eta)||_{w} \leqslant c \hspace{0.05 cm} e_{0}^{\frac{1}{2} - \epsilon}
\hspace{0.05 cm} e^{-\kappa_{7}\tau(\mathcal{C})}. \nonumber
\end{equation}
\textit{Proof} From the definition (5.7)  
\begin{equation}
\begin{aligned}
||K^{\#} (\mathcal{C}, \eta)||_{w} &\leqslant 
\sum_{\cup_{i}\{\text{X}_{i}\} \cup_{k}\{\tilde{\mathrm{Q}}_{k}\} \cup_{m}\{\text{Y}_{m}\} = \mathcal{C}} 
\prod_{i} ||f(\text{X}_{i}, \eta)||_{w} \hspace{0.1 cm} 
 \prod_{m} ||K^{\prime}(\text{Y}_{m}, \eta)||_{w}
\\ & \hspace{1 cm} \prod_{k} ||\int \mathcal{D}\Phi_{\mathrm{Q}_{k}} \mathcal{D}\bar{\psi}_{\tilde{\mathrm{Q}}_{k}} 
\mathcal{D}\psi_{\tilde{\mathrm{Q}}_{k}} 
\sum_{v \hspace{0.05 cm} \text{on} \hspace{0.05 cm} \tilde{\mathrm{Q}}_{k} : dv=0}
 \rho(\tilde{\mathrm{Q}}_{k}, \eta)||_{w} \\
&\leqslant  \sum_{\cup_{i}\{\text{X}_{i}\} \cup_{m}\{\text{Y}_{m}\} \cup_{k}\{\tilde{\mathrm{Q}}_{k}\} = \mathcal{C}} 
\hspace{0.1 cm} \prod_{i} c \hspace{0.05 cm} e_{0}^{2-2\epsilon} \hspace{0.05 cm} e^{- \kappa^{\prime} \tau(\text{X}_{i})} \hspace{0.1 cm}
 \prod_{m} c \hspace{0.05 cm} e_{0}^{\frac{1}{2} - \epsilon} \hspace{0.05 cm} e^{-\kappa_{5} \tau(\text{Y}_{m})}
\\ & \hspace{1 cm} \prod_{k} c \hspace{0.05 cm} e_{0}^{2} \hspace{0.05 cm} e^{-\frac{3}{8} p_{\lambda} \tau(\mathrm{Q}_{k})} \hspace{0.05 cm}    
\int  \mathcal{D}\Phi_{\tilde{\mathrm{Q}}_{k}} \mathcal{D}\bar{\psi}_{\tilde{\mathrm{Q}}_{k}} 
\mathcal{D}\psi_{\tilde{\mathrm{Q}}_{k}} \hspace{0.05 cm} e^{- (m_{\text{min}} \gamma_{3} - p_{\lambda}^{-1} + c)
 ||\Phi^{\prime}||_{\tilde{\mathrm{Q}}_{k}}^{2}} 
\end{aligned}
\end{equation} 
Let $\kappa_{6} = \text{min}(\kappa^{\prime}, \kappa_{5}, \frac{3}{8} p_{\lambda})$.
Since $\{\text{X}_{i}\}, \{\tilde{\mathrm{Q}}_{k}\}, \{\text{Y}_{m}\}$ overlap, 
$\tau(\text{X}_{i}) + \tau(\tilde{\mathrm{Q}}_{k}) + \tau(\text{Y}_{m}) \geqslant \tau(\mathcal{C})$,
we can extract a factor of $e^{-\kappa_{6} \frac{\tau(\mathcal{C})}{2}}$.
As large field regions are rare we only extract a factor of $e_{0}^{\frac{1}{2} - \epsilon}$
from the dominant small field part and do the integration of the last term.
\begin{equation}
\begin{aligned}
 ||K^{\#} (\mathcal{C}, \eta)||_{w} 
 &\leqslant e_{0}^{\frac{1}{2} - \epsilon} \hspace{0.05 cm} e^{- \frac{\kappa_{6}}{2} \tau(\mathcal{C})}
\sum_{\cup_{i}\{\text{X}_{i}\} \cup_{m}\{\text{Y}_{m}\} \cup_{k}\{\tilde{\mathrm{Q}}_{k}\} \subset \mathcal{C}} 
\hspace{0.1 cm} \prod_{i} c \hspace{0.05 cm} e_{0}^{2-2\epsilon} \hspace{0.05 cm} 
e^{- \frac{\kappa^{\prime}}{2} \tau(\text{X}_{i})} \hspace{0.1 cm}
\\ & \hspace{0.1 cm} \prod_{m} c \hspace{0.05 cm} 
e_{0}^{\frac{1}{4} - \frac{\epsilon}{2}} \hspace{0.05 cm} e^{- \frac{\kappa_{5}}{2} \tau(\text{Y}_{m})}  
  \prod_{k} c \hspace{0.05 cm} e_{0}^{2} \hspace{0.05 cm} e^{-\frac{3}{16} p_{\lambda} \tau(\mathrm{Q}_{k})} \hspace{0.05 cm}  
 \\ &  \hspace{0.1 cm} \int  \mathcal{D}\Phi_{\tilde{\mathrm{Q}}_{k}} \mathcal{D}\bar{\psi}_{\tilde{\mathrm{Q}}_{k}} 
\mathcal{D}\psi_{\tilde{\mathrm{Q}}_{k}}  \hspace{0.05 cm}  e^{- \langle\bar{\psi}, \psi \rangle_{\tilde{\mathrm{Q}}_{k}}}
 e^{- (m_{\text{min}} \gamma_{3} - p_{\lambda}^{-1} + c) ||\Phi^{\prime}||_{\tilde{\mathrm{Q}}_{k}}^{2}} \\
&\leqslant  e_{0}^{\frac{1}{2} - \epsilon} \hspace{0.05 cm} e^{- \frac{\kappa_{6}}{2} \tau(\mathcal{C})}
\Big(\frac{\pi}{m_{\text{min}} \gamma_{3} - p_{\lambda}^{-1} + c}\Big)^{\frac{|\mathcal{C}|}{2}} \hspace{0.1 cm}
 \Big[\sum_{\cup_{i}\{\text{X}_{i}\} \subset \mathcal{C}} 
 \prod_{i} c \hspace{0.05 cm}  e_{0}^{2- 2\epsilon} \hspace{0.05 cm} e^{- \frac{\kappa^{\prime}}{2} \tau(\text{X}_{i})}\Big]  
 \\ & \Big[\sum_{\cup_{m}\{\text{Y}_{m}\} \subset \mathcal{C}} \prod_{m}  c \hspace{0.05 cm} e_{0}^{\frac{1}{4} - \frac{\epsilon}{2}}
  \hspace{0.05 cm} e^{- \frac{\kappa_{5}}{2} \tau(\text{Y}_{m})}\Big] 
\hspace{0.1 cm} \Big[\sum_{ \cup_{k}\{\mathrm{Q}_{k}\} \subset \mathcal{C}} \prod_{k} c \hspace{0.05 cm} 
e_{0}^{2} \hspace{0.05 cm} e^{-\frac{3}{16} p_{\lambda} \tau(\mathrm{Q}_{k})}\Big].
\end{aligned}
\end{equation} 
Rewrite $\{\text{X}_{i}\}, \{\text{Y}_{m}\}, \{\mathrm{Q}_{k}\}$ as an ordered collection, $(\text{X}_{1}, \cdots, \text{X}_{n_{1}})$,
$(\text{Y}_{1}, \cdots, \text{Y}_{n_{2}})$ and \\
 $(\mathrm{Q}_{1}, \cdots, \mathrm{Q}_{n_{3}})$ %and using$|\mathcal{C}| \leqslant 81 (\tau(\mathcal{C}) + 1)$

\begin{equation}
\begin{aligned}
||K^{\#} (\mathcal{C}, \eta)||_{w} &\leqslant 
e_{0}^{\frac{1}{2} - \epsilon} \hspace{0.05 cm} e^{- \frac{\kappa_{6}}{2} \tau(\mathcal{C})}
 \\ & \hspace{0.1 cm}
 \Big[\sum_{n_{1}=1}^{\infty} \frac{1}{n_{1}!} \sum_{(\text{X}_{1}, \cdots, \text{X}_{n_{1}}) \subset \mathcal{C}} 
 \prod_{i=1}^{n_{1}} c \hspace{0.05 cm} e_{0}^{2-2\epsilon} \hspace{0.05 cm} e^{- \frac{\kappa^{\prime}}{2} \tau(\text{X}_{i})}\Big] \\ 
 & \Big[\sum_{n_{2}=1}^{\infty} \frac{1}{n_{2}!}  \sum_{(\text{Y}_{1}, \cdots, \text{Y}_{n_{2}}) \subset \mathcal{C}} 
 \prod_{m=1}^{n_{2}} c \hspace{0.05 cm} e_{0}^{\frac{1}{4} - \frac{\epsilon}{2}} \hspace{0.05 cm} 
 e^{- \frac{\kappa_{5}}{2} \tau(\text{Y}_{m})}\Big] \\ &
\hspace{0.1 cm} \Big[\sum_{n_{3}=1}^{\infty} \frac{1}{n_{3}!} \sum_{(\mathrm{Q}_{1}, \cdots, \mathrm{Q}_{n_{3}}) \subset \mathcal{C}} 
\prod_{k=1}^{n_{3}} c \hspace{0.05 cm} e_{0}^{2} \hspace{0.05 cm} e^{-\frac{3}{16} p_{\lambda} \tau(\mathrm{Q}_{k})}\Big].
\end{aligned}
\end{equation} 
Following lemma 25, \cite{D1} (there exists a constant $b$, such that 
$\sum_{\text{X} \cap \mathcal{C} \neq \oslash} e^{-a \tau(\text{X})} \leqslant b$), 
the first bracketed term is bounded by $e^{c \hspace{0.05 cm} e_{0}^{2-2\epsilon} b}$, the
second bracketed term is bounded by $e^{c  e_{0}^{\frac{1}{4} - \frac{\epsilon}{2}} b}$ and the third bracketed term is bounded
by $e^{c \hspace{0.05 cm} e_{0}^{2} b)}$. 
Then for some $\kappa_{7} = \frac{\kappa_{6}}{2}$  
\begin{equation}
||K^{\#} (\mathcal{C}, \eta)||_{w} \leqslant c \hspace{0.05 cm} e_{0}^{\frac{1}{2} - \epsilon} 
 \hspace{0.05 cm} e^{-\kappa_{7}\tau(\mathcal{C})}.
\end{equation}
Since $K^{\#}(\mathcal{C}, \eta)$ is small and has an exponential decay, we can use the standard procedure, 
see for example \cite{D1}, and write
\begin{equation}
\sum_{\{\mathcal{C}_{l}\}} \prod_{l} K^{\#}(\mathcal{C}_{l}, \eta) = e^{\sum_{\mathcal{C}} E(\mathcal{C}, \eta)}
\end{equation}
where
\begin{equation}
 E(\mathcal{C}, \eta) = \sum_{n=1}^{\infty} \sum_{\{\mathcal{C}_{1}, \cdots, \mathcal{C}_{n}\} : \cup_{l}\mathcal{C}_{l} = \mathcal{C}} 
 \rho^{T}(\mathcal{C}_{1}, \cdots, \mathcal{C}_{n}) \prod_{l} K^{\#}(\mathcal{C}_{l}, \eta)
\end{equation}
where $\rho^{T}$ is a function having a property that
$\rho^{T}(\mathcal{C}_{1}, \cdots, \mathcal{C}_{n}) = 0$ if $\mathcal{C}_{l}$ can be divided into disjoint sets and 
from \cite{D1}
\begin{equation}
||E(\mathcal{C}, \eta)||_{w} \leqslant  c \hspace{0.05 cm} e_{0}^{\frac{1}{2} - \epsilon} 
\hspace{0.05 cm} e^{-\kappa_{7}\tau(\mathcal{C})}.
\end{equation}
Rewrite the generating functional  
\begin{equation}
\mathrm{Z}[\eta] = \Big(\frac{2\pi}{e_{0}}\Big)^{-|\mathbb{T}| + 1}
\Big( \frac{\text{det}\hspace{0.1 cm}C^{\frac{1}{2}}}{\text{det}\hspace{0.1 cm}S}\Big)
\hspace{0.1 cm} \mathrm{Z}_{0}(\mathbb{T}) \hspace{0.1 cm} e^{\sum_{\mathcal{C}} E(\mathcal{C}, \eta)}
\end{equation}
and take the logarithm
\begin{equation}
\text{log} \hspace{0.05 cm} \mathrm{Z}[\eta] = \sum_{\mathcal{C}} E(\mathcal{C}, \eta) +
\text{log} \hspace{0.05 cm} \Big[\Big(\frac{2\pi}{e_{0}}\Big)^{-|\mathbb{T}| + 1}
\Big( \frac{\text{det}\hspace{0.1 cm}C^{\frac{1}{2}}}{\text{det}\hspace{0.1 cm}S}\Big)
\hspace{0.1 cm} \mathrm{Z}_{0}(\mathbb{T})\Big]. 
\end{equation}

\section{Mass Gap}

\subsection{Correlation functions}
The truncated correlation function is defined as
\begin{equation}
\begin{aligned}
\langle \bar{\psi}_{\beta}(y_{2}) \psi_{\alpha}(y_{1}) \rangle - 
\langle  \bar{\psi}_{\beta}(y_{2})\rangle \langle \psi_{\alpha}(y_{1}) \rangle
&= \frac{1}{e_{0}^{2}} \left. \frac{\delta}{\delta \bar{\eta}_{\alpha}(y_{1})} \text{log}\hspace{0.05 cm} 
\mathrm{Z} [\eta,\bar{\eta}] \frac{\delta}{\delta \eta_{\beta}(y_{2})} \right\rvert_{\eta = \bar{\eta} = 0} \\
&= \frac{1}{e_{0}^{2}} \left. \frac{\delta}{\delta \bar{\eta}_{\alpha}(y_{1})} \sum_{\mathcal{C}} 
E(\mathcal{C}, \eta,\bar{\eta}) \frac{\delta}{\delta \eta_{\beta}(y_{2})} \right\rvert_{\eta = \bar{\eta} = 0}
\end{aligned}
\end{equation}
$E(\eta,\bar{\eta}) = \sum_{\mathcal{C}} E(\mathcal{C}, \eta,\bar{\eta})$.
There are no external bosonic sources. All the bosonic fields are integrated out.  
Thus, we have a power series of $E(\eta,\bar{\eta})$ as
\begin{equation}
E(\eta,\bar{\eta}) = b_{0} + \sum_{z_{1},z_{2}} \eta(z_{1}) \hspace{0.05 cm} 
b(z_{1}, z_{2}) \hspace{0.05 cm} \eta(z_{2}) + \text{higher order terms}
\end{equation}
with
\begin{equation}
||E(\eta,\bar{\eta})||_{w} = 
\sum_{m \geqslant 0}   
\max\limits_{x \in X} \max\limits_{1 \leqslant j \leqslant 2m} 
\sum_{\substack{ z_{1}, \cdots, z_{2m}  \\  z_{j} = x}} e^{\kappa t(z_{1}, \cdots, z_{2m})} \hspace{0.05 cm} |b(z_{1}, \cdots, z_{2m})|.
\end{equation}
Note that
\begin{equation}
\begin{aligned}
\sum_{z_{1},z_{2}}  \eta(z_{1}) \hspace{0.05 cm} b(z_{1}, z_{2}) \hspace{0.05 cm} \eta(z_{2})
= \sum_{y_{1},y_{2}} & \sum_{\alpha,\beta}  \bar{\eta}_{\alpha}(y_{1}) \hspace{0.05 cm} 
b_{\alpha,\beta} (y_{1},y_{2}) \hspace{0.05 cm} \eta_{\beta}(y_{2})  \\
& + \sum_{y_{1},y_{2}} \sum_{\alpha,\beta} \eta_{\beta}(y_{1}) \hspace{0.05 cm} 
b_{\beta,\alpha} (y_{1},y_{2}) \hspace{0.05 cm}  \bar{\eta}_{\alpha}(y_{2}).
\end{aligned}
\end{equation}
The right hand side depends on $\eta(z_{1}), \eta(z_{2})$ with $\{z_{1},z_{2}\} \in \mathcal{C}$.
When we take the derivative term by term only those terms with $\{z_{1},z_{2}\} \in \mathcal{C}$
contribute. Thus,
\begin{equation}
\begin{aligned}
\left. \frac{\delta}{\delta \bar{\eta}_{\alpha}(y_{1})}  
E(\eta,\bar{\eta}) \frac{\delta}{\delta \eta_{\beta}(y_{2})} \right\rvert_{\eta = \bar{\eta} = 0}
&= \left. \frac{\delta}{\delta \bar{\eta}_{\alpha}(y_{1})} \sum_{\mathcal{C}\supset \{y_{1},y_{2}\}} 
E(\mathcal{C}, \eta,\bar{\eta}) \frac{\delta}{\delta \eta_{\beta}(y_{2})} \right\rvert_{\eta = \bar{\eta} = 0} \\
&= \sum_{\mathcal{C}\supset \{y_{1},y_{2}\}} b_{\alpha,\beta} (y_{1},y_{2}) - b_{\beta,\alpha} (y_{1},y_{2})
\end{aligned}
\end{equation}
Since $\eta = \bar{\eta} = 0$ only one term survives; sum over $\alpha,\beta, y_{1}, y_{2}$ collapses to 
a single term for every $\mathcal{C}$.  

We have shown that 
 \begin{equation}
||E(\mathcal{C}, \eta)||_{w} \leqslant 
c \hspace{0.05 cm} e_{0}^{\frac{1}{2} - \epsilon} \hspace{0.05 cm} e^{-\kappa_{7}\tau(\mathcal{C})}. 
\end{equation}
Let $\mathcal{C} \supset \{y_{1}, y_{2}\}$ and $t(y_{1}, y_{2})$ denotes the length of the
shortest tree joining sites $y_{1}, y_{2}$ such that $t(y_{1}, y_{2}) = |y_{1}-y_{2}|$. 
Then by definition $\tau(\mathcal{C}) = [r_{\lambda}]^{-1} |y_{1} - y_{2}|$.
Therefore, for $\mathcal{C} \supset \{y_{1}, y_{2}\}$,  
\begin{equation}
||E(\eta, \bar{\eta})||_{w} \leqslant 
c \hspace{0.05 cm} e_{0}^{\frac{1}{2} - \epsilon} e^{-\frac{\kappa_{7}}{[r_{\lambda}]} |y_{1}-y_{2}|}.  
\end{equation}
From (6.3) and (6.7) with $\kappa = 0$
\begin{equation}
|b( y_{1}, y_{2})| \leqslant  
e^{- \kappa t(y_{1}, y_{2})} ||E(\eta,\bar{\eta})||_{w} 
\leqslant c \hspace{0.05 cm} e_{0}^{\frac{1}{2} - \epsilon} \hspace{0.05 cm} e^{-\frac{\kappa_{7}}{[r_{\lambda}]}|y_{1}-y_{2}|}.
\end{equation}
Therefore,
\begin{equation}
\begin{aligned}
|\langle \bar{\psi}_{\beta}(y_{2}) \psi_{\alpha}(y_{1}) \rangle - 
\langle  \bar{\psi}_{\beta}(y_{2})\rangle \langle \psi_{\alpha}(y_{1}) \rangle|
&=  \frac{1}{e_{0}^{2}} |\sum_{\mathcal{C}\supset \{y_{1},y_{2}\}} b_{\alpha,\beta} (y_{1},y_{2}) - b_{\beta,\alpha} (y_{1},y_{2})| \\
&\leqslant  \frac{1}{e_{0}^{2}} \sum_{\mathcal{C}\supset \{y_{1},y_{2}\}}  
c \hspace{0.05 cm} e_{0}^{\frac{1}{2} - \epsilon}  \hspace{0.05 cm} e^{-\frac{\kappa_{7}}{[r_{\lambda}]} |y_{1} - y_{2}|} \\
&\leqslant  c \hspace{0.05 cm} \frac{1}{e_{0}^{2}} e_{0}^{\frac{1}{2} - \epsilon} \hspace{0.05 cm} 
e^{-\frac{\kappa_{7}}{2 [r_{\lambda}]} |y_{1} - y_{2}|} \sum_{\mathcal{C}\supset \{y_{1},y_{2}\}} 
e^{-\frac{\kappa_{7}}{2[r_{\lambda}]} |y_{1} - y_{2}|} \\
&\leqslant c \hspace{0.05 cm} e_{0}^{-\frac{3}{2} - \epsilon}  \hspace{0.05 cm}  e^{-\frac{\kappa_{7}}{2 [r_{\lambda}]} |y_{1} - y_{2}|} 
\hspace{0.05 cm} \sum_{\mathcal{C}\supset \{y_{1},y_{2}\}} e^{- \frac{\kappa_{7}}{2} \tau(\mathcal{C})} \\
&\leqslant c \hspace{0.05 cm} e_{0}^{-\frac{3}{2} - \epsilon}  \hspace{0.05 cm}  e^{-\frac{\kappa_{7}}{2 [r_{\lambda}]} |y_{1} - y_{2}|}
\end{aligned}
\end{equation}
where in the fourth line we have used $\tau(\mathcal{C}) = [r_{\lambda}]^{-1} |y_{1} - y_{2}|$.

\textbf{Acknowledgement} I would like to thank Jonathan Dimock for helpful comments and suggestions.

\end{document}